\documentclass[structabstract]{aa} 
\usepackage{graphicx}
\usepackage{txfonts}
\usepackage{natbib}
\begin{document}

%
   \title{A study of galaxies near low-redshift quasars.}

   \subtitle{$Master$ $thesis$ $in$ $astronomy.$}

   \author{Beatriz Villarroel}

   \institute{Uppsala Astronomical Observatory\\
              SE-751 20 Uppsala, Sweden\\
              \email{beatriz.villarroel.rodriguez@gmail.com}
             }
 
  \abstract
   {The impact of quasars on their galaxy neighbors is an important factor
   in the understanding of the galaxy evolution models.}
   {The aim of this work is to characterize the close 
   environments of quasars at low redshift (z$<$0.2) with the most
   statistically complete sample up to date using the seventh data release of the 
   Sloan Digital Sky Survey.}
   {We have used 305 quasar-galaxy associations with spectroscopically
   measured redshifts within the projected distance
   range of 350 kpc, to calculate how surface densities of galaxies,
   colors, star-formation rates, oxygen abundances,
   dust extinction and ionization changes as a function of the distance
   to the quasars. We also identify and exclude the AGN from our main galaxy sample 
   and calculate surface density of different galaxy types.
   We have done this in three different quasar-galaxy redshift difference ranges   
    $|\Delta$z$|<$ 0.001, 0.006, and 0.012.}
   {Our results suggest that there is a significant
   increase of the galaxy surface density around quasars, indicating 
   quasar formation by merging scenario. We also observe
   a gap in the surface densities of AGN and blue galaxies at a distance of 
   approximately 150 kpc from the quasar. We see no significant
   changes of star formation rate, color and ionization as function of distance
   from the quasar. From this investigation we cannot see any effects from quasars
   on their galaxy companions.}
   {}


   \maketitle
%
\tableofcontents

\section{Introduction}
The extremely luminous nature of quasars is believed to be driven by an active galactic 
nucleus due to accretion of material upon a supermassive black hole \citep[e.g.][]{Rees1984,LyndenBell1969}. 
Feeding such a black hole should affect the formation of galaxies in proximity to quasars 
\citep{SR1998,Gunn1979}, and this could play a big role in the structure formation scenarios derived from 
the Standard Cosmology. 

Could quasars be the results of mergers according to a hierarchical structure formation scenario 
predicted by Cold Dark Matter Models? For many years there have been controversial indications that 
both nuclear activity and star formation may be triggered or enhanced by interactions between galaxies. 
For instance, the presence of post-starburst populations in close quasar companions \citep{Canalizo1997}, 
morphological asymmetries with high star formation rates near the quasar host \citep{Gehren1984,Hutchings1984} 
and possible quasar companion galaxies \citep{Canalizo2001}. Ionized gas is also observed around 
quasars \citep[e.g.][]{Boroson1985}, and some authors have suggested that the quasars could even ionize the 
galactic medium up to several megaparsecs away \citep{Babul1991,Rees1988}. 
	
However, all these evidence-supported models and theories need much more substance if one wishes to raise 
their current status from speculation to fact. A great deal of work has gone into the study of close 
galaxy-galaxy pairs (less than 30 kpc) \citep[e.g.][]{Rogers2009} and of the surrounding environments of 
quasars at larger radii (less than 1Mpc). Many studies of the quasar environment using the Sloan Digital Sky Survey (SDSS)
database used galaxies with photometrically determined redshifts having large redshift uncertainties $\delta$z $>$0.025. 
These allowed more quantitative studies than qualitative, which underlines the importance of careful quasar-galaxy 
association studies with spectroscopically defined redshifts $\delta$z $<$ 0.001.
	
I have used a non-volume-limited sample drawn from the seventh data release (DR7) of 
the SDSS, containing more than 100,000 quasars, to study how quasars influence 
their nearby galaxies at low redshifts. In my thesis I examine how colors, surface densities, star formation rates, 
ionization of the galactic medium and certain spectral lines of galaxies near quasars change depending on the distance 
between the quasar and the neighbor. I have also investigated the active galactic nuclei (AGN) activity in host galaxies
near quasars. This study is done for a sample consisting of 305 quasar-galaxy associations at redshifts 
0.03 $<$ z $<$ 0.2 using Standard Cosmology ( $\Omega_{\Lambda}$=0.70, $\Omega_{M}$=0.30, H$_{o}$=70 km/s/Mpc.)
I hope to see in what ways quasars may impact surrounding galaxies, which is of major 
importance to understand AGN-feedback models and the formation of the most luminous structures in our Universe. 

\section{Background Theory}

\subsection{A Quasi Stellar Object}
The most luminous galaxies in the Universe were discovered in 1963 \citep{Schmidt1963}. 
Their point-like appearance made them easily confused with ordinary stars for astronomers 
while their spectra were clearly different. They got the name QSO ('Quasi Stellar Object') 
from their stellar appearance and those associated with strong emission
were initially called 'Quasar' ('QUAsi StellAr Radio source'), a name that 
later stuck with all types of QSO. The first quasar to be observed was 3C273 
at a redshift of z=0.158 and showed an optical spectrum similar to Seyfert 1 galaxies. Ten percent (10\%) 
of quasars are radio-loud and are often located in elliptical or interacting galaxies 
and many also display evidence of radio jets. The radio-quiet variant is instead located both in spiral and elliptical hosts. 
About 50\% of all QSO host galaxies show signs of tidal interactions and many of them have a close companion \citep{Canalizo1997}.

Most of the strong quasars can be found at very large redshift yet remain 
still observable to us. This means they have extremely large luminosities. 
Their bolometric luminosity can reach $L=10^{48}$ erg / s, which is one hundred times 
stronger than the luminosity of our own Milky Way galaxy. Moreover, quasars shine 
not only in optical but also in infrared, x-rays and gamma rays. Their blackbody spectrum peaks in the ultraviolet, 
but due to their physical structure (see section 2.2) and the far distances from us, we can detect them in 
optical wavelengths. The luminosity of many quasars is highly variable and we can see changes of luminosity happen on 
time-scales as short as only a few hours. This suggests that most of this energy output comes from a tiny region 
comparable to the size of our own Solar System.
 
The highest density of quasars can be found at redshift z $=$ 2-3, while now there are no strong 
quasars with $L=10^{48}$ erg / s left in our Local Universe (z$=$0). At higher redshift, not only the 
number density was higher because the scale of the Universe itself was much smaller, but also because the 
number of strong quasars was much larger than today.

Because of their luminosity and ability to be observed at huge distances, quasars are excellent probes for 
testing the cosmic reionization that happened somewhere around the z $>$ 6. The epoch of reionization was 
the period when the neutral hydrogen in the intergalactic medium got ionized by the first photons in the 
Universe and in this way terminated the previous Dark Ages. Traces from this ending period of neutral 
intergalactic medium should be observed as absorption of light ('Gunn-Peterson effect') in the
quasar spectra. It is even hypothesized that the quasars themselves could have caused 
the reionization \citep{Kramer}. Quasars are also interesting for the study of structure formation in the Universe since 
they are extremely luminous objects with mysterious evolution. But to understand 
what potential role they could have in our Universe it is important to first discuss the details of the quasar phenomenon. 

\subsection{The AGN engine}

Quasars are thought to contain the most energetic active galactic nuclei
observed in nature. Other types of AGN will be described in section 2.2.3. 
That AGN are small in size can be deduced first from the point-like appearance on 
many telescope images, but also from the argument about the high variability 
of many AGN.

	If one assume changes of luminosity happens everywhere at an astronomical 
	sphere simultaneously, one can expect that the light from the end closest 
	to us will reach our eyes first. The light from the side furthest from us 
	has a longer path to travel and will thus reach us a bit later. This sets a 
	size constraint on the area of luminosity change corresponding to R=c $\delta$t, 
	where $\delta$t is the variation in time, c the speed of light and R, the radius 
	of the source. The typical values of an AGN is of the same scale as our 
	Solar System.  An AGN hides a huge luminosity in this very small volume and different 
	photon energies seem to appear in regions of different size.
	
	The model with the best observational evidence has an accreting supermassive black hole (SMBH) as
	the engine of AGN. A standard sized AGN SMBH has a mass close to $10^{8}  M_{\sun}$ while 
	the black hole itself is surrounded by an event horizon located at the so-called 
	Schwarzschild radius \citep{Hawking}. Any light inside the event horizon will 
	not be able to escape.  
	
	Outside the SMBH event horizon is an accretion disk. The accretion disc is 
	created when a number of gas clouds around the SMBH accelerates towards it 
	because of the gravity of the black hole. These gas clouds collides one with 
	another and in the collision they lose kinetic energy which goes into heat. 
	Since they also slow down with the loss of kinetic energy, they fall into 
	smaller circular orbits near the event horizon. As the gas clouds continue to lose angular
	 momentum they spiral inward toward the SMBH itself. Viscous gas 
	 clouds in orbits around the black hole will experience a friction between each other and this 
	 causes further heating and redistribution of clouds into smaller and smaller 
	 orbits around the black hole. An accretion disc is suddenly visible around the black hole. 
	 The temperature of the gas keeps increasing and the newly formed accretion disc becomes the 
	 luminous source of an AGN. The luminosity of the accretion disc will depend on the rate 
	 of infalling matter into the black hole.
	 
	Does this argument lead to the notion that a SMBH can increase in size, the accretion disk getting 
	hotter and hotter, and hence an AGN will be indefinitely luminous? The answer is no. 
	The limit amount of power that can be produced around such an ultramassive 
	black hole is called the Eddington limit. The limit comes from the fact that when a SMBH gets thicker and hotter 
	the accretion disk exerts a radiation pressure that grows larger the more luminous the AGN is. 
	At the Eddington limit the radiation pressure outward from the accretion disk balances 
	the flow of infalling matter. Therefore accretion of material on to the SMBH stops, and the AGN cannot get fueled.
	
	Some quasars (and radio galaxies, see section 2.5) also produce radio jets. Why some produce jets, while 
	others do not is a question yet poorly understood \citep[e.g.][]{Kadler}. Neither does one know what the 
	ejected material is, even though it is agreed that they have to be overall electrically neutral. 
	The jets that are beamed in opposite direction are assumed to be aligned with the axis of rotation 
	of the accretion disk. At the end of the jets (at least the expected jets), one can see dominant 
	radio lobes often accompanied by regions with enhanced star formation. Sometimes we see only one jet, 
	while other times two jets are observed. This is suggested to be a phenomenon of so-called 'relativistic beaming', 
	that is dependent on angle and rate of jet. The jets that arise from AGN are extremely powerful plasma jets with
	velocities close to the speed of light. The relativistic effect can change the apparent luminosity of a jet, 
	and a jet close to our line of sight could appear having hundreds of times higher instrinsic luminosity, 
	while its counterjet would seem much weaker. 
	
\subsection{A Basic AGN}

The central engine (SMBH, accretion disc and possible jets) were thus just described. 
Now let us look in more detail what else an AGN needs to have. 

We think there is a donut-like dust torus around the engine with a hole in the middle. 
This dust torus is essential to explain how the emitted UV or x-ray emission from AGN appear as  
IR emission actually observed from many AGN. 
	
The existence of both broad-line regions (BLR) and narrow-line regions (NLR) in different types 
of AGN makes it necessary to include these into the model of an AGN. The BLR region, is found in the hole 
of the torus, while the NLR is found outside the engine region. 
What determines if a region will show narrow lines or broad lines, and presence or absence of 
forbidden lines, is the density (that determines the forbidden line presence) and the 
motion of the gas clouds which affect spectral line widths. BLR typically have dense fast-moving 
clouds that cause Doppler broadening of the lines, while NLR consist of slow low-density clouds. The faster clouds are naturally therefore 
considered to be those located nearest the black hole. The BLR has a temperature of the magnitude 
T$=10^{4}$ K and orbital speeds of larger than 1000 km/s. When viewing a galaxy face-on one should 
observe the BLR. 
	
The NLR has much lower orbital speeds between 200-900 km/s and is the region in which the torus is 
embedded, spanning several kiloparsec around the AGN engine. The NLR is therefore always in view
no matter how we view the galaxy.
	 
The typical lines seen in an AGN spectrum originates from ordinary hydrogen gas clouds in the 
NLR and BLR, that upon illumination with UV light (or x-rays) create both Balmer lines and 
forbidden lines like [\ion{N}{ii}]6585 and [\ion{O}{iii}]5007, that are typical emission lines in galaxies
with ionized gas. Since an AGN is a strong source of ionization, these emission lines are particularly 
strong in AGN.

\subsection{Other types of AGN}

Before we continue we shall have a look at the other type of AGN to be able to 
easier understand the concept of AGN unification that follows in section 2.6 
We list here short descriptions of other types of AGN.

\subsubsection{Seyfert Galaxies}

Carl Seyfert discovered this class of active galaxies in the early 1940s \citep{Seyfert}.
What he saw was bright point-like sources in some spiral galaxies. We now know 
that the Seyfert galaxies are one type of AGN that shows a variability in luminosity and 
excess IR radiation. There are two major categories of Seyfert galaxies: Seyfert I and Seyfert II. They differ 
in the presentation of their respective BLR and NLR regions (see 2.3 for the concepts). 
The Seyfert I galaxies, show prominent broadening of H$\alpha$, H$\beta$ and other Balmer lines 
arising from hydrogen clouds that have been illuminated by UV-radiation or X-rays. Also low-probability 
of occurrence lines, so-called 'forbidden lines' such as [\ion{O}{iii}]5007 and [\ion{N}{ii}]6585 are strong in spectra of Seyfert Is. 
These 'forbidden lines' are narrow in their spectra. Seyfert Is are typically efficient x-ray emitters. 
Seyfert II galaxies have also narrow 'forbidden' lines like Seyfert Is, but differ since the 
permitted Balmer lines instead are narrow here with full-width half maximas (FWHM) around 500 km/s, in comparison to 
Balmer lines in Seyfert Is that has FWHM around 1000-5000 km/s. 
No x-rays are seen in Seyfert IIs. There are also many Seyfert types in between these two categories, 
such as Seyfert 1.5. It is commonly thought that Seyfert I and Seyfert 
II galaxies are the same type of galaxy with different viewing angles. The Seyfert I could be 
described as a view where the broad line region is visible together with x-rays, contrary to the situation in Seyfert IIs.

\subsubsection{Radio galaxies}
These AGN are very bright at radio wavelengths, and especially in the nucleus. 
They are much less common than Seyfert galaxies in the nearby Universe. There seem to be two types of radio galaxies:
\begin{enumerate}
	\item Those that emit radio jets from extended radio lobes, often several hundred kpc away from the AGN core.
	\item Those that emit radio waves from the compact core or the galaxy halo. They are also believed 		
	to contain small radio jets.
\end{enumerate}
Furthermore, there exist both Broad Line Radio Galaxies (BLRG) as well as Narrow Line Radio 
Galaxies (NLRG). The BLRGs are often found in 'N galaxies' that are radio galaxies with bright 
star-like nuclei, while the NLRGs are found in giant elliptical galaxies. 

\subsubsection{Blazars}
A luminous, but highly variable class of AGN are the blazars that 
have high polarization and display 'superluminal' features of motion \citep{Biretta}. The 
blazars appear to be radio loud quasars with their jet along our line of sight. 
They are also located in elliptical galaxies like radio galaxies and radio-loud quasars. 
A major subclass are called Optically Violent Variables (OVV). 
They have slightly stronger emission lines and are often located at higher redshifts.

\subsubsection{BL Lac objects}

BL Lacs take their name from BL Lacertae which was thought to be a variable star constellation. 
The spectra later revealed that it was not a star, but rather a low power radio galaxy. 
Their emission lines are weak or non-existent. BL Lacs are also located at 
fairly low redshifts in massive and spheroidal host galaxies. In AGN unification they are 
thought to be a weaker sort of blazar.

\subsubsection{LINERs}

Low-ionization nuclear emission-line regions (LINERs) occur in a relatively large fraction of nearby galaxies
of different morphologies and luminosities. These regions have an emission-line spectrum with low-ionization
states. The higher ionizationed emission lines are weak or absent. LINERs are subjects of scientific debate,
since it unsure whether these nuclear emission regions arise from AGN activity or from star-forming
regions. Neither is the mechanism behind this low ionization known, and both shockwaves and UV light
are argued to be the main reason. LINERs are commonly referred to as AGN in scientific literature.

\subsection{AGN unification}

To easier understand how different AGNs might be related scientists have tried to make a unified model in 
order to combine distinct classes of AGN in accordance to certain principles. The basic principle in 
AGN unification is that all AGN are the same type of object, but they might have different accretion rates, 
different masses and viewing angle. 

There are two facts that serve as strong evidence for the model of AGN unification. The first is the 
comparison of hydrogen emission lines from broad and narrow line regions in AGN. If we consider the 
hydrogen ionization be caused by continuum radiation of AGN, the luminosity of the H$\alpha$ line 
should be proportional to the luminosity of the featureless continuum at the wavelength 4800 \AA\
since both should have the same origin. When the values of these two parameters for QSOs, Seyferts and radio galaxies are plotted against 
each other (on a logarithmic scale) the same relation appears for all types of objects \citep{Carroll}.

The second fact, is the discovery of Antonucci and Miller in 1985, where they saw that a particular Seyfert 2 galaxy observed with 
polarized light had a spectrum very similar to Seyfert 1 galaxies. This implied that the galaxy they 
observed was something very close to a Seyfert 1 galaxy, but that had to pass through optically 
thick material on the way, thus making the viewing angle something crucial on which the unification 
models are based.
	
The unification models cannot explain the differences between radio-loud and radio-quiet objects, but  
separates these into two classes. For the radio quiet AGNs one has proposed that type 2 Seyferts are similar 
to type 1 Seyferts, but having different viewing angle. Radio-quiet QSO are similar to the type 1 
Seyferts, but have a much larger accretion rate. For the radio-loud objects like blazars 
we see the torus face on and look directly into the jet, making it appear much more variable.

\subsection{Theories of Quasar formation}

There are two major pathways we can follow when discussing quasar formation. The first path, leads to the idea of 
primordial black holes formed early (at z$>$ 10 - 15) in the Universe \citep[e.g.][]{HaimannLoeb2001,Schneider2002} when population 
III stars collapsed within dark matter halos. Some claim \citep{Schneider2002} that these seed could already slowly accrete 
material at rates close to Eddington rate and at the furthest redshifts we present day can observe, already have the 
masses M$\backsim 10^{8} M_{\sun}$.

The second path, follows the idea that mergers between smaller objects form supermassive 
black holes and that those quasars we observe beyond redshift z $>$ 6 and closer came 
from multiple mergers between gas-rich galaxies.  Indeed have 50\% 
of all quasar host galaxies some features that show previous mergers. For many years 
there have been controversial indications that both nuclear activity and star formation 
may be triggered or enhanced by interactions between galaxies. For instance, the presence 
of post-starburst populations in close quasar companions \citep{Canalizo1997}, morphological 
asymmetries with high star formation rate around the quasar host \citep{Gehren1984,Hutchings1984} 
and possible quasar companion galaxies \citep{Canalizo2001}. 
	
\subsubsection{Modeling the number density}
There are analytical alternatives to N-body simulations when we wish 
to explore whether mergers of normal galaxies in the past could reproduce 
the number density of observed quasars, which also means reproducing the 
decrease of bright quasars from z $=$ 2 towards higher redshifts. These models also need 
to agree with our current observations.

According to the Standard View nowadays, the gravitational growth of structures 
came originally from small, primordial density perturbations during the matter-radiation 
equality era around z$\backsim$3800. Depending on the size of these perturbations, dark matter halos 
could form where gas could collapse from Jeans instability and form objects inside. 
Press-Schechter (1974) did an analytical model that could predict the fraction of the 
volume of the Universe that collapse into objects with minimum mass M. The advantage of using 
Press-Schechter formalism is that it can spit out number densities of virialized objects with 
certain masses M by counting on the fraction of matter that is trapped into dark matter halos 
with the mass M'. The formalism was tested and agreed with the much more time-demanding N-body 
simulations. The only problem was that it underrepresents the number densities by a factor of two, 
which usually is corrected by simply multiplying the calculated results by two. Different structure 
formation scenarios and different redshifts do need to be taken into account when using this formalism.
	
Carlberg (1990) implemented the Press-Schechter formalism to estimate the number density of optically 
bright quasars \citep{Carlberg1990}. The idea he tried to explore was whether quasars could be a bi product of 
mergers. One reason for the decline after z$>$ 2-3 would be that at much higher redshift the 
galaxies are so rare that the probability of relevant mergers occurring is low. At the peak z$\backsim$2-3 
the merger rate is at its highest, while at more recent times the 
Universe is larger with the objects further apart. Further, only a small 
fraction of mergers can create a quasar since not all mergers will include black holes or enhanced 
possibilities for fuel transport or fuel supply. Carlberg's model also required certain assumptions concerning 
the relations between the mass of the host galaxy and the luminosity of the quasar as well as relations between 
the mass of the host and the life-time of the quasar. All this would be needed in order to predict the quasar 
luminosity function accurately.
	 
The rate of change would increase when smaller objects merge into halos, but also 
decrease when these objects merge into even larger structures. This is a hierarchical 
model scenario, and would be able to continue as long as the newly created halos also can cool down in time. 
	
Carlberg's model made fairly good predictions or the evolution of high-luminosity quasars, but the great uncertainties 
in the modelling parameters resulted in that the paper itself "only" could put constraints on quasar evolution via merger scenario,
rather than answer any direct questions.

\subsection{Evolution and Death of a Quasar}

The quasar population had its greatest dominance at a redshift z $=$ 2-3 when the highest number density of
high-luminosity quasars existed. Considering the large numbers of SMBH of masses M=$10^{8} M_{\sun}$ residing inside the galaxies, astronomers 
should be able to observe many galaxies with relic SMBH. This is because we believe that once a black hole
has formed it will
take very long time to evaporate (see articles on Hawking radiation). Indeed, we think many of the normal elliptical 
galaxies and spirals today are dead AGN. But how did they die? The most striking 
clue of evolution in quasars is the luminosity evolution, which shows quasars can reach extreme luminosities and numbers 
around the peak z $=$ 2-3, down to $M_{r}<$-32. Still, many selection effects such as an unseen faint quasar population of unknown number, 
and number evolution at high redshift can complicate the puzzle.

Three major elements are responsible for making the quasar shine. The first is the mass of the 
black hole, that is strongly connected to the star formation in the bulge. There is a very tight correlation shown between the 
mass of the black hole, and the stellar velocity dispersion in the galactic bulge. This shows the building of a host galaxy 
at the same time as black hole formation \citep{Granato2001}. While a supermassive black hole is able to accrete 
material efficiently, the same gas is used for star formation around the AGN. The second thing needed is fuel, that is gaseous 
components. The third factor is that the gas is able to fall inwards towards the black hole by loss of angular momentum, 
which is measurable by the galaxy's accretion rate. The common view of quasar evolution is that quasars fuel themselves 
with surrounding gas-rich galaxies, making their SMBH heavier and heavier. At a point where the Eddington limit is reached, the radiation pressure from the 
SMBH heats up the gas until no new gas infall is possible-- making the transport impossible. When the gas cannot
fall in, the quasar starts losing luminosity independently of the size of the black hole, and that gas cannot any longer 
be used to form stars in the host galaxy. Since the surrounding region takes a very long time to cool this means the gas in the accretion 
disc blows away and the accretion disc stops emitting very energetic photons. When the region is cold enough to allow new material to infall, 
our universe will have expanded so much that either all field galaxies will be too far away from each other
to merge, or the galaxies will be trapped into growing clusters of galaxies. 

In the non-quasar AGN population cosmic downsizing has been noticed, where the number density of low x-ray 
luminosity AGN is only high at low redshift. A similar peak for stronger x-ray AGN lies at considerably higher redshifts. 
This could mean many quasars still experience some SMBH growth  \citep{Fan2001}. Lowered accretion could be 
the possible cause for this AGN downsizing, but some groups have reported no evolution in other 
x-ray properties than x-ray luminosity \citep{Vignali2005} which means no changes in accretion rate. 
That means that AGN downsizing must either be driven by a decrease in accretion rate, or by a 
decrease of the characteristic mass scale.
	
From observing the most distant quasars one can learn how the metallicity in those 
objects has changed over time. The proper timer to study the early star formation history is given 
by the ratio between iron versus $\alpha$ elements \citep{HaimannFerland1993} that is proportional to the ratio between 
materials emitted in core-collapse and Type Ia supernovae. Plotting the Fe / Mg ratio in quasars against redshift 
has shown that the metallicity increases with redshift, counter-intuitive as it is \citep{Iwamuro2002}. 
Other groups claim to show no evolutionary difference in metallicity \citep[e.g.][]{Dietrich2002}. 
However, this can be interpreted in the light of how star formation history relates to stellar mass growth, where the 
highest mass galaxies are also those with the highest star formation \citep[e.g.][]{Noeske2007}, and also the highest supernovae rate 
resulting in very fast metal enrichment. The high-redshifts objects were thus not only luminous because of 
the AGN accretion, but also due to a much higher star formation rate in massive galaxies in the past.

\subsection{AGN feedback}

AGN feedback is a mechanism introduced to explain some peculiar observations. 
One hot topic within galaxy physics is how the red, giant elliptical galaxies 
at high redshift were formed. This can be approached by a proposed AGN feedback mechanism. 
There are essentially two different populations of normal, non-AGN galaxies; 
 red mostly quiescent galaxies with early-type morphology and blue star-forming galaxies 
 \citep{Strateva2001,Brinchmann2004}. These have made astronomers wonder whether there might exist some 
 mechanism that is quenching the star formation in galaxies, resulting in the proposal of various models 
 of feedback from an active galactic nuclei \citep[e.g.][]{TB1993,Hopkins2006,Sijacki2006}. This is an 
 alternative to the more conservative explanations,such as mergers between late-type galaxies e.g. 
 \citep{Barnes1992,Toomre1972} or the loss of gas due to ram pressure stripping \citep{Gunn1972}.
  
Also, why are smaller objects formed at lower redshifts? This 'downsizing' can be explained with AGN feedback. 
For x-ray quasars one has shown that the growth of the SMBH peaks at redshift z = 2  \citep{Silverman2007} and 
the fact that we observe more low-luminosity quasars at lower redshift supports 
the need of a mechanism that quenches both the star formation and the black hole growth at lower 
redshift. In both problems, the feedback assumes the quenching of star formation around an AGN.
Optimal feedback has been proposed to happen during early stages of massive spheroid formation 
\citep{Croton2006} where the AGN can quench star formation. Two modes have been suggested for AGN feedback: 
quasar mode feedback and radio mode feedback. In the quasar mode when the supermassive 
black hole reaches the Eddington limit it can blow out the surrounding gas and this way 
quench the star formation in the host galaxy or nearby galaxies. This will only happen 
when the mass of the SMBH gets saturated. Since there also is a relation between the star formation 
in the bulge and the mass of black hole for black holes below the Eddington limit, reaching the limit will 
stop the black hole growth but also the star formation when it blows away the surrounding gas. 
This mode could account for the dominant old population of massive galaxies.
	
The 'radio mode'  is a bit more ambiguous in its behavior. It allows the AGN to stop 
star formation in the galaxy, but can also trigger it initially. Close to the AGN 
a jet-driven cocoon arises from the overpressure of interstellar clouds creating an overpressure around
the protodisk and the protospheroid, which further increases the pressure and causes gravitational 
instability and thermal bubbles. This gravitational instability might initially be at redshifts 
larger than z$ > $3 leading to an increased star-formation rate. However, as the gas gets hotter 
with increased pressure it will get too hot around the AGN to form stars. While it is not directly 'quenching' the 
star-formation rate it inhibits new star formation. 
This includes AGN with accretion rates considerably smaller than galaxies near the Eddington limit have.
So while the 'quasar mode' should quench the star formation, the 'radio mode' initially increases it 
and later inhibits new stars from forming around an AGN of any luminosity.
	  
The AGN feedback could be important in the evolution of ordinary galaxies as it allows 
blue star-forming galaxies via a 'green' transition valley with AGN, transform into heavy red, 
old, early-type galaxies.

\subsection{The role of Quasars in Structure Formation}

To understand what role a quasar might have in structure formation we must briefly discuss 
the cosmological scenarios in which it could be involved. The cosmology we assume 
in our study is the Hot Big Bang Theory model. Some of its key points to have in mind is that 
the Hot Big Bang Theory model predicts the Universe to have a finite age and that there is a 
cosmic expansion (that however is having a different rate at different stages of the Universe). 
For our structure assembly discussion we assume that in a very early stage the Universe was isotropic 
and that the temperature of matter and density was almost perfectly uniform.
 
The small amount of non-uniformity led to density fluctuations. Sometimes these regions with enhanced density simply 
grew together with the expansion of the Universe while others, when similar regions also had a sufficient size, 
instead collapsed into structures with baryons and dark matter. 
	
Furthermore, the collapsed enhanced-density region contained not only baryons, but for the most part 
consisted of non-baryonic dark matter. This collapsed region made up a body with baryonic 
mass in the centre, surrounded by a dark matter halo. This body has the same mass as the galaxy 
that soon formed from it. This type of collapse is called monolithic collapse. 
	
From models of Cold Dark Matter one can predict that the first galaxies had masses corresponding 
to$~10^{6} M_{\sun} $(compare to local galaxy masses of the order M=$10^{11}$ $M_{\sun}$). The hierarchical scenario predicts 
the creation of larger structures such as heavy galaxies, groups and clusters from the 
merging of smaller building blocks. Some elliptical galaxies are thought to be results of such mergers 
as there are good indications, such as the 'clock indicator' [$\alpha$/Fe], that the star formation in ellipticals
was initiated very quickly through mergers. Also, simulations have shown that merging smaller galaxies gives 
morphologies and stellar velocities similar to those of elliptical galaxies. Some models \citep{Hopkins2009} 
predict an evolutionary sequence for forming ellipticals from two star-forming galaxies that merge. 
Right before the merger they could be seen as a double-cored ultraluminous infrared galaxies (ULIRGs) and thereafter
merge into a quasar that blows away all gas. The end product after most gas has been consumed by star formation 
will be a red elliptical galaxy.
	
But how can you explain the anti-hierarchy observed in AGN and early-type galaxies over time, what we
refer to as downsizing? At higher redshifts we find more luminous and massive galaxies while at lower redshifts, 
many dwarf galaxies are observed and too few massive galaxies. There are two competing scenarios downsizing. 
The first one is through gravitational heating \citep{Dekel} that affects the galaxies with 
the most massive halos. The accretion of baryons into dark matter halos above a threshold mass
does that the baryons condense at the bottom of the potential well of the halo. This makes
the halo hot and infalling gas cannot cool and form stars. The star formation terminates through 
this gravitational heating, which leads to downsizing. The second way is via AGN feedback where 
an AGN is able terminate gas cooling and thus star formation in massive halos such as its host galaxy, 
or in a nearby galaxy.	

\subsection{Earlier studies of Quasar Environments}
Many earlier studies of field quasar environments have been performed and most of 
them seem to agree that even though quasars are not likely to reside in 
cluster-dense regions, small overdensities around brighter AGN or quasars are likely to be found. 
This is often argued to support the merging scenario behind quasar fueling \citep[e.g.][]{Serber2006}. Many of these studies use the Sloan Digital 
Sky Survey since it is currently one of the most powerful resources for the study of galaxies and other objects
on the sky. 

\subsubsection{Low-redshift studies}

Coldwell \& Lambas (2006) used a sample from the Sloan Digital Sky Survey (SDSS) DR5 \footnote{SDSS Data Release 5} 
of 2070 quasars whose redshift distribution they matched with a similar sample of galaxies. They analyzed the environments of 
quasars, clusters and ordinary galaxies. They compared different properties such as colors and star formation rate of 
galaxies as function of projected distance from the quasar, ranging from 0 to 3 Mpc. They also checked 
whether there are more AGN in environments of quasar compared to in other environments such as clusters 
or field galaxies. They found no difference, claiming that AGN activity is not affected by quasar 
presence. With the help of 'eClass' \footnote{A function based on Principal Component Analysis that can be used as
morphology indicator, see section 3.5} and concentration indices in the SDSS, they found a higher relative fraction of blue galaxies with 
disc-type morphology, indicating more late-type objects around 
quasars. They also showed that closer than 0.5 Mpc a higher star formation rate is found in quasar environments, 
compared to the environments of normal galaxies and clusters of galaxies \citep{CL2006}.

Canalizo and Stockton (1997) used the Keck telescope to investigate three QSO companions at low redshifts, 
ranging from 0.2 $<$ z $<$ 0.29. They made age determination of the stellar populations of these three QSO 
companions, and saw either present or passed starburst populations \citep{Canalizo1997}

Yee (1987) drew a small sample of 37 quasars of radio-quiet galaxies from the Palomar Bright Quasar Survey 
with redshifts 0.05 $<$ z $<$ 0.3. He concluded that 40\% of the quasars had a 
close companion (r $<$ 100 kpc) with a magnitude larger than $M_{r}$ $<$ -19. The magnitude distributions 
of these companions were in agreement with simulated distributions of arbitrary field galaxies, which showed that the proximity of radio-quiet quasars had no effect on companions. 
He could also demonstrate that the companions had no effect on the radio-quiet quasars \citep{Yee1987}.

Serber et al. (2006) drew a sample of 20000 quasars from the SDSS with magnitudes M$_{i}<$-22 and redshifts z$<$0.4. He then matched 10$^{5}$ galaxies from the photometric catalog with same average 
redshift and limit. They did not care about excluding foreground or background galaxies and simply assumed an 
average redshift of z $=$ 0.13. They investigated the projected distance range from 0 to 1 Mpc 
and found around the brightest quasars an overdensity within 1 Mpc that monotonically 
declined at larger distances. This has been argued to be one of the strongest supporting studies 
of merger-driven scenarios for quasar activity because of the discovery
of over-density of galaxies around quasars \citep{Serber2006}.

Li et al. (2008) used the SDSS to check for AGN with close neighbors from the photometric catalog. 
Using two point-angular correlation functions, they showed that even though the SFR was increased in AGN 
with close companions this was not associated to any enhanced nuclear activity due to the close companions, 
but simply to tidal interactions between the AGN and neighbors. They used the L([\ion{O}{iii}])/M$_{BH}$-ratio that 
is a good measure for black hole accretion rate or nuclear activity (when corrected for dust extinction)
to study whether the nuclear activity of the AGN was influenced by presence
of close companions or not. No influence was observed. The number of AGN and galaxies were not 
mentioned in the paper, nor upon which criteria the neighbors were selected \citep{Li2008}.

Strand et al. (2008) had a different approach to the quasar environment problem. They divided
up the quasar sample into two major subsamples: type I quasars and type II quasars
according to the AGN unification scheme (see section 2.5) and with the assumption that quasars
can be divided into two major types (type I and type II). They used 4234 spectroscopically confirmed type I quasars within the redshift range 
0.11 $<$ z $<$ 0.6. They also obtained a sample in the redshift limit 0.3 $<$ z $<$ 0.6 of 160 type 2 quasars drawn from a 
sample by Zakamanska et al. (2003). To these quasar targets, they matched a sample of galaxies with 
similar redshift distribution and applied a redshift cut $|\Delta z|$  = 0.05 concerning the difference between quasar 
and galaxy. What they wanted to do, is to compare the densities of galaxies with photometric redshifts around 
type I and type II quasars, and compare these to the densities around type I and type II AGN, out to scales of 2 Mpc. Among their findings was an 
increased overdensity for type II AGN compared to type I AGN, while in the upper redshift range 
(0.3 $<$ z $<$  0.6), the type 2 quasars and brighter type I quasars experienced the same overdensity on small distance relative to the
density at larger distance. More interestingly, was the use of the $|\Delta z|$  = 0.05 where
they nearly doubled their overdensities compared to if they did not take any individual redshift
differences between quasar and galaxy into account. This indicates the need of proper redshifts in order to 
observe overdensities around these brighter objects \citep{Strand2008}.

\subsubsection{High-redshift studies}

Hu et al. (1991) did use a narrow-band filter survey to investigate three radio-loud quasars with 
Ly-$\alpha$ companions at redshifts z $>$ 2.9. All three companions showed remarkable similarities in size, 
luminosity, morphologies and nebulosity. All three companions showed excessive ionization, suggested to come from the 
quasar and same continuum and emission nebulosity as low-redshift quasar companions. Hu et al. attributed the 
non-evolution of the metallicities in nebulae over redshift to be understandable in terms of quasar fueling as a result 
from tidal interactions of normal galaxies with the quasar \citep{Hu1991}.

Shen et al. (2008) used the DR5 quasar catalog to compare clustering of different types of quasars at 
redshifts up to z $<$2.5. They used in total 38208 quasars. They saw that in general there was no 
luminosity dependence of quasar clustering, that the clustering was independent of virial black
hole mass, and clustering was independent of the quasar color. On the other hand they saw that
radio-detected quasars are more clustered than radio-quiet.

\section{Data and methods}

The data for all quasar-galaxy associations was taken from the Sloan Digital Sky Survey (SDSS). 
For all objects we downloaded spectral line information, luminosities and redshifts. We also
downloaded all angular distances between quasar and galaxy for each quasar-galaxy association.

\subsection{The Sloan Digital Sky Survey}

The Sloan Digital Sky Survey is a database, accessible from anywhere in the world via a web browser, 
that has mapped a large number of objects on the Sky, imaging over 11663 deg$^{2}$ of the sky where 9380 deg$^{2}$ 
also have spectroscopy.

The SDSS uses a 2.5 m optical telescope at Apache Point Observatory, New Mexico. It has around 30 photometric CCDs with 
2048$x$2048 pixels in each imaging camera, along with two spectrographs. The photometric redshift survey has a 
photometric system with five filters: u,g,r,i,z, each detecting light at an average wavelength of 
3551 \AA, 4686 \AA, 6115 \AA, 7491 \AA, 8931 \AA.

The optical fibers used in the fiber spectroscopy are 3 arcseconds wide and must be separated by at least 55 arcseconds. 
This means that objects situated closer than 55'' cannot be detected by the spectrographs unless the image is located
at a fiber plate overlap area. The SDSS seventh data release (DR7) \footnote{The latest data release as of January 2010.}
has already mapped around 930000 galaxies (with a median redshift of 0.1) and 120 000 quasars.

The SDSS spectrograph can detect galaxies with a Petrosian apparent magnitude r$<$17.77. Quasars
are modelled with a point-spread function (PSF) and can reach a PSF magnitude i $<$ 19.1 (or PSF magnitude i $<$ 20.1 for objects at  z $>$ 2.3). 
On average quasars have been detected much further away than normal galaxies, up to redshifts z $>$ 5 - 6.
Galaxies can reach Petrosian r $<$ 17.77. The advantage of using spectroscopic redshifts is the small error of 
the redshift estimation, $\delta$z=0.001, compared to photometric redshift errors of $\delta$z=0.025.

The photometric catalogue of the SDSS has different magnitude limits from the spectroscopic catalogue. The photometric 
redshifts are estimated from colors. 
Colors of galaxies and quasars depend on redshifts. At the same time, 
synthetic colors can be modeled from spectral energy distributions (SED) of galaxies and quasars which can also provide information 
on spectral type and k-corrections. 
The photometric redshift errors are larger due to uncertainties in the SED and other systematic errors, and lies around $\delta$z=0.024. 
The advantages of using the photometric catalogue is that one can reach apparent 
magnitudes of Petrosian r $<$ 22 and thus maps many more objects \citep{Oyaizu} and further out in the space.

We have used the casjobs interface of the SDSS DR7 \footnote{http://casjobs.sdss.org} 
to search for quasar-galaxy associations. 

\subsection{Observed properties}

We list the properties that could be directly downloadable from the SDSS.

\subsubsection{Luminosities}

SDSS uses five filters: u,g,r,i,z. The apparent magnitudes in the SDSS are estimated by the use of model magnitudes.
These model magnitudes are calculated in the SDSS by first adjusting the surface brightness of the image of disk 
galaxies to an exponential profile. For elliptical galaxies one instead uses a de Vaucouleurs profile to calculate 
the magnitudes. The apparent magnitudes can further be corrected for galactic extinction \citep{Finkbeiner} and these are called 'dered' magnitudes. We downloaded the dered magnitudes for both our galaxies and our quasars.

\subsubsection{K-corrections}

Since light that gets redshifted also moves between filters an additional correction called 
"k-correction" needs to be adjusted for in our magnitude calculations. This term can for
ordinary galaxies be obtained from a templace fitting procedure performed to get photometric 
redshifts. Luckily, this is automatically done in the SDSS. From the PhotoZ table, we downloaded the 
k-corrections for our galaxy sample through the casjobs interface. The k-corrections for the quasar 
sample are calculated in next chapter.

\subsubsection{Spectral lines}

In the SDSS line profile information is available on spectral lines. These spectral values one obtains by 
first substrating the continuum flux from the spectra and after that fitting Gaussians to specific spectral lines. 
The dispersion $\sigma$ of the Gaussian, together with the height of the fitted Gaussian (h) and continuum
value(cont.), can be used to calculate the so-called equivalent width of the Gaussian (EW), 
that gives the line strength relative to the continuum. The equivalent width can be calculated as 

\begin{equation}
  Ew = \frac{\sqrt{2\pi}\sigma h}{cont.}
\end{equation}

The line flux (F), that we need for calculations such as the star-formation rates and the oxygen abundance, is calculated
as

\begin{equation}
  F = \sqrt{2\pi} \sigma h  
\end{equation}

\subsubsection{Redshifts}

Redshift is of great use for distance determination in astrophysics and to determine the movements of astronomical 
objects far away. They arise when electromagnetic radiation loses energy and gets shifted to longer wavelengths 
due to the Doppler effect, scattering or gravitational effects. This means that the light emitted from a particular 
source that moves away appears redshifted when it reaches the observer.

The redshift can be estimated as 
\begin{equation}
  z = \frac{\lambda_\mathrm{obs}-\lambda_\mathrm{emit}}{\lambda_\mathrm{emit}}
\end{equation}

where $\lambda_\mathrm{obs}$ is the wavelength the observer perceives, and $\lambda_\mathrm{emit}$ 
the rest-frame emission wavelength.

Different kinds of redshifts are expected to appear from general relativity theory such as 
relativistic Doppler effect (for light sources moving at a speed close to light), 
gravitational redshift (near ultramassive objects such as a black hole) and the cosmological 
redshift that arises due to the expansion of the Universe.

For objects near us the simple Doppler shift can be used together with Hubble's law to calculate the 
distance. When one is interested in objects further away or in more precise calculations 
one should calculate the luminosity distance using the cosmological redshift formula \citep{Kennefick2008}, 
that is dependent upon the cosmological model. In our study, we have assumed the $\Lambda$CDM-model. 

\bigskip

  $d_L(z;\Omega_M,\Omega_\Lambda,H_0) =$ 

\begin{equation}
= \frac{c(1+z)}{H_0}\int_0^z \left[ (1+z')^2(1+\Omega_M z')-z'(2+z')\Omega_\Lambda \right]^{-\frac{1}{2}} \mathrm{d}z'\\
\end{equation}

Our SDSS-objects had spectroscopic redshifts with confidence levels $z_\mathrm{conf} > 0.95$. In the SDSS the objects 
with spectroscopic redshifts have errors of $\delta z = \pm 0.001$. 

\begin{figure*}[htb!]
\includegraphics[width=16cm,height=10.5cm]{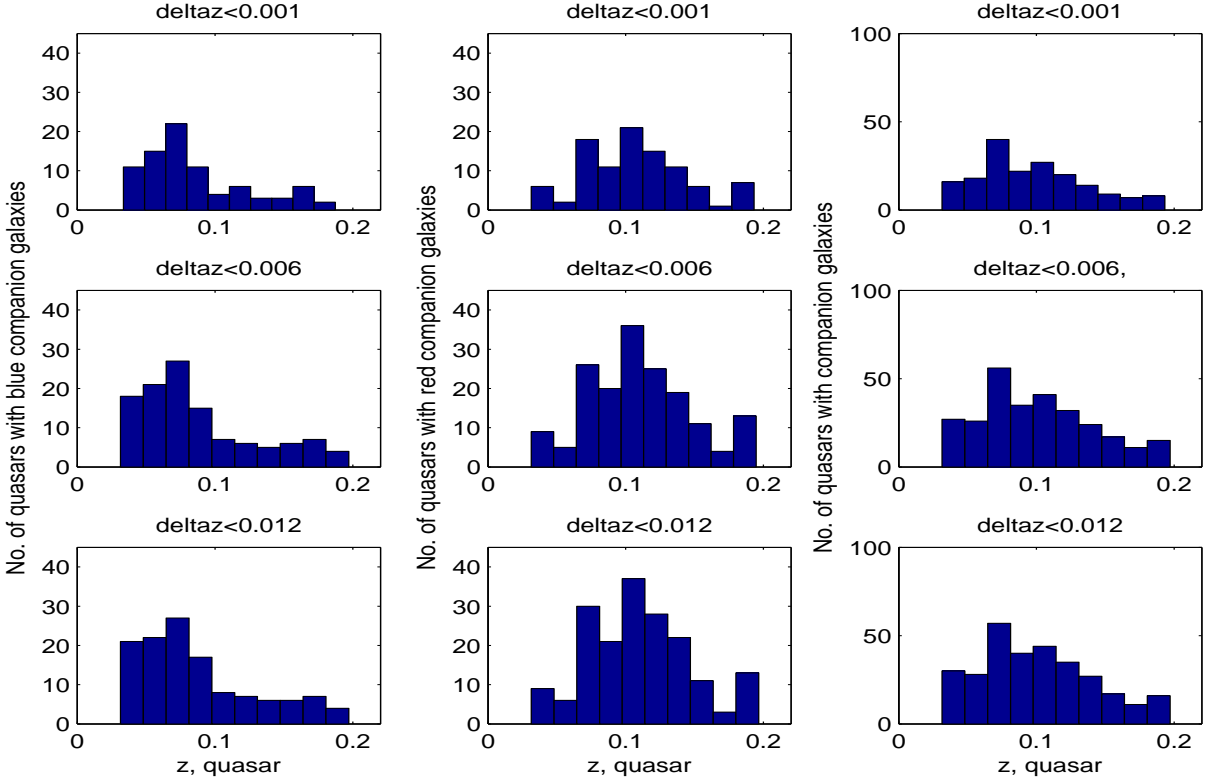}
\caption{Redshift distribution of our objects. Left panel shows the redshift distribution
of quasars with blue companion galaxies. Middle panel shows the redshift distribution
of quasars with red companion galaxies. Right panel shows redshift distribution of all quasars. 
Three different redshift difference cuts are pictured for the pairs.}
\label{NFig82}
\end{figure*}

\subsubsection{Distance between quasar and neighbor galaxy}

The casjobs-interface of SDSS has a function called fDistanceArcMin(ra1,dec1,ra2,dec2). 
This function calculates the distance in arcminutes between two objects 1 and 2 on the Sky, given 
their right ascension (ra) and declination (dec).  We used this function in the following way: 
for the QSO neighbor queries we divided the sample up in two parts (see appendix): a low redshift 
part $0.03 < z < 0.1$ where we searched for neighbouring galaxies up to fDistanceArcMin $< 10$ arcminutes 
(which at $z=0.03$ corresponds to a projected distance of 363 kpc) and a high-redshift part $0.1 < z < 0.2$ 
where we instead looked up to fDistanceArcMin $< 3$ arcminutes around the quasar (which at $z=0.10$ corresponds to 363 kpc). 
We limited the redshift difference between the quasar and the galaxy to be less than $|\Delta z|$  $< $0.012  in the queries. 
We chose this redshift cut because it is about half of the error on photometric redshifts in the SDSS ($\delta z < 0.025$), 
which makes it comparable to many other studies. It corresponds to a physical distance of 52 Mpc between galaxy and quasar.

Knowing both the physical distance to the quasar-galaxy association and their angular distance, we can easily calculate 
the distance between the objects, or rather the projected distance. Since Hubble's law can be used for objects 
with redshift $z < 0.2$, we used it to first calculate the distance in kpc to the quasar-galaxy association, assuming 
both galaxy and quasar to be at the redshift of the quasar. We take this approximation into consideration, see result 
section description.

\subsection{Quasar sample}

The quasars were extracted from the SDSS quasar catalogue, seventh data release (DR7). 
The casjobs table we used was QSOCatalogAll which contains all objects that 'smell' like a quasar \footnote{QSOCatalogAll includes all objects that have been flagged as quasars in the SDSS.} and 
includes the list of confirmed quasars. The quasars in the SDSS catalogues are selected such  that they have at least one emission line with full width at half maximum (FWHM) larger than 
1000 km/s and have highly reliable redshifts \citep{Schneider2003}. We flagged these for brightness \footnote{There is a problem of very bright objects in the SDSS. The very strong objects with r$<$17.5 get their properties
measured twice, the second time as fainter objects. This leads to duplicate entries of the same objects, why one should avoid objects that are too bright.},
blending \footnote{Blended objects are composite objects in the SDSS that are detected with multiple peaks within them. SDSS tries to deblend these objects, but when the attempts fail we get another multiple entries of the same object in the database.} and saturation \footnote{Saturated objects are those that had a poor photometry and that contain more than one saturated pixel. These can be galaxies that have superimposed stars on them.}and demanded the redshift reliability zconf$>$0.95.
 
The quasars we picked were within the redshift range $0.03 < z < 0.2$.

\subsection{Galaxy sample}

The neighbor galaxies were found with the SDSS casjobs Galaxy-view. This view contains objects from the PhotoPrimary table and includes all objects that are photometrically shown to be a galaxy. The galaxies 
were selected to have a projected distance smaller than 363 kpc from their quasar partner. This was done 
with the help of the function fDistanceArcmin. The galaxies with measured redshifts in the SDSS have all a Petrosian magnitude $r < 17.77$. We also flagged them for brightness, blending and saturation, with the same zconf $> 0.95$. For the galaxies we downloaded spectral line information, dered magnitudes, k-correction and eCoefficients 
for morphology  (see section 3.5). We demanded the redshift difference between a quasar and the galaxy association to be smaller than 
$|\Delta z|$  $<$ 0.012.

We obtained 366 quasar-galaxy associations from the SDSS from which we had to exclude many repetitions of the same associations. 
We ended up with 305 quasar-galaxy pairs.

\begin{figure}[htb!]
	\centering
		\includegraphics[scale=.6]{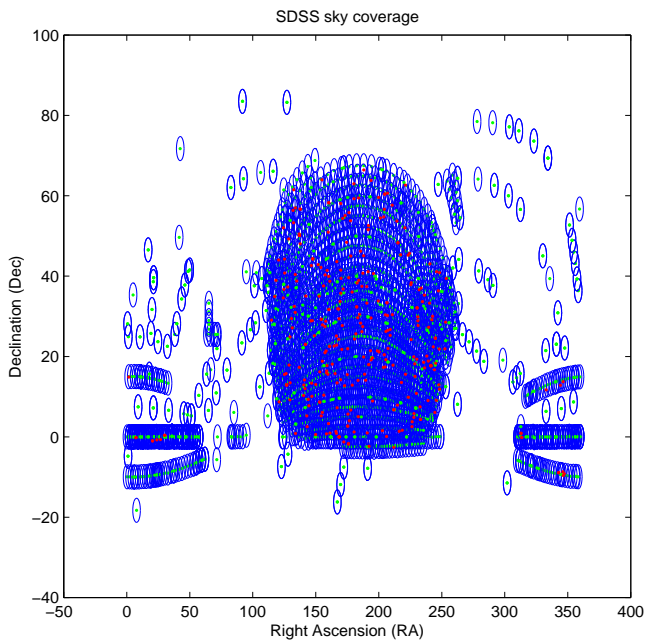}
	\caption{SDSS sky coverage. The ellipses are representing the spectroscopic plates.
	The green dots the middle of each plate. The red dots are representing the quasars.}
	\label{circle}
\end{figure}

Figure \ref{circle} shows where our quasar-galaxy associations were found in the SDSS on the sky. Figure 
shows a small area. The green rings represent the 
overlapping plates with their coordinates, while the red dots are our quasar-galaxy associations.

\begin{figure}[htb!]
	\centering
	\includegraphics[scale=.6]{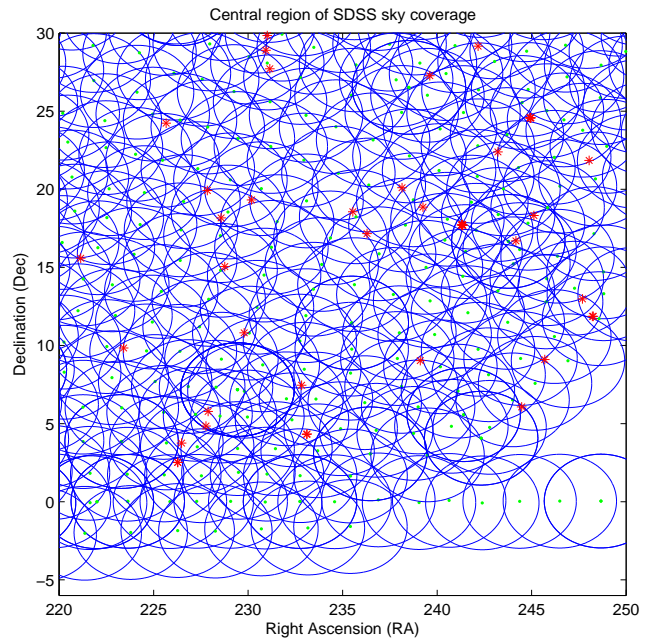}
	\caption{The central region of the SDSS sky coverage. The ellipses are representing the spectroscopic plates. The dots are showing
	the middle of each plate. The stars are representing our quasars.}
	\label{circle4}
\end{figure}

\subsubsection{AGN}

AGN usually exhibit strong [\ion{O}{iii}]-emission compared to normal galaxies due to the photo ionization 
caused by the active nuclei, as well as other forbidden emission lines such as [\ion{N}{ii}], [\ion{Ne}{v}], [\ion{S}{ii}] that also can arise from shock-heating. While star-forming galaxies have many young OB-stars and much HII-gas, the photons emitted by starburst galaxies still cannot produce the same strong presence of forbidden lines in the same way as for instance a Seyfert galaxy can. The AGN shines due to the very high energy photons, the high densities and large amounts of the gas that flows with high velocities into the accretion disk. The broad range of ionized states arises naturally as a consequence from when the accelerated electrons in the magnetic field in the accretion disk emit photons with a broad range of energies reaching even the UV-region, much more energetical than what can produced by hot stars in gas rich regions. This means also that the effective temperatures of the gas is very high in an AGN, and at the very high temperatures the photoionization of heavy ions will dominate the opacity. For the project, we wish to separate the AGN neighbor galaxies from the normal (or star forming) neighbor galaxies. The emission lines that are used here to separate AGNs from starburst galaxies, are the [\ion{O}{iii}], [\ion{O}{ii}], the H$\beta$ line, arising from Lyman continuum photons and the H$\alpha$-line. Galaxies with strong [\ion{O}{iii}] and [\ion{N}{ii}]-lines relative 
to H$\beta$ and H$\alpha$ tend to have photoionization caused by a strong radiation field, compared to the radiation field young stars can create and are thought to be AGN. 

We use Baldwin-Phillips \& Terlevich (1981) line-ratio diagrams \citep{BPT1981} combined with
Kauffman et al. (2003) criteria to select which of our neighbor galaxies are AGN:

\begin{equation}
log([\ion{O}{iii}]/H\beta) > 0.61/(log[\ion{N}{ii}]/H\alpha)-0.05)+1.3 \label{eq,BPT}
\end{equation}\label{BPTequation}

Further, we need to also include the type 1-AGN that are dominated by Doppler broadened Balmer lines
to ensure a proper exclusion, and therefore exclude galaxies with $\sigma$(H$\alpha) <$15 \AA.\  
We detected 69 AGN among the quasar neighbor galaxies.


\begin{figure*}[htb!]
 \centering
  \includegraphics[scale=.6]{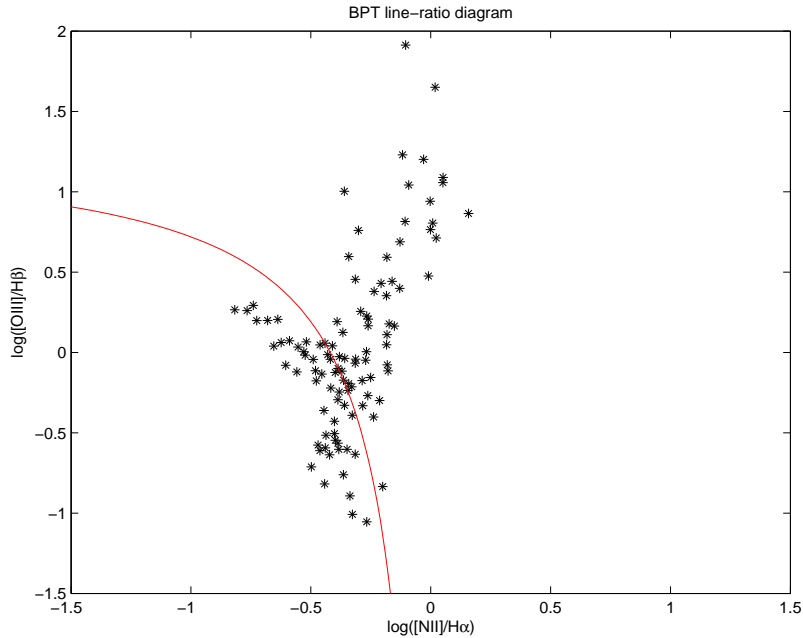}
     \caption{BPT diagram for galaxies with measurable H$\alpha$, H$\beta$, [\ion{O}{iii}] and
     [\ion{N}{ii}]. The division line separates star-forming galaxies from AGN. Those galaxies
     below the line are star-forming, while those above the line are AGN. The galaxies are represented by the stars in the plot.}
               \label{BPT}
     \end{figure*}

\subsubsection{Non-AGN neighbors}

Among our neighbor galaxy sample we should have a certain fraction of AGN. For our studies of 
how quasar can influence the SFR or ionization of galaxies we would actually like to separate 
the influence from the quasar from the influence of the presence of a potential AGN itself. 

To be more clear: let us assume a hypothetical scenario where all galaxies close to quasars
turn into an AGN. If we would observe star formation in the neighbor galaxy as a function 
of projected distance to the quasar, we would perhaps also observe an increase in the SFR. 
But would this really come as an effect from the quasar or rather as a effect of the AGN 
inside the neighbor galaxy? For this very reason we exclude the 69 AGN we found and reduce our 
sample of primary interest to 236 quasar-galaxy associations.

\subsection{Corrections}

The Balmer emission line fluxes and equivalent widths were corrected for underlying stellar 
absorption by assuming average absorption line strengths corresponding to 2.5 \AA\ 
in $EW$ for H$\alpha$ and 4 \AA\ for H$\beta$. Further were internal extinction corrections for blue galaxies with 
$EW$(H$\beta$) $>$ 2 \AA \ done, since the lines with H$\beta$ lower than this value are too noisy
to permit a reliable extinction correction. The spectral lines were internal extinction-corrected 
following a standard interstellar extinction curve \citep[e.g.][]{Osterbrock2009,Whitford}.

\begin{figure*}[htb!]
 \centering
  \includegraphics[scale=.6]{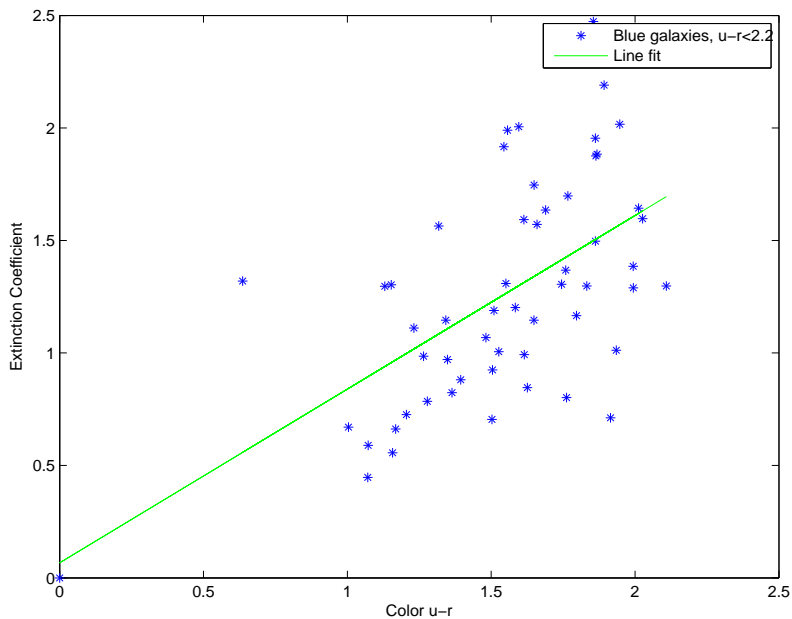}
     \caption{Extinction coefficient versus u-r color of galaxies. The stars are showing the blue galaxies. Their extinction coefficients are calculated according to Osterbrock. The line shows the fitted trend.}
               \label{extinction}
     \end{figure*}

For the blue galaxies whose line fluxes could not be corrected according to this method (by not
fulfilling the chosen criteria), we plotted the extinction coefficient versus the u-r color for the blue galaxies 
with properly calculated extinction coefficients, where a trend could be seen. We made 
a line fit to this trend, which we used to estimate the extinction coefficients for all the other
galaxies in the blue sample depending on their u-r color, see figure \ref{extinction}.

The colors were corrected for extinction by taking the inclination of the observed galaxies into account.
This because the internal extinction varies depending on if we see a galaxy edge-on or face-on. The colors
were corrected for all galaxies fulfilling the criteria described for the method using the derived
analytical expressions in a recent study \citep{CP2009}. This means we corrected the colors for all galaxies having
0$<$ u-r $<$ 4, a H$\alpha$ line width ranging from 0 to 200 \AA\, 
an absolute magnitude -21.95 $ < M_{r} < $-19.95 
and finally a concentration index C in the range 1.74 $< C <$ 3.06.

\subsection{Morphology classification}

The companion galaxies in the sample underwent a morphological classification in order to make a 
comparative analysis between disk type and spheroidal galaxies. The method used was the Principal
Component Analysis (PCA) for spectroscopic objects (Connolly \& Szalay 1999). The spectral classification 
coefficient is generally calculated by using two out of five eigencoefficients that build up the expansion 
of the eigentemplates of a galaxy's spectrum. These eigencoefficients can be extracted from the 
SDSS galaxies and are defined as

\begin{equation}
eClass= \mathrm{arctan} (-eCoeff2/eCoeff1).  
\end{equation}

Early-type and late-type galaxies eClass values range from -$0.35 <eClass < 0.55$. Early-type galaxies have an 
eClass value $-0.35 < eClass < 0$, while late-types have $0 < eClass < 0.55$.

\subsubsection{Results of test sample}

Via the SkyServer and some python-scripts available to make this easy \footnote{The SkyServer is the website where you can access the SDSS database. It includes tools to both view and download the SDSS data. One can use a user-friendly window for data searches, but also the SQL-based searches that can be more user-adapted and efficient. The homepage is located at http://cas.sdss.org/dr7/en/ }, we downloaded the eCoefficients eCoeff1 and eCoeff2 for the original 305 companion galaxies.  Out of the 305 galaxies, 183 could fit into the eClass as late-type and 34 as early-type.

\begin{table}[ht]
\centering
\begin{tabular}{c c c}
\hline\hline
Galaxy type & eClass results & Visual inspection \\ [0.5ex]
\hline
Total & 183 & 183 \\
Late-type & 183 & 70 \\
Early-type & 0 & 113 \\
or indefinable &  & \\ [1ex]
\hline
\end{tabular}
\label{eClass}
\caption{Visual inspection of late-type galaxies detected by eClass. Out of the originally
305 companion galaxies, 183 galaxies were predicted to be late-type galaxies by the eClass
function and underwent visual inspection. The right column shows the results from the visual inspection of these eClass-defined
late-type galaxies. The 41 remaining early-type galaxies predicted from eClass, did not undergo
any visual inspection and are therefore not included in the table.}
\end{table}

\begin{table}[ht]
\centering
\begin{tabular}{c c c c}
\hline\hline
Galaxy type&eClassresults&GalaxyZoo&Visual inspection \\ [0.1ex]
\hline
Total&172&172&172\\
Late-type&172&30&25\\
Early-type&0&142&139\\
or indefinable &  & \\ [1ex]
\hline
\end{tabular}
\label{GalaxyZoo1}
\caption{Visual inspection of late-type galaxies detected by eClass and comparision
to GalaxyZoo results. Out of the 275 companion galaxies that had GalaxyZoo information available, 
172 galaxies were predicted to be late-type galaxies by the eClass function. The right column shows the results from the visual inspection and Galaxy Zoo-data of these eClass-defined late-type galaxies.}
\end{table}

We did a visual morphology inspection out of the 183 eClass-defined late-type galaxies, see table 1. Out of these
only 70 turned out to be a morphological late-type galaxy, while the rest appeared either undefinable
or as early-types. 70 late-types confirmed out of 183 eClass-defined galaxies is a success rate of 38\%.
This is as many late type galaxies as one would have obtained if one simply picked 183 galaxies randomly from SDSS at 
redshifts 0.03$<$ z$ <$0.02. Therefore, we decided to skip further morphological analysis using eClass.

We also did a second morphology test and compared eClass to results from GalaxyZoo project
\citep{Lintott,Lintott2010}, see table 2. For our 305 galaxies, 
only 275 had been classified in the GalaxyZoo project, and via the SDSS casjobs one can download the results from the
classification. For this sample, eClass yielded that 172 galaxies should have late-type morphology and 27 
should be early-types. For the 172 late-type galaxies, we investigated which could were defined as Spirals, Ellipticals and Unidentified with the GalaxyZoo.
GalaxyZoo results yielded 30 late-type galaxies, 30 early-type galaxies and 112 unidentified.

Of those 30 galaxies selected by the GalaxyZoo as late-type galaxies, 22 were late-types by visual
inspection, and 8 were not possible to identify. Of the 30 galaxies selected by the Galaxy Zoo as
early-type galaxies, 28 were early-type galaxies and two were unidentifiable. Galaxy Zoo 
provides a very good tool for morphological identification.

\subsection{Biases, uncertainties and other problems}

Any experiment that can bring out interesting results, also can have potential problems and limitation in
the reliability of these results. Sometimes the errors in a measurement cannot be properly quantified, but sometimes
we also have selection effects in our sample. The selection effects might be hard to notice but they are important to quantify in order not to misinterpret the results. I list here observational biases
that are specific to our observations and I briefly mention some also spectroscopic and photometric issues that are 
general for the SDSS. Finally, I will discuss potential biases in our data analysis and error propagations.

\subsection{Observational biases}
What problems could arise from our way of selecting galaxies?

\begin{enumerate}

\item Our sample is non-volume-limited. Volume-limited samples are good to use because
galaxies of different luminosities dominate at different redshifts. At the furthest redshifts
only the brightest objects are seen, which means that the lowest observable galaxy luminosity
increases with redshift. To make a volume-limited sample, one would have to make a minimum luminosity cut
at the maximum used redshift and throw away all galaxies fainter than this minimum luminosity limit.
This would have decreased our sample a lot, and weakened our analysis. To avoid this problem
we have decided not to investigate the magnitudes of quasars and galaxies in any way, and prefer to work 
with their colors.

\begin{figure*}[htb!]
 \centering
  \includegraphics[scale=.6]{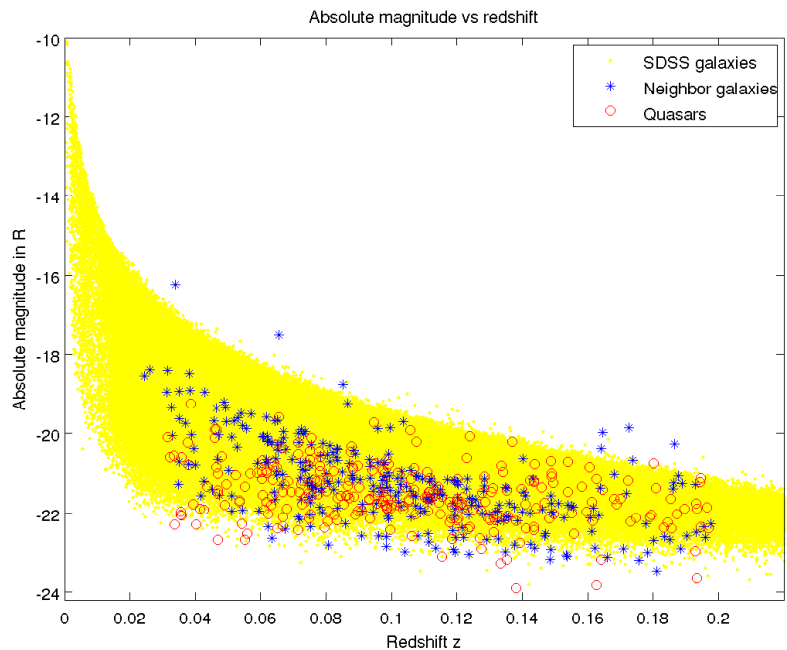}
     \caption{A non-volume limited sample. Our quasars are represented by circles. The companion galaxies are marked with stars. The sea of dots represents all the SDSS galaxies in the database.}
               \label{NFig84}
     \end{figure*}

\item We have used three different spectroscopic redshift intervals between quasar and galaxy $|\Delta z|$  $<$ 0.001, 0.006 and 0.012. The spectroscopically defined
redshifts with zconf $>$ 0.95 in the SDSS should have redshift estimation errors smaller than 
$\delta z<$ 0.001. A large part of the quasars and galaxies lying close to each other on the sky even with $|\Delta z|$  $>$0.001 are thought to be
quasar-galaxy associations. To perform studies with a huge number of galaxies lying closer than $|\Delta z|$  $<$ 0.001 to the quasar is
with present day galaxy catalogs difficult due to the very limited number of galaxy-quasar pairs detected.
At the same time, many of our companions could be simply background and foreground galaxies and totally unassociated to the quasar.
To sort out the effect of the number of background and foreground galaxies in our sample we introduce the redshift
cuts $|\Delta z|$  $<$ 0.001, 0.006 and 0.012.

\item Our sample size is fairly small. Given the number of galaxies of different properties, it might get difficult
to observe any but the most obvious trends, especially if the individual errors are large.

\end{enumerate}

\subsection{Spectroscopic problems}

\begin{enumerate}
\item The optical fiber separation in SDSS makes it not possible to observe pair of objects that are 
closer than a separation s$<$55'' from each other. Many of the pairs will be seen as images in the SDSS
but lacking redshift for one of the objects. This is especially noticed for low redshifts where
object pairs cannot be resolved because of this 55'' problem. 

\begin{figure*}[htb!]
\centering
 \includegraphics[scale=.6]{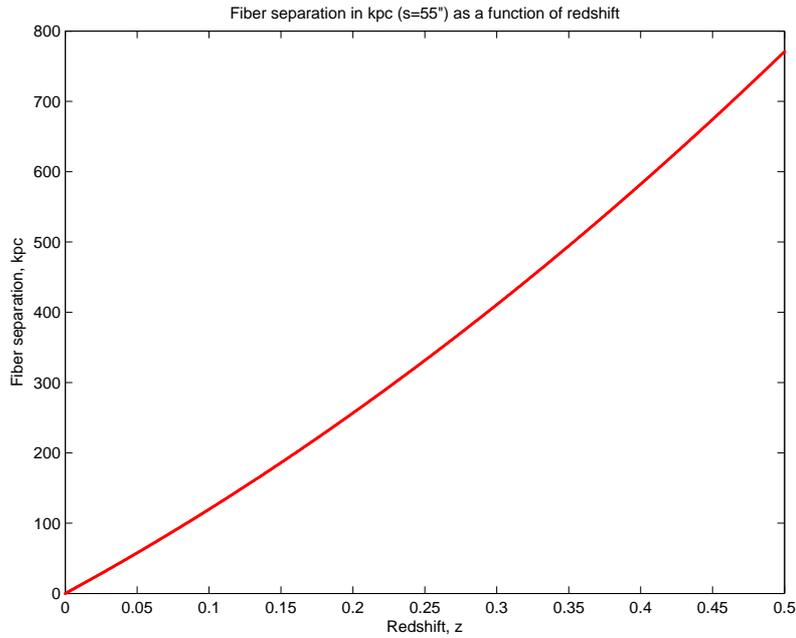}
    \caption{Fiber separation 55'' transformed into natural projected distance as function of redshift.}\label{FiberSeparation}
    \end{figure*}

As seen in figure \ref{FiberSeparation}, with these fiber separations pairs closer than 100 kpc from each other at redshift z=0.1 will not be seen. For close quasar-galaxy pairs to be seen they have to reside in overlap areas of fibers.

\item Fixed aperture diameter (3'') of the fiber spectrograph shows that different areas of galaxies are covered at different redshifts. At close redshifts emission line fluxes might be measured in the middle of a galaxy, while further away they might be an average value of the emission line flux of the whole galaxy.
\end{enumerate}
\subsection{Photometric problems}

\begin{enumerate}
\item We chose to 'flag' all our objects for brightness. This means that objects that are too bright
(m$_r >$ 15; m$_g >$ 15; m$_i >$ 14.5;) were not targeted, otherwise they could enlight too 
much of the surrounding sky and objects. This is a limit of major importance when dealing with 
quasar environments to avoid light contamination.

\item All spectroscopic targets in the SDSS have Petrosian magnitudes r $<$ 17.77. Faint objects will
not be detected. The absolute magnitude of the brightest object observed always increases with the redshift.

\item There is a red leakage in the u-band around 7100 \AA\ in the SDSS. This is a detector effect with much variability
in the magnitude of the red contamination \citep{Stoughton}. 

\end{enumerate}

\subsection{Analysis biases}
Even though one gets a relatively unbiased sample, one can still bias the results in the data analysis.
Here I list some factors that could have an influence on our results.

\begin{enumerate}

\item When we separate the type II AGN from the non-galaxy-neighbor sample we use the BPT-Kauffmann criterion. This requires good measurements of Balmer lines and forbidden lines. The type I AGN
we separate by removing those galaxies having a $\sigma$(H$\alpha$) $>$ 15 \AA\, namely AGN with Doppler broadened Balmer lines. Yet, we can have more AGN (such as LINERs) hiding in the sample where the criterions we used were not sufficient to exclude them.

\item We do not separate radio-loud quasars from radio-quiet. Due to the different suggested
modes of AGN feedback ('radio type' and 'quasar type') a mixing could decrease the probability 
of detection of any mode.

\item We have a very small sample of 236 non-AGN companion galaxies. This might not be sufficiently large enough to give a good picture of the reality.

\item Many of the objects present large error bars in their fluxes, which can hide eventual trends in the data.

\end{enumerate}

\subsection{Error propagation}

The errors in individual measurements of each line are calculated in the following way.
We have errors dh for the height h and d$\sigma$ for the width for each line. 

If the flux (F) is calculated as 

\begin{equation}
  F = \sqrt{2\pi}\sigma h
\end{equation}

the commonly used Taylor expansion for appoximating the flux error dF, would be

\begin{equation} \label{Taylor1}
  \left(\frac{dF}{F} \right)^2 = \left(\frac{d\sigma}{\sigma} \right)^2 + \left(\frac{dh}{h} \right)^2 + 2\frac{dh d\sigma}{h\sigma}\rho_{\sigma,h}
\end{equation}

where $\rho_{\sigma,h}$ is the correlation coefficient of $\sigma$ and h. We now face the problem 
that we do not know what is the correlation coefficient. Since h and $\sigma$ 
are the height and width of the $same$ emission line, there must exist a correlation between these 
two parameters and their errors (which in turn depends on how they were measured) meaning
we cannot put it to zero if we wish to avoid underestimation of error. Our solution is therefore 
to use the alternative way, the linearized approximation with partial derivates, yielding an upper
limit of the flux error

\begin{equation} \label{smart1}
dF= \frac{\delta F}{\delta\sigma}d\sigma+\frac{\delta F}{\delta h}dh
\end{equation}

or more concisely

\begin{equation} \label{smart2}
dF= \sqrt{2\pi}( \sigma\ dh +h d\sigma)
\end{equation}

which we use for calculating the errors of the emission line fluxes.

To calculate the errors in the ratio of two emission line fluxes, as in F([\ion{O}{iii}]/[\ion{O}{ii}])
we can use a Taylor expansion, but this time with a correlation coefficient corresponding 
to $\rho_{OIII,OII}$=0, since the two lines are measured independently and therefore
have no correlation.

To finally calculate the error dP (P for parameter) in every bin in our binned plots, we use

\begin{equation} \label{smart3}
dP= \sqrt{\frac{\sum w_{i} \sigma_{i}^2}{\sum w_{i}}}
\end{equation}

where w$_{i}$ is the individual weight. Since we chose not to use weights for the binned plots, 
we set w$_{i}$ =1, which reduces to the following equation:

\begin{equation} \label{smart4}
dP= \sqrt{\frac{\sum \sigma_{i}^2}{n}}
\end{equation}

where $\sigma_{i}$ here is the individual error in the investigated parameter (flux, flux ratio)
and n is the number of measurements in the bin.

\subsection{Bootstrapping}

Bootstrapping is a resampling method in statistics. It allows to estimate the statistics of a sample and calculate medians, variances and percentiles. The
main idea of bootstrapping is to use subsets of data with random data points. After generating large numbers of data subsets one calculates the statistics of each one of them. 
After this, one combines the gathered information about median values and variances of all these data subsets. For instance the median value calculated from bootstrapping, will 
be the median value in a list containing all the calculated median values in the different subsets (bootstraps) from the bootstrapping. 

We used bootstrapping to estimate the error bars in our surface density calculations. First was the
companion galaxies binned up into 10 pieces each of 35 kpc (to cover the whole projected distance
between quasar and companion galaxy) for each bootstrapping round. We used annular surface areas, i.e. the number of galaxies as a function of distance between quasar and galaxy were divided up in different 'annuli' (rings) around the quasars. When this was done, we performed bootstrapping on our sample 100 times to calculate surface densities in each bin and accompanying error bars. 

The error bars in the surface densities are plotted in fig
\ref{endNFig85} with a 68.3\% confidence level, corresponding to 1 $\sigma$.
\section{Results}
In this chapter I describe the diagnostic tools we have been using.
\subsection{Absolute magnitudes}

Knowing the apparent magnitudes of galaxies and the distance from us will allow us to actually calculate their 
rest-frame luminosity or their absolute magnitude. The luminosity distance
we calculate from redshift. We downloaded the dereddened apparent magnitudes from the SDSS 
for both quasars and galaxies. 

The luminosity distance was calculated using a 
LCDM model with H$_{o}$=70 km/s/Mpc, matter density parameter $\Omega_{M}$=0.30 and dark energy density
parameter $\Omega_{\Lambda}$=0.70.
The absolute magnitudes of both galaxies and quasars could be calculated as

\begin{equation}
M_{abs}=m_{obs} - 5log(D_L/10 pc) - K(z)
\end{equation}\label{eq2}

where D$_{L}$ is the luminosity distance. The k-correction term K(z) differ for quasars and galaxies.
For quasars we calculated it, but for galaxies we could download it (see section 3.2.2).

\subsubsection{K-correction for quasars}
The K-correction for our quasars was calculated assuming a universal
power-law spectral energy distribution (SED) with mean optical spectral index $\alpha$=-0.5. 
Since most of our quasars are at low redshift the K-correction is very small
while the spectral index barely has changes in this range \citep{Kennefick2008}. Not much evolution has 
been observed in the spectral energy distribution for quasars between
redshifts $0 < z <0.2$.

The K-correction for the quasars can thus be calculated as

\begin{equation}
K(z)=-2.5 \alpha log(1+z) - 2.5 log(1+z)
\end{equation}\label{eq1}

\subsubsection{Magnitude distributions of quasars and galaxies}

\begin{figure*}[htb!]
\centering
\includegraphics[width=16cm,height=10.5cm]{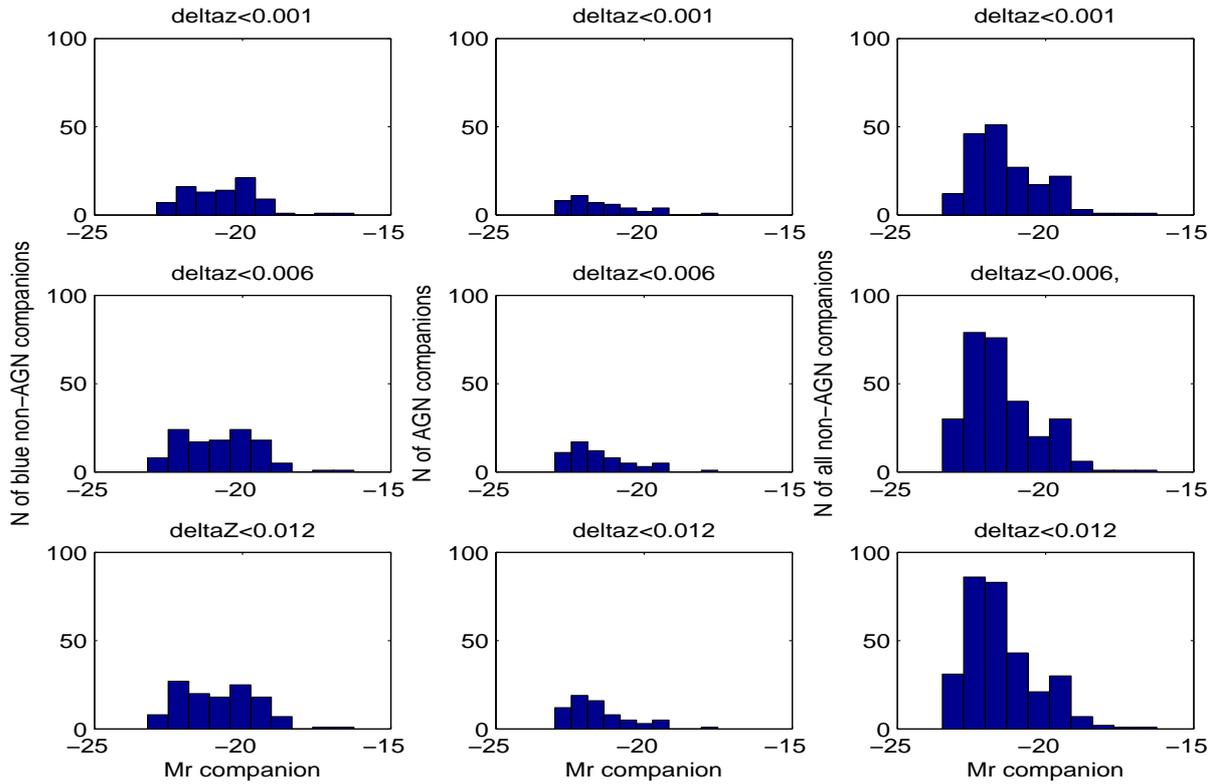}
\caption{Magnitude distribution of neighbor galaxies. Left panel
shows how the magnitude distribution looks for blue non-AGN companion galaxies to quasars.
Middle panel, the magnitude distribution of AGN companion galaxies. Right panel, of all non-AGN companions, irrespective
of their color.}
\label{NFig80}
\end{figure*}

\begin{figure*}[htb!]
\centering
\includegraphics[width=16cm,height=10.5cm]{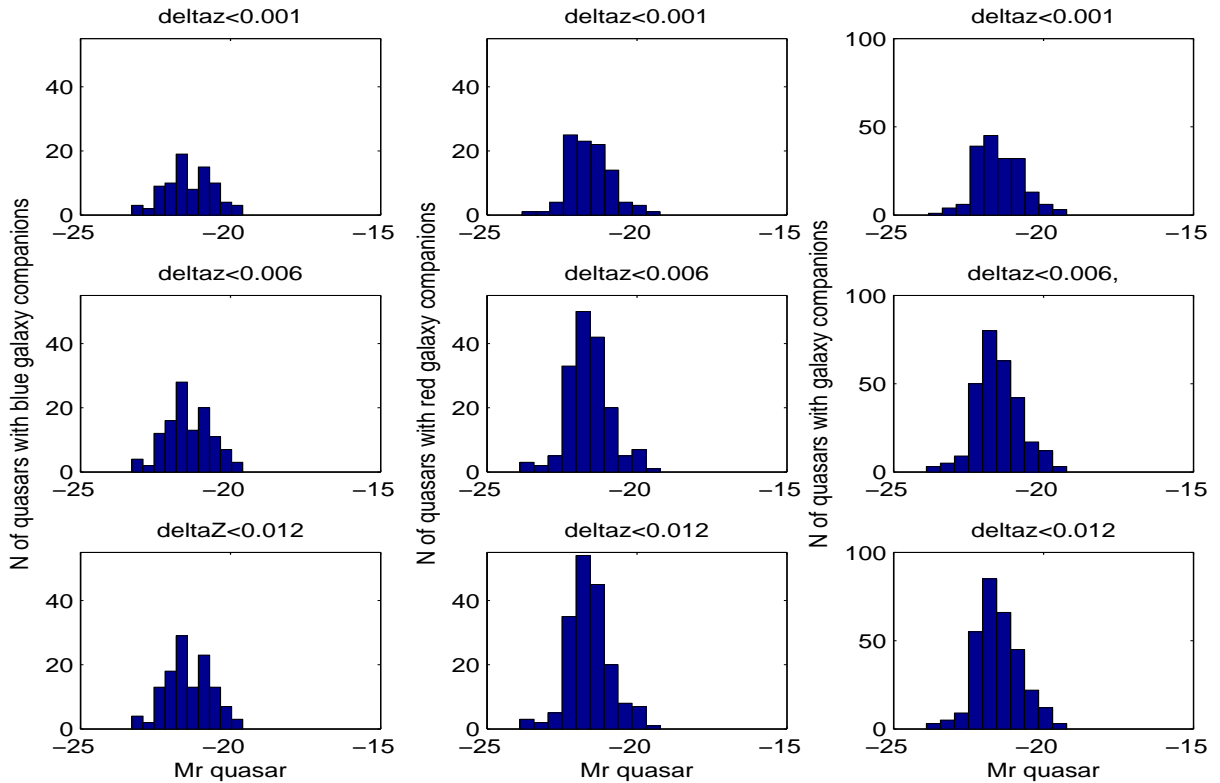}
\caption{Magnitude distribution of quasars around specific types
of companion galaxies. The magnitude distribution of quasars with
blue neighbor galaxies are found in left panel. Those with red companions
in middle panel. The overall magnitude distribution of quasars
with companion galaxies are found in right panel. AGN are not excluded.}
\label{NFig81}
\end{figure*}

Since we use a non-volume limited sample one must realize that magnitudes are likely to be highly biased in our sample. Two figures
of the magnitude distributions of our quasars and galaxies are plotted here. Fig. \ref{NFig80} shows the 
distribution of galaxy magnitudes depending on color, $|\Delta z|$  and in total, while fig. \ref{NFig81} 
shows the corresponding quasar magnitude distribution. Blue galaxies are defined as those with color $U_{e}-R_{e}$ $<$ 2.2, while red galaxies $U_{e}-R_{e}$ $>$ 2.2 (see following section). Note the Poisson form of the galaxy magnitude distribution, that is different from the quasar magnitude distribution (that looks more Gaussian).

Fig. \ref{NFig81} also shows that the quasars have the same magnitude distribution irrespectively whether they have a blue or red companion, which means that the nature of the companion is very unlikely to affect quasar luminosity. This has been earlier confirmed by Howard Yee \citep{Yee1987}.

\subsection{Colors of non-AGN neighbor galaxies}

Colors of galaxies can reveal certain things about a galaxy. For instance, the color usually
depends on the continuum energy distribution and the strength of various spectral lines.
Galaxies with a greater amount of young, massive stars and higher star-formation are bluer in their nature while 
older star populations are redder. But metallicity and dust can also change the color of a galaxy. 
	
The SDSS galaxy colors show a bimodal distribution \citep{Strateva2001} where the two peaks correspond 
to the blue galaxies (u-r $<$ 2.2) and the red galaxies (u-r $>$ 2.2). The blue one usually represents disk type galaxies with high star formation rate and this blue color population has a large scattering on the color-magnitude 
diagram. The more luminous blue galaxies are often also less blue \citep{Tully1998}. The brighter blue galaxies usually are slightly redder, having higher metallicity and are typically somewhat older. One typically assigns the blue peak to consist of mostly "late type" galaxies. The red peak, on the other hand, has a much stronger color-luminosity relation and usually consists of early-type galaxies. Red galaxies are typically old and with a very low star formation rate. 
	
The difficulty is to connect the blue star-forming galaxies with the red ones. In color-magnitude diagrams for mixed galaxies, three regions mainly appear: a blue region, a red region and the binding valley, the "green" valley. Much research goes into understanding how blue young galaxies transform into red, old ellipticals. Two paths are usually discussed. One way, is through the existence of AGN feedback. Another popular theory is that winds from 
supernovae in heavily star-forming regions quenches the local star formation. These winds would 
then be formed in very massive galaxies, that in turn could be formed in mergers between galaxies. For instance, 
the most star-forming galaxies we know today, the very red ultraluminous infrared galaxies (ULIRGs), 
have double cores which clearly shows a history of mergers. Also many 
poststarburst galaxies, that often have a spectrum combining two stellar populations of different ages, show 
redder colors.
	
Thus we can see that colors are very important in order to understand the evolution and star formation history 
of galaxies. We will also investigate how the colors of the neighbor galaxies 
of the quasars changes and the distribution of galaxies of different colors around the quasars. 
	
Rest-frame colors are easily calculated by applying extinction corrections and k-corrections in different filters. 
In our study we used the u-r color which is supposed to have low noise, is widely used and therefore
good for comparing to other literature. 
	
\begin{figure*}[htb!]
\centering
\includegraphics[width=16cm,height=10.5cm]{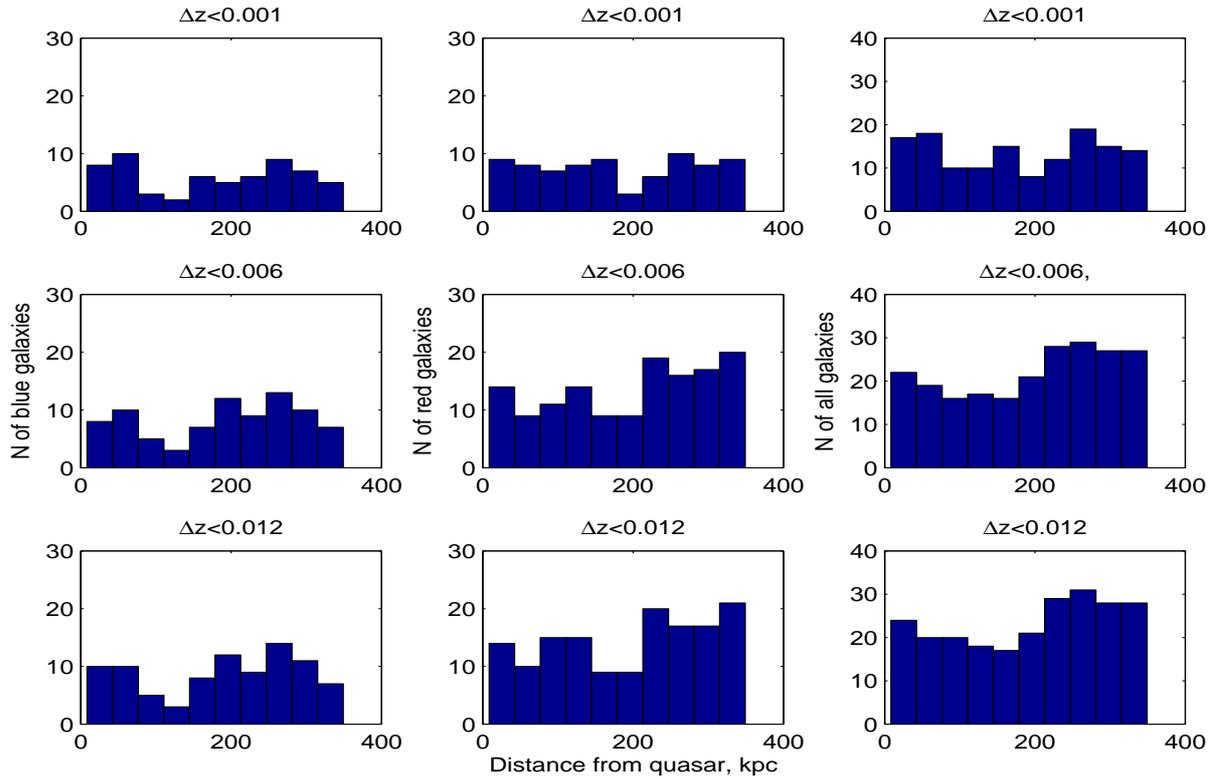}
\caption{Number distribution of galaxies around the quasars. 
Left panel shows blue galaxies (89). Middle panel shows red galaxies (147). Right panel shows all galaxies,
with AGN (69) excluded in all cases.}
\label{AGNNDtot}
\end{figure*}

Fig \ref{AGNNDtot} shows the distribution of non-AGN-galaxies in our sample as a function of distance from quasars. The
data is binned up in 10 bins with 35 kpc each. We see that we have a greater number of red galaxies among our companions.
A better understanding of the distribution of red galaxies will reach the reader when he compares with the 
surface densities in section 4.7, fig. \ref{endNFig85}. In this plot the increase in the number of galaxies as a function of distance has to do with the bin, that spans 35 kpc in radii, covering a greater area the further 
away from the quasar we go. More galaxies should be observed in the more distant bins. 
	
Also to be noticed is the drop of blue galaxies in the bin around 150-200 kpc and sudden increase to more or 
less the same number of blue galaxies as red galaxies in the closest bin to the quasar. This sudden drop 
could have three explanations: a) Unknown physics with gas-rich galaxies. b) Chance. A low number 
of blue galaxies could by chance make the blue galaxies appear to have a  drop at this position, while a 
larger number would smooth out any strangenesses and make the noise lower. c) Bad binning. When we 
have samples that are at least three or four times larger than our blue galaxy sample, we can use more robust histogram 
binning programs \citep{Knuth} to calculate the optimal bin sizes and sort out any potential noise. Even though we used the standard bin size (10) of Matlab, a bad choice can make noise appear as some interesting trend. This emphasizes the importance to redo this study when many more companion galaxies to quasars with spectroscopic redshifts will be available in upgraded data releases in the SDSS. We wish to separate potentially interesting physics from poor histogram binning. 
	
\begin{figure*}[htb!]
 \centering
   \includegraphics[width=16cm,height=10.5cm]{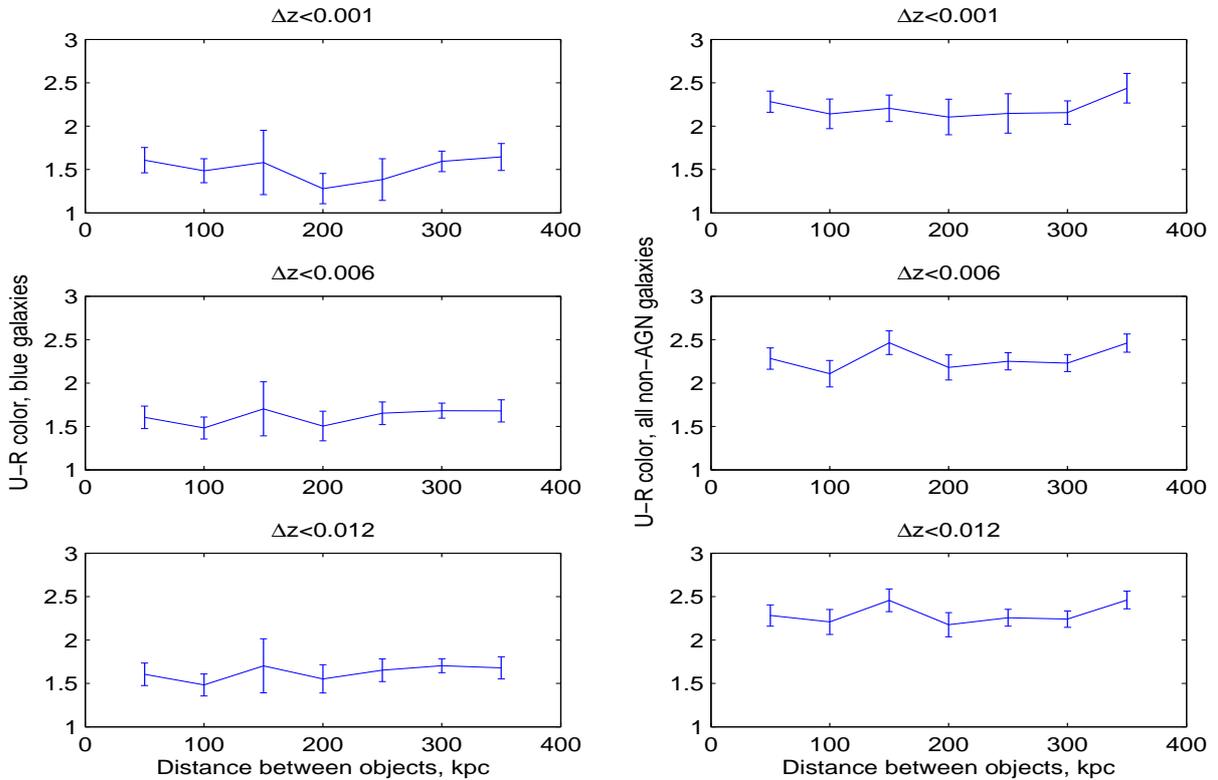}
    \caption{$U_{e}-R_{e}$ color of galaxies versus distance to quasar. In the left diagrams, are the $U_{e}-R_{e}$ colors of the bluer companion galaxies with $U_{e}-R_{e}$$<$2.2. In the right diagrams, all the non-AGN galaxy companions.}
               \label{Color}%
     \end{figure*}

Figure \ref{Color} shows how the $U_{e}-R_{e}$ color of the companion galaxy changes as a function of the distance between the quasar and the galaxy companion. We have chosen to perform binning of the data (50 kpc width), and calculate the 
average value of the color in each bin since this simplifies the visualization of the results. We can see minor 
changes where the blue ($U_{e}-R_{e}$ $<$ 2.2) galaxies tend to be somewhat bluer close to the quasar, while the red ones 
undergo a negligible transition towards the redder at smaller separations. There might be a slight tendency 
of bluer blue galaxies closer to the quasar, but the trend appears so small that it can hardly be statistically significant, considering the overlap between the error bars. But if this would be true, it 
could mean either an increase in SFR close to the quasar for a galaxy or that we simply see a larger number 
of blue, star-forming galaxies closer to the quasar, not necessarily with a higher star formation rate. Those galaxies can also be simply smaller, so that we their average star formation rate might appear higher than the SFR in the center of larger galaxies.

\subsection{Star formation in non-AGN neighbor galaxies}

Star formation rate is the measure of how much gas forms into new stars per time unit. 
These star births occur in diffuse nebulae ('diffuse', meaning they have no defined structure or boundaries) 
that are higher density regions of cold giant molecular clouds, consisting of mainly molecular hydrogen.

Typically different shock waves can cause collapse of such a molecular cloud, via winds from supernovae, collisions between molecular clouds or magnetic interactions. This results in sudden star formation, 
further ionization of the medium of the new born stars, followed by expansion of the cloud, resulting in
HII regions that we often see in spiral and irregular galaxies.  Usual sites of star formation 
are found in the spiral arms of late type galaxies, bridges, knots, bars that make the gas flow easier, nuclei and 
tidal tails. Much initial star formation in disk type galaxies will come from the gravitational 
instability in the disk itself. At the same time, stellar winds from O and B stars or winds from 
supernovae, may have the opposite effect. Instead they might heat the gas needed for star formation, thus 
expanding it so that it might not later collapse to form new stars. 

Already in the mid-1970's, mergers and tidal interactions were predicted to cause increased star-formation 
in galaxies \citep{Bushouse1987}. During mergers large galaxy disks may get disrupted by tidal effects, causing 
large gas flows and thus fueling strong star formation in the central regions \citep{Barnes1996}. Also more 
recent simulations  \citep{Martig2008} have suggested that interactions between two galaxies occuring near 
a big tidal field, for instance near a group of galaxies, could increase the star-formation rate in the merger. 
Many groups have observed an increased star-formation rate during mergers \citep[e.g.][]{Nikolic} and also in the neighbourhoods of AGN \citep[e.g.][]{CL2006} . 

There are several indicators of high star formation. From the UV-heated dust around O-stars comes high 
Far Infrared (FIR) emission that is easy to measure since it is free of any extinction. The radio continuum 
around 1.4 GHz also has this advantage, which makes it a convenient probe. Less direct methods are also useful: 
estimation of the number of O-stars necessary for the observed ionisation, via either measuring the forbidden lines, 
or the free-free thermal continuum radiation. These methods instead suffer from extinction and problems when trying 
to separate the thermal emission from non-thermal emission. Also polycyclic aromatic hydrocarbons 
(PAHs) can be used as starformation indicators. These PAHs fluoresce especially strongly around 7.7 $\mu$m and 
could not survive the harsh environment of an AGN, since hard x-ray scattering would destroy their structure.

The most common way of calculating the star formation rate is with the H$\alpha\ \lambda$ 6563 emission line. 
In star-forming galaxies the H$\alpha$ emission originates from photoionization by massive, 
short-lived stars. The advantage of the H$\alpha$ line is that it is the spectral line that suffers least from 
extinction and depends least on the conditions of the ionized gas \citep[e.g.][]{Moustakas2006}.

We have used the measured H$\alpha$ emission from the spectroscopic catalog of the SDSS to estimate the star 
formation rate in our galaxy sample. We use the Bergvall-R\"onnback calibration \citep{Bergvall}:

\begin{equation}\label{eq4}
L(H\alpha)= SFR *1.51*10^{34}
\end{equation}
and

\begin{equation}\label{eq5}
L(H\alpha)=4 \pi D_L^{2} \sqrt{2\pi} \sigma h 10^{-20}
\end{equation}

where $\sigma$ and $h$ are width and height of the H$\alpha$ emission line, $D_L$ is the luminosity
distance in Mpc and the emission line luminosity is expressed in watts.

As an alternative way of measuring the star formation rate we used the [\ion{O}{ii}]- emission line. 
[\ion{O}{ii}] $\lambda$3727 is mainly useful when trying to calculate the star formation rate
in galaxies at redshifts higher than z=0.4 where the H$\alpha$ emission line is redshifted
beyond the red limit of the SDSS spectrograph. The [\ion{O}{ii}]-emission line is more affected by the metallicity and internal dust extinction in galaxies \citep{Charlot2002}. To reduce our reddening 
and metallicity effects, we use the K98 calibration \citep{Kewley2002}
for calculating the SFR from the [\ion{O}{ii}]- emission line:

\begin{equation}\label{Kewley}
SFR=(1.4 \pm 0.4) *10^{-41} L([\ion{O}{ii}])
\end{equation}

where the L([\ion{O}{ii}]) is given in ergs/s.

\begin{figure*}[htb!]
 \centering
   \includegraphics[width=16cm,height=10.5cm]{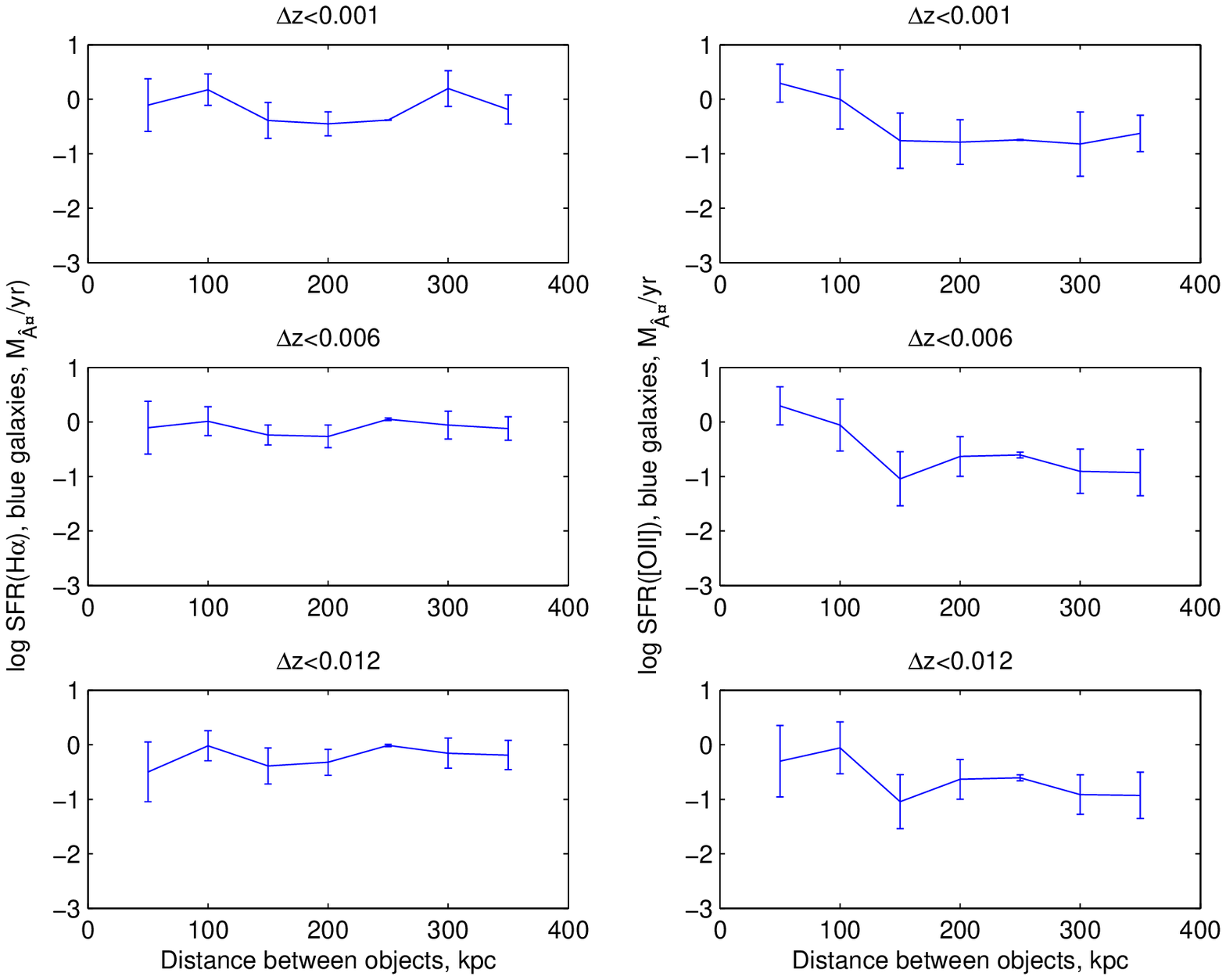}
     \caption{Star formation rate in blue companion galaxies in solar masses per year, log10(SFR). Left panel shows SFR from from H$\alpha$-flux. Right panel shows SFR calculated from [\ion{O}{ii}]-flux. Three different redshift difference cuts are pictured for the pairs. Galaxies with SFR$>$100 and individual errors in SFR $>$1000 are excluded. The individual errors are unweighted in the calculations of mean values and error bars. Only objects with used emission line fluxes that are greater than three times the error in the emission line flux are included.}
               \label{SFR}%
     \end{figure*}

Figure \ref{SFR} shows the average star-formation rate in blue companion galaxies calculated from H$\alpha$ flux versus distance from the quasars (left panel), as well as SFR calculated from [\ion{O}{ii}] (right panel). According to the figure, it seems like SFR from H$\alpha$ remains unaffected over distance, while star formation rate from the [\ion{O}{ii}]-line increases closer to the quasar. What could this inconsistency depend on? This could arise in case the metallicity or dust extinction, would vary as well over distance. From the previous section, we can see for that dust extinction, it seems not to be the case. Another reason might be that we have a contamination of AGN that sneak through the BPT and H$\alpha<$15 \AA\ criteria near the quasar, and therefore create a false trend in the SFR from [\ion{O}{ii}].

\begin{figure*}[htb!]
 \centering
   \includegraphics[width=16cm,height=10.5cm]{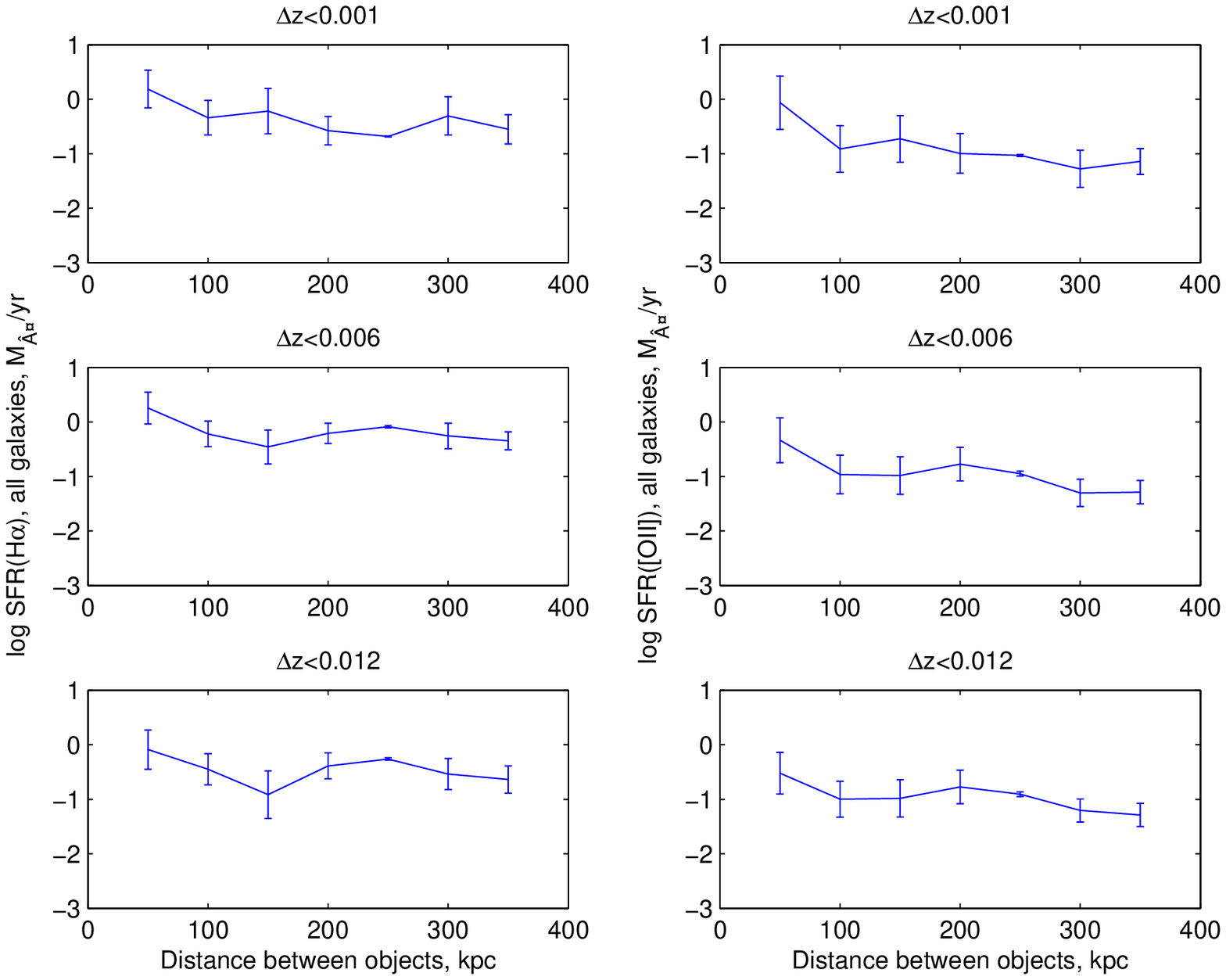}
     \caption{Star formation rate in all non-AGN companion galaxies, log10(SFR). Left panel shows SFR from from H$\alpha$-flux. Right panel shows SFR calculated from [\ion{O}{ii}]-flux. Three different redshift difference cuts are pictured for the pairs. Galaxies with SFR$>$100 and individual errors in SFR $>$1000 are excluded. The individual errors are unweighted in the calculations of mean values and error bars. Only objects with used emission line fluxes that are greater than three times the error in emission line flux are included.}
               \label{SFRall}%
     \end{figure*}

Figure \ref{SFRall} shows the average star-formation rate in all the non-AGN companion galaxies.
Here we do see that the SFRs from [\ion{O}{ii}] and H$\alpha$ both decrease as a function of 
distance from the quasar, while the correlation in the SFR from [\ion{O}{ii}] still is clearer.

We could not use a specific SFR (SFR/$M_{r}$) because we use a non-volume limited sample that could bias all our results.

\subsection{Dust in non-AGN neighbor galaxies}

Dust extinction is a considerable problem when one wishes to look at different emission lines
since it often does not let photons pass through and is a wavelength-dependent phenomenon.
It is important to correct emission lines for dust extinction, but in our case we are 
also interested in seeing how the dust content in the galaxies varies over the 
distance from the quasars. One standard way of measuring the dust
content is by using the H$\alpha$/H$\beta$ emission line flux ratio. It means, we could only 
use galaxies with fluxes greater than $f$ $>$0 in H$\alpha$ and H$\beta$.

The theoretical value of dust content in galaxies with the dust only heated from stars is 
F(H$\alpha$/H$\beta$)=2.8. Young galaxies with small dust content usually keep 
their F(H$\alpha$/H$\beta$) emission line flux ratio at this average theoretical value. Older galaxies with smaller dust content 
would have lower value, while higher F(H$\alpha$/H$\beta$)-ratio depends on higher dust content. 

\begin{figure*}[htb!]
 \centering
   \includegraphics[width=9cm,height=12cm]{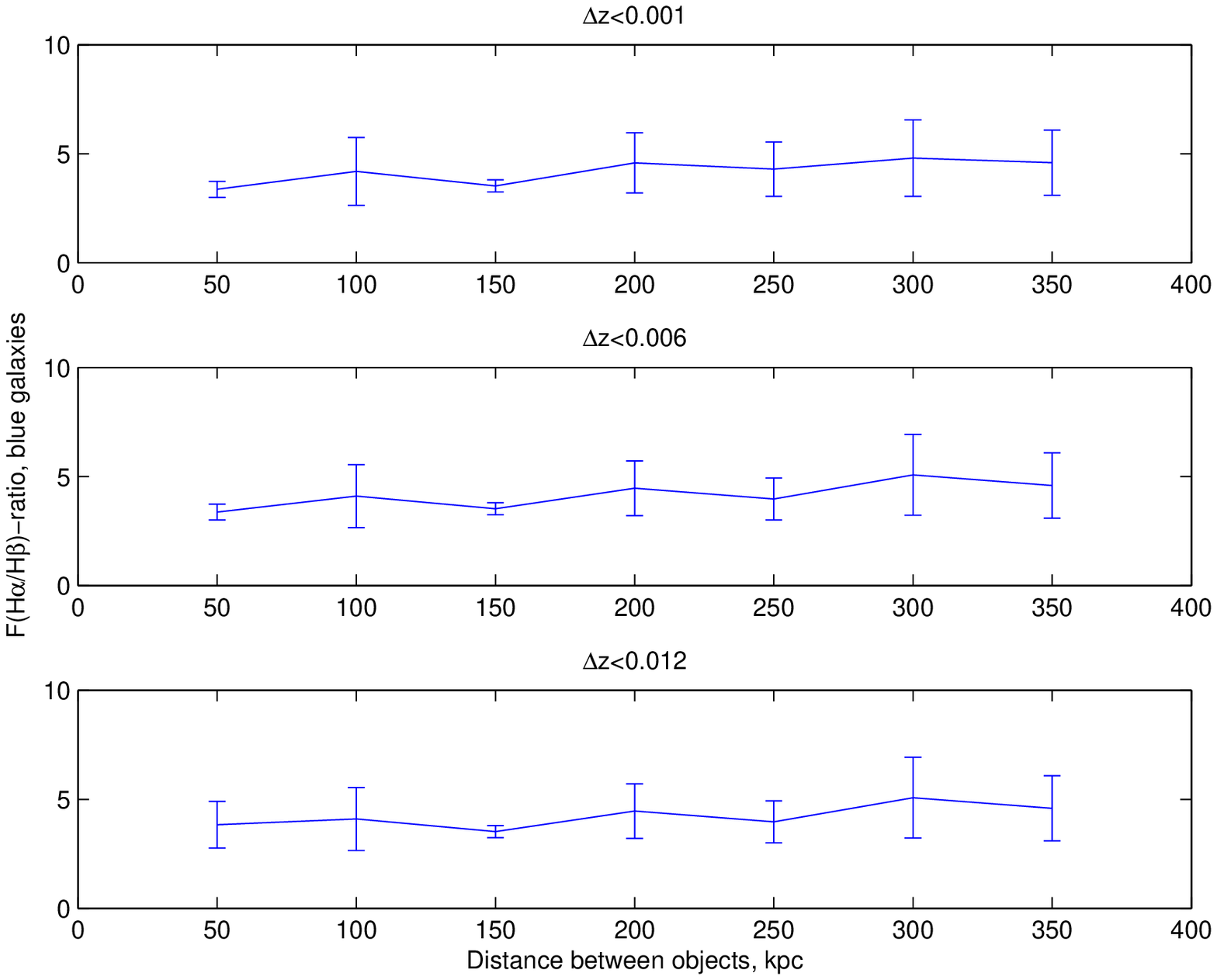}
     \caption{Dust extinction: the F(H$\alpha$/H$\beta$ )-ratio in the blue galaxies with $U_{e}-R_{e}$$<$2.2. Three different redshift difference cuts are pictured for the pairs. The individual errors are unweighted in the calculations of mean values and error bars. Only objects with used emission line fluxes that are greater than three times the error in emission
line flux are included.}
               \label{NFig70}%
     \end{figure*}

In our sample, we can see that there is no significant difference in the H$\alpha$/H$\beta$ emission line 
flux ratio in the galaxies depending on distance from the quasars, see Figure \ref{NFig70}. The most 
sensitive plot with $|\Delta z|$  $<$ 0.001, shows the average dust extinction ratio closest to the theoretical value 
and seems to be least supportive of any changes in the closest bin. This suggests that no significant changes in the dust extinction ratio are seen as a function of distance between the galaxies and the quasars.

\subsection{Ionization in non-AGN neighbor galaxies}

The flux ratios F([\ion{O}{iii}]/[\ion{O}{ii}]) and F([\ion{O}{iii}]/H$\beta$) are useful since they provide information on the ionization state of the gas in the galaxies and high values could occur both in the presence of an AGN and in areas dominated by high star formation activity from massive stars. Usually one needs to subtract the contributions from the stellar populations in order to study the possible contribution of an AGN \citep{Li2008}. The cause of increased degree of ionization can be the large number of highly energetical ionizing photons coming from hot, young stars, or the non-thermal emission from an AGN. AGN have since long been suggested to ionize intergalactic medium at distances up to several Mpc. It has been suggested that when the gas has a higher excited state it also implies a lower metallicity \citep[e.g.][]{Donizelli}. Fewer metals can act as coolants which in turn will result in a higher [\ion{O}{iii}]-emission. Also, shock excitation could increase the F([\ion{O}{iii}]/[\ion{O}{ii}])-ratio \citep{SuthDop}.

\begin{figure*}[htb!]
 \centering
  \includegraphics[width=16cm,height=10.5cm]{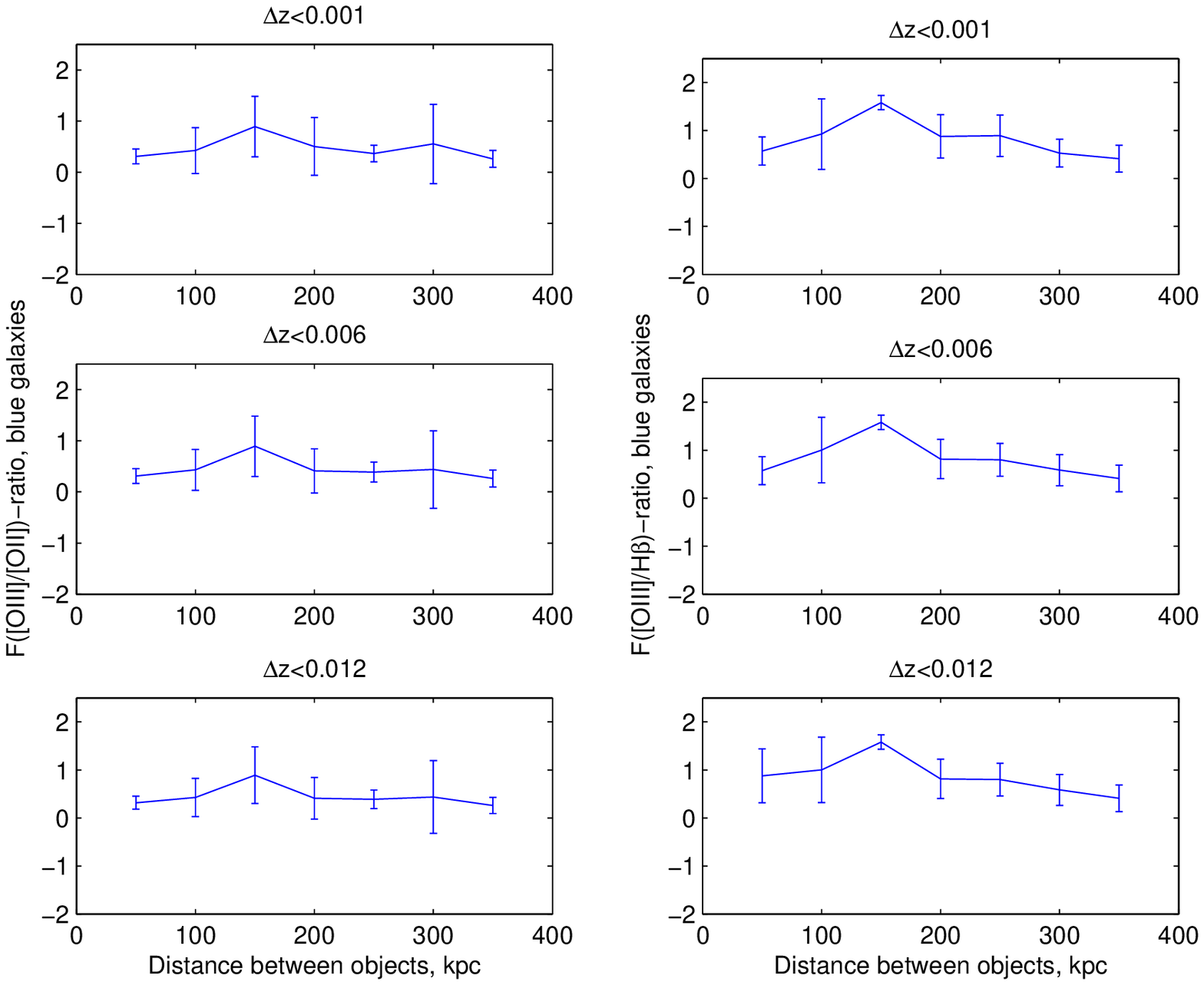}
     \caption{Ionization in blue companion galaxies. Left panel shows
     F([\ion{O}{iii}]/[\ion{O}{ii}]) ratio. Right panel shows F([\ion{O}{iii}]/H$\beta$). 
     Three different redshift difference cuts are pictured for the pairs. Galaxies with ionization$>$7
are not included since they are believed to be AGN that sneaked into the sample despite exclusion. Also
are galaxies with individual errors$>$1000 in ionization are excluded. The individual errors are unweighted in the calculations of mean values and errors. Only objects with used emission line fluxes that are greater than three times the error in emission line flux are included.}
               \label{slutIonization}
     \end{figure*}

In an interaction an increased F([\ion{O}{iii}]/[\ion{O}{ii}])-ratio could also support the idea of 
higher formation of massive stars as the ionization could arise from the stronger radiation field 
that more heavy stars might have. The F([\ion{O}{iii}]/[\ion{O}{ii}])-parameter shows no change in ionization in the galaxy companions with distance from the quasar for the blue galaxies (see fig. \ref{slutIonization}).



\begin{figure*}[htb!]
 \centering
   \includegraphics[width=16cm,height=10.5cm]{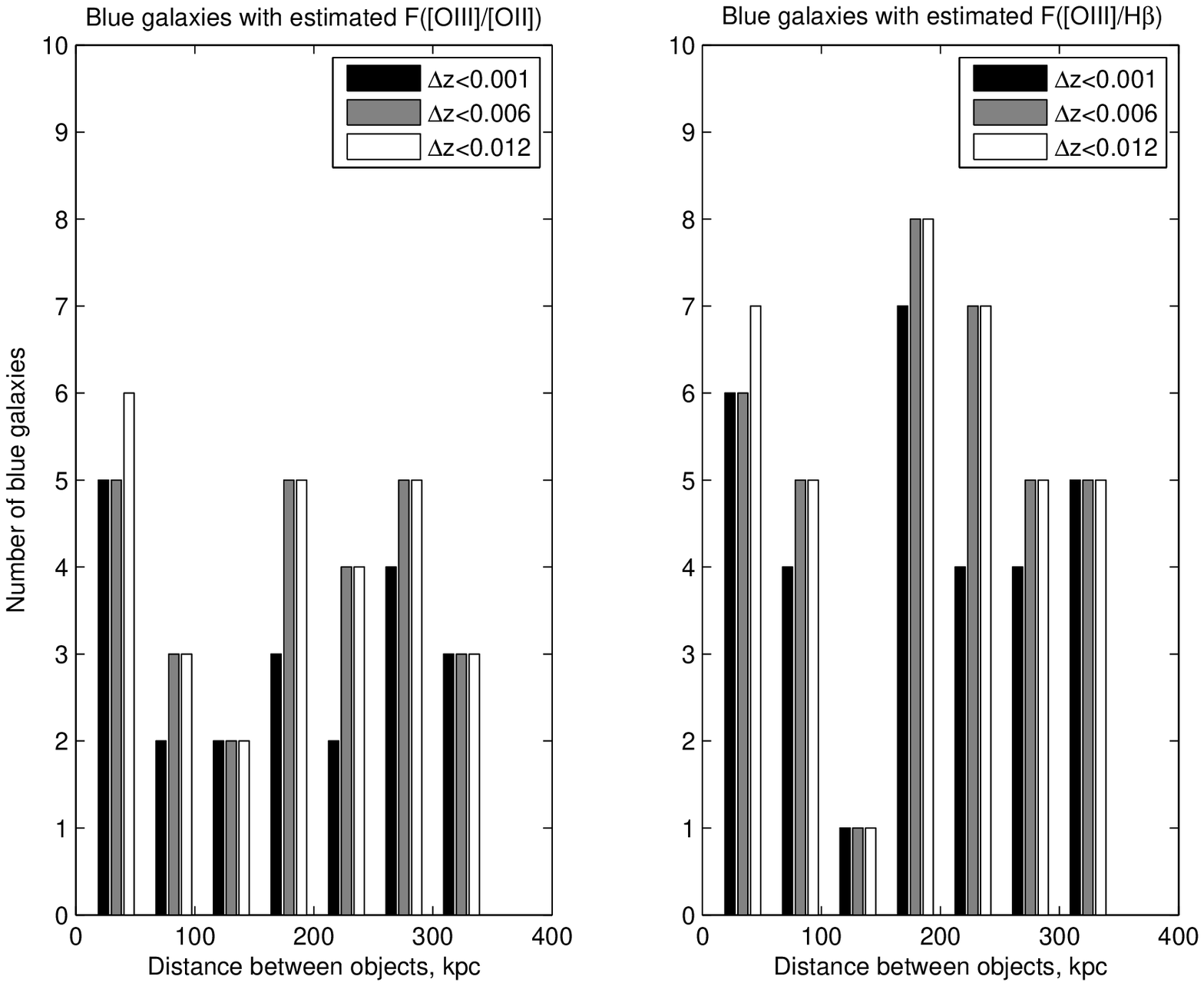}
     \caption{Left panel a) Number of galaxies with [\ion{O}{iii}] and [\ion{O}{ii}] measured.  
Right panel b) Number of blue galaxies with [\ion{O}{iii}] and H$\beta$ measured. The three $|\Delta z|$ are represented by different bar colors. Galaxies with ionization$>$7 are not included since they are believed to be AGN that sneaked into the sample despite exclusion. Also are galaxies with individual errors$>$1000 in ionization are excluded. The individual errors are unweighted in the calculations of mean values and error bars. Only objects with used emission line fluxes that are greater than three times the error in emission line flux are included.}
               \label{NFig75}
     \end{figure*}

The F([\ion{O}{iii}]/[\ion{O}{ii}])-parameter shows no changes in the degree of ionization in the blue galaxy companions with distance from the quasar, and neither does the F([\ion{O}{iii}]/H$\beta$)-ratio, see figure \ref{slutIonization}. In
the figure \ref{NFig75} the number of galaxies with measurable ionization is plotted.

\subsection{Oxygen abundance in non-AGN neighbor galaxies}
Metallicity is an important feature to investigate since it tells us a lot about the previous 
star-formation history, potential mass-loss and possible gas flows. In disk galaxies such as spirals, 
metallicity has turned out to be directly correlated with the mass of the galaxy. Interestingly, some
studies report that the metallicity in a disk galaxy actually changes during a merger which has been interpreted as a result of gas flows \citep[e.g.][]{Shields}. 

One method of measuring the metallicity is by measuring the oxygen abundance. 
Earlier studies have shown that there is a linear relation between the [\ion{N}{ii}]$\lambda$6583 line 
and the oxygen abundance \citep{Raimann2000,Storchi} that otherwise can be directly 
inferred from electron temperature measurements:

\begin{equation}
12+log[O/H)]=9.12 + 0.73 log (L([\ion{N}{ii}])/ L(H\alpha))
\end{equation}\label{Eq6}

This method, the "N2-method", has been shown to be a useful indicator to calculate metallicities for galaxies with 
oxygen abundance 12+log(O/H) $<$ 8.5, compared to the R23-index that is good for metal-poor galaxies
with oxygen abundances 12+log(O/H) $<$ 7.9 \citep{Yin2008}. We calculated the oxygen abundance
using the N2-method for our non-AGN galaxies.

\begin{figure*}[htb!]
 \centering
   \includegraphics[width=16cm,height=10.5cm]{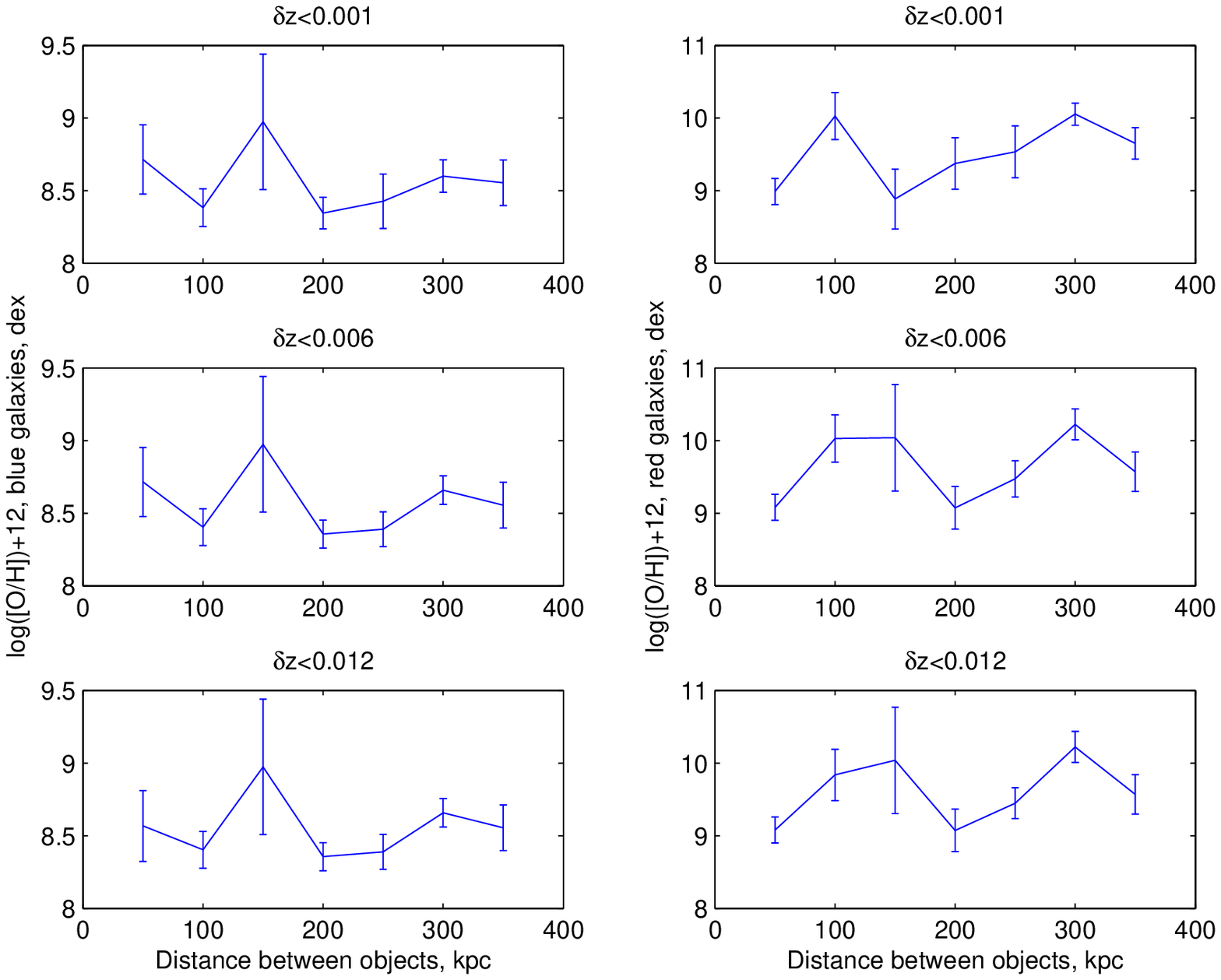}
     \caption{Oxygen abundance in companion galaxies, N2-method. Left panel shows blue companion galaxies.
     Right panel shows red companion galaxies. Three different redshift difference cuts are pictured for the pairs. Galaxies with individual errors in oxygen abundance$>$1000 are excluded. The individual errors are unweighted in the calculations of mean values and error bars since their relative errors could not be determined accurately. Only objects with H$\alpha$ emission line fluxes that are greater than three times the error in emission line flux are included.}
               \label{Figure16}
     \end{figure*}

Figure \ref{Figure16} shows that the scattering in the plots with oxygen abundance is too large
to tell us anything.

\subsection{Least square approximations of SFR, dust extinction and ionization}

As a comparative method of analyzing our data, we performed least square approximations (LSAs) of SFR, dust extinction, ionization and oxygen abundance in all non-AGN galaxies as well as in a sample of blue non-AGN galaxies, as a function of distance from the quasars. Numerical results from differently weighted LSAs are presented in a table in this section, together with a few representative plots from the LSAs with suitable weights w$\sim$ 1/($\mu$+$\sigma)^{2}$, where $\mu$ is the median error in the sample.

In the least square approximations, we did $not$ exclude galaxies where the flux of an emission line is smaller than three times the flux error, contrary to the case of the binned plots. The reason why we did not exclude the galaxies with poor signal, is that we wish to set limits of any observable correlations in our sample. By comparing the LSAs with different weights and different qualities of spectras, we make it possible for other authors with different approaches in their method to easier compare with our results and deduce their expected average value for galaxies at a certain distance from quasars. The LSAs using the weighting w$\sim$ 1/($\sigma)^{2}$ do in principle "eliminate" the galaxies with big errors in the line fluxes from the LSAs. For w$\sim$ 1/($\sigma)^{2}$, w$\sim$ 1/($\mu$+$\sigma)^{2}$ and unweighted LSAs for SFR, dust extinction and ionization as well as unweighted LSAs for color and oxygen abundance, we tried to estimate the greatest signal in the parameters as function of distance. The reason why we kept the LSAs for color and oxygen abundance unweighted, is because since the relative errors in the extinction corrected colors and the oxygen abundances not could be accurately determined, the results would risk to get severely biased by weighted LSAs.

Interestingly, the individual SFRs determined from the [\ion{O}{ii}]-flux display a greater range of SFRs in the plots of the LSAs. Even if one would take away those with bad spectra, the range stays as large in the SFR from [\ion{O}{ii}]-flux, while the three galaxies with the lowest SFR in $H\alpha$ flux disappear in the plot with only blue galaxies. Also in the LSA, we can see an up to 2$\sigma$ increase of SFR from [\ion{O}{ii}]-flux. This makes us wonder, why is there a difference in the SFRs calculated from [\ion{O}{ii}] and $H\alpha$-lines? [\ion{O}{ii}]-flux is used in cases at high redshifts where the $H\alpha$-line cannot be used, but how reliable is it in the light of our results? Could the difference between the measurements of SFR be a consequence of hidden AGN at close distance to the quasars?

But what about the ionization in galaxies near quasars? In case of the F([\ion{O}{iii}]/[\ion{O}{ii}] in the blue galaxies, the maximum 
trend is a 1$\sigma$ increase with w$\sim$ 1/($\mu$+$\sigma)^{2}$, and a decrease with w$\sim$ 1/($\sigma)^{2}$. Same problem
arises for F([\ion{O}{iii}]/H$\beta$), where the w$\sim$ 1/($\sigma)^{2}$ gives a decrease. However, the two other weights
give an increase of the trend. But overall, we might have indications for an increased ionization in galaxies near quasars.

It looks also, like there might be a decrease of dust, and a slight shift towards bluer color. All indications taken
together, are we observing an increased AGN activity in some galaxies near quasars? A decrease of dust, slight shift of color,
increase of ionization F([\ion{O}{iii}]/H$\beta$) and F([\ion{O}{iii}]/[\ion{O}{ii}], and primarily the deviating
SFRs could support this scenario. 
\onecolumn
\begin{table*}
\centering
\begin{tabular}{c c c c c c}
\hline\hline
Parameter & $|\Delta z|$ & weight & $\alpha$ & $\sigma(\alpha)$ & trend size $\alpha$/$\sigma(\alpha)$\\
\hline
SFR-H$\alpha$ & 0.001 & 1/$\sigma^{2}$ & -0.00082 & 0.00133& 0 \\
- & 0.006 & - & -0.00003 & 0.00134 & 0 \\
- & 0.012 & - & -0.00020 & 0.00134& 0 \\
- & 0.001 & 1/($\mu$+$\sigma)^{2}$ & - & - & - \\
- & 0.006 & - & - & - & - \\
- & 0.012 & - & - & - & - \\
- & 0.001 & none & -0.00024 & 0.00131 & 0 \\
- & 0.006 & - & -0.00004 & 0.00118 & 0 \\
- & 0.012 & - & 0.00015 & 0.00128 & 0 \\
SFR-[\ion{O}{ii}] & 0.001 & 1/$\sigma^{2}$ & -0.00408 & 0.00164 & 2$\sigma$ \\
- & 0.006 & - & -0.00206 & 0.00160 & 1$\sigma$ \\
- & 0.012 & - & -0.00216 & 0.00158 & 1$\sigma$ \\
- & 0.001 & 1/($\mu$+$\sigma)^{2}$ & - & - & - \\
- & 0.006 & - & - & - & - \\
- & 0.012 & - & - & - & - \\
- & 0.001 & none & -0.00374 & 0.00215 & $<$ 2$\sigma$ \\
- & 0.006 & - & -0.00380 & 0.00186 & 2$\sigma$ \\
- & 0.012 & - & -0.00367 & 0.00180 & 2$\sigma$ \\
F([\ion{O}{iii}]/[\ion{O}{ii}] & 0.001 & 1/$\sigma^{2}$ & 0.00042 & 0.00034 & 1$\sigma$ \\
- & 0.006 & - & 0.00004 & 0.00042 & 0 \\
- & 0.012 & - & 0.00003 & 0.00041 & 0 \\
- & 0.001 & 1/($\mu$+$\sigma)^{2}$ & -0.00018 & 0.00043 & 0 \\
- & 0.006 & - & -0.00027 & 0.00040 & 0 \\
- & 0.012 & - & -0.00026 & 0.00039 & 0 \\
- & 0.001 & none & 0.00001 & 0.00080 & 0 \\
- & 0.006 & - & -0.00017 & 0.00071 & 0 \\
- & 0.012 & - & -0.00014 & 0.00069 & 0 \\
F([\ion{O}{iii}]/H$\beta$) & 0.001 & 1/$\sigma^{2}$ & 0.00140 & 0.00070 & 1$\sigma$ \\
- & 0.006 & - & 0.00140 & 0.00067 & 1$\sigma$ \\
- & 0.012 & - & 0.00139 & 0.00067 & 1$\sigma$ \\
- & 0.001 & 1/($\mu$+$\sigma)^{2}$ & -0.00029 & 0.00080 & 0 \\
- & 0.006 & - & -0.00043 & 0.00077 & 0 \\
- & 0.012 & - & -0.00049 & 0.00077 & 0 \\
- & 0.001 & none & -0.00134 & 0.00080 & $>$1$\sigma$ \\
- & 0.006 & - & -0.00149 & 0.00076 & 2$\sigma$ \\
- & 0.012 & - & -0.00178 & 0.00075 & 2$\sigma$ \\
F([H$\alpha$/H$\beta$) & 0.001 & 1/$\sigma^{2}$ & 0.00031 & 0.00065 & 0 \\
- & 0.006 & - & 0.00031 & 0.00062 & 0 \\
- & 0.012 & - & 0.00028 & 0.00062 & 0 \\
- & 0.001 & 1/($\mu$+$\sigma)^{2}$ & 0.00270 & 0.00168 & $<$ 2$\sigma$ \\
- & 0.006 & - & 0.00340 & 0.00169 & 2$\sigma$ \\
- & 0.012 & - & 0.00293 & 0.00166 & $<$ 2$\sigma$ \\
- & 0.001 & none & 0.00270 & 0.00158 & $<$ 2$\sigma$ \\
- & 0.006 & - & 0.00340 & 0.00169 & 2$\sigma$ \\
- & 0.012 & - & 0.00293 & 0.00166 & $<$ 2$\sigma$ \\
Color $U_{e}-R_{e}$ & 0.001 & none & 0.00024 & 0.00058 & 0 \\
- & 0.006 & - & 0.00057 & 0.00050 & 1$\sigma$ \\
- & 0.012 & - & 0.00059 & 0.00048 & 1$\sigma$ \\
Oxygen abundance & 0.001 & none & 0.00026 & 0.00088 & 0 \\
- & 0.006 & - & 0.00040 & 0.00085 & 0 \\
- & 0.012 & - & 0.00058 & 0.00083 & 0 \\
\hline
\end{tabular}
\label{LSAppBlueA}
\caption{1a. Blue neighbor galaxies. AGN excluded. $\mu$ is the median error in the sample.}
\end{table*}
\twocolumn


\onecolumn
\begin{table*}
\centering
\begin{tabular}{c c c c c c}
\hline\hline
Parameter & $|\Delta z|$ & weight & $\alpha$ & $\sigma(\alpha)$ & trend size $\alpha$/$\sigma(\alpha)$\\
\hline
SFR-H$\alpha$ & 0.001 & 1/$\sigma^{2}$ & -0.00024 & 0.00131 & 0 \\
- & 0.006 & - & -0.00004 & 0.00118 & 0 \\
- & 0.012 & - & 0.00015 & 0.00128 & 0 \\
- & 0.001 & 1/($\mu$+$\sigma)^{2}$ & - & - & - \\
- & 0.006 & - & - & - & - \\
- & 0.012 & - & - & - & - \\
- & 0.001 & none & -0.00024 & 0.00131 & 0 \\
- & 0.006 & - & -0.00004 & 0.00118 & 0 \\
- & 0.012 & - & 0.00015 & 0.00128 & 0 \\
SFR-[\ion{O}{ii}] & 0.001 & 1/$\sigma^{2}$ & -0.00274 & 0.00188 & 1$\sigma$ \\
- & 0.006 & - & -0.00293 & 0.00190 & 1$\sigma$ \\
- & 0.012 & - & -0.00294 & 0.00178 & $>$ 1$\sigma$ \\
- & 0.001 & - & - & - & - \\
- & 0.006 & - & - & - & - \\
- & 0.012 & - & - & - & - \\
- & 0.001 & none & -0.00274 & 0.00188 & $>$ 1$\sigma$ \\
- & 0.006 & - & -0.00293 & 0.00190 & $>$1$\sigma$ \\
- & 0.012 & - & -0.00294 & 0.00178 & $>$$\sigma$ \\
F([\ion{O}{iii}]/[\ion{O}{ii}] & 0.001 & 1/$\sigma^{2}$ & -0.00078 & 0.00019 & 4$\sigma$ \\
- & 0.006 & - & -0.00072 & 0.00019 & $>$ 3$\sigma$ \\
- & 0.012 & - & -0.00073 & 0.00019 & $<$ 4$\sigma$ \\
- & 0.001 & 1/($\mu$+$\sigma)^{2}$ & -0.00038 & 0.00028 & 2$\sigma$ \\
- & 0.006 & - & -0.00042 & 0.00026 & 2$\sigma$ \\
- & 0.012 & - & -0.00037 & 0.00029 & 2$\sigma$ \\
- & 0.001 & none & -0.00074 & 0.00051 & 1$\sigma$  \\
- & 0.006 & - & -0.00137 & 0.00055 & 2$\sigma$ \\
- & 0.012 & - & -0.00116 & 0.00055 & 2$\sigma$  \\
F([\ion{O}{iii}]/H$\beta$) & 0.001 & 1/$\sigma^{2}$ & 0.00133 & 0.00068 & 2$\sigma$ \\
- & 0.006 & - & 0.00130 & 0.00063 & 2$\sigma$ \\
- & 0.012 & - & 0.00131 & 0.00062 & 2$\sigma$ \\
- & 0.001 & 1/($\mu$+$\sigma)^{2}$ & -0.00032 & 0.00076 & 0 \\
- & 0.006 & - & -0.00048 & 0.00072 & 0 \\
- & 0.012 & - & -0.00049 & 0.00071 & 0 \\
- & 0.001 & none & -0.00133 & 0.00075 & 2$\sigma$ \\
- & 0.006 & - & -0.00148 & 0.00068 & 2$\sigma$ \\
- & 0.012 & - & -0.00164 & 0.00068 & 2$\sigma$ \\
F([H$\alpha$/H$\beta$) & 0.001 & 1/$\sigma^{2}$ & 0.00021 & 0.00063 & 0 \\
- & 0.006 & - & 0.00026 & 0.00062 & 0 \\
- & 0.012 & - & 0.00022 & 0.00061 & 0 \\
- & 0.001 & 1/($\mu$+$\sigma)^{2}$ & 0.00288 & 0.00157 & $<$ 2$\sigma$ \\
- & 0.006 & - & 0.00402 & 0.00152 & $>$ 2$\sigma$ \\
- & 0.012 & - & 0.00355 & 0.00148 & $>$ 2$\sigma$ \\
- & 0.001 & none & 0.00288 & 0.00157 & $<$ 2$\sigma$ \\
- & 0.006 & - & 0.00402 & 0.00152 & $>$ 2$\sigma$ \\
- & 0.012 & - & 0.00355 & 0.00148 & $>$ 2$\sigma$ \\
Color $U_{e}-R_{e}$ & 0.001 & none & 0.00061 & 0.00060 & 1$\sigma$ \\
- & 0.006 & - & 0.00048 & 0.00044 & 1$\sigma$ \\
- & 0.012 & - & 0.00044 & 0.00043 & 1$\sigma$ \\
Oxygen abundance & 0.001 & none & 0.00095 & 0.00100 & 1$\sigma$ \\
- & 0.006 & - & 0.00146 & 0.00095 & 1$\sigma$ \\
- & 0.012 & - & 0.00138 & 0.00093 & 1$\sigma$ \\
\hline
\end{tabular}
\label{LSAppAllA}
\caption{A. All neighbor galaxies. AGN excluded.}
\end{table*}
\twocolumn

\onecolumn
\begin{table*}
\centering
\begin{tabular}{c c c c c c}
\hline\hline
Parameter & z & weight & $\alpha$ & $\sigma(\alpha)$ & trend signal $\alpha$/$\sigma(\alpha)$\\ 
\hline
SFR-H$\alpha$ & 0.001 & 1/($\sigma^{2}$) & 0.00166 & 0.00192 & 1$\sigma$ \\
- & 0.006 & - & 0.00155 & 0.00132 & 1$\sigma$ \\
- & 0.012 & - & 0.00137 & 0.00214 & 0 \\
SFR-[\ion{O}{ii}] & 0.001 & 1/($\sigma^{2}$) & 0.00050 & 0.00206 & 0 \\
- & 0.006 & - & 0.00132 & 0.00206 & 0 \\
- & 0.012 & - & 0.00096 & 0.00198 & 0 \\
F([\ion{O}{iii}]/[\ion{O}{ii}] & 0.001 & 1/($\sigma^{2}$) & -0.00018 & 0.00030 & 0 \\
- & 0.006 & - & 0.00086 & 0.00039 & 2$\sigma$ \\
- & 0.012 & - & 0.00094 & 0.00036 & $>$ 2$\sigma$ \\
F([\ion{O}{iii}]/H$\beta$) & 0.001 & 1/($\sigma^{2}$) & -0.00003 & 0.00082 & 0 \\
- & 0.006 & - & -0.00044 & 0.00086 & 0 \\
- & 0.012 & - & -0.00038 & 0.00084 & 0 \\
F(H$\alpha$/H$\beta$ ) & 0.001 & 1/($\sigma^{2}$) & 0.00118 & 0.00096 & 1$\sigma$ \\ 
- & 0.006 & - & 0.00152 & 0.00090 & $>$ 1$\sigma$ \\ 
- & 0.012 & - & 0.00169 & 0.00084 & 2$\sigma$ \\ 
Color $U_{e}-R_{e}$ & 0.001 & none & -0.00048 & 0.00082 & 0\\
- & 0.006 & - & -0.00061 & 0.00070 & 1$\sigma$\\
- & 0.012 & - & 0.00005 & 0.00064 & 0\\
Oxygen abundance & 0.001 & none & -0.00162 & 0.00118 & 1$\sigma$\\
- & 0.006 & - & -0.00162 & 0.00118 & 1$\sigma$\\
- & 0.012 & - & -0.00162 & 0.00118 & 1$\sigma$\\
\hline
\end{tabular}
\label{LSField}
\caption{All non-AGN field galaxies.}
\end{table*}
\twocolumn

\onecolumn
\begin{table*}
\centering
\begin{tabular}{c c c}
\hline\hline
Parameter & galaxy type & trend signal A\\
\hline
SFR-H$\alpha$ & blue & no\\
SFR-H$\alpha$ & all & no\\
SFR-[\ion{O}{ii}] & blue & 1$\sigma$ $<$ A $<$ 2$\sigma$ (-)\\
SFR-[\ion{O}{ii}] & all & 1$\sigma$ (-)\\
F([\ion{O}{iii}]/[\ion{O}{ii}] & blue & 0$<$ A $<$ 1$\sigma$ (?)\\
F([\ion{O}{iii}]/[\ion{O}{ii}] & all & 1$\sigma$ $<$ A $<$ 4$\sigma$ (-)\\
F([\ion{O}{iii}]/H$\beta$) & blue & 1$\sigma$ $<$ A $<$ 2$\sigma$ (?)\\
F([\ion{O}{iii}]/H$\beta$) & all & 1$\sigma$ $<$ A $<$ 2$\sigma$) (?)\\
F(H$\alpha$/H$\beta$ ) & blue & 0$<$ A $<$ 2$\sigma$ (+)\\ 
F(H$\alpha$/H$\beta$ ) & all & 0 $<$ A $<$ 2$\sigma$ (+)\\ 
Color $U_{e}-R_{e}$ & blue & 0$<$ A $<$ 1$\sigma$ (+)\\
Color $U_{e}-R_{e}$\ & all & 1$\sigma$ (+)\\
Oxygen abundance & blue & no\\
Oxygen abundance & all & 1$\sigma$ (+)\\
\hline
\end{tabular}
\label{TrendSize}
\caption{First column describes the observed parameter. Second column the type of quasar neighboring galaxies. Third column describes the range of the detected trend signal depedning on used weighting, and if its an increase (-) or decrease (+) with shortened distance to the quasars. The question mark indicates that the trend even flips sign depending on the weighting model. All the AGN are excluded. The table is a summary of the tables introduced in the appendix, where details can be found.}
\end{table*}
\twocolumn

\begin{figure*}[htb!]
 \centering
   \includegraphics[width=16cm,height=10.5cm]{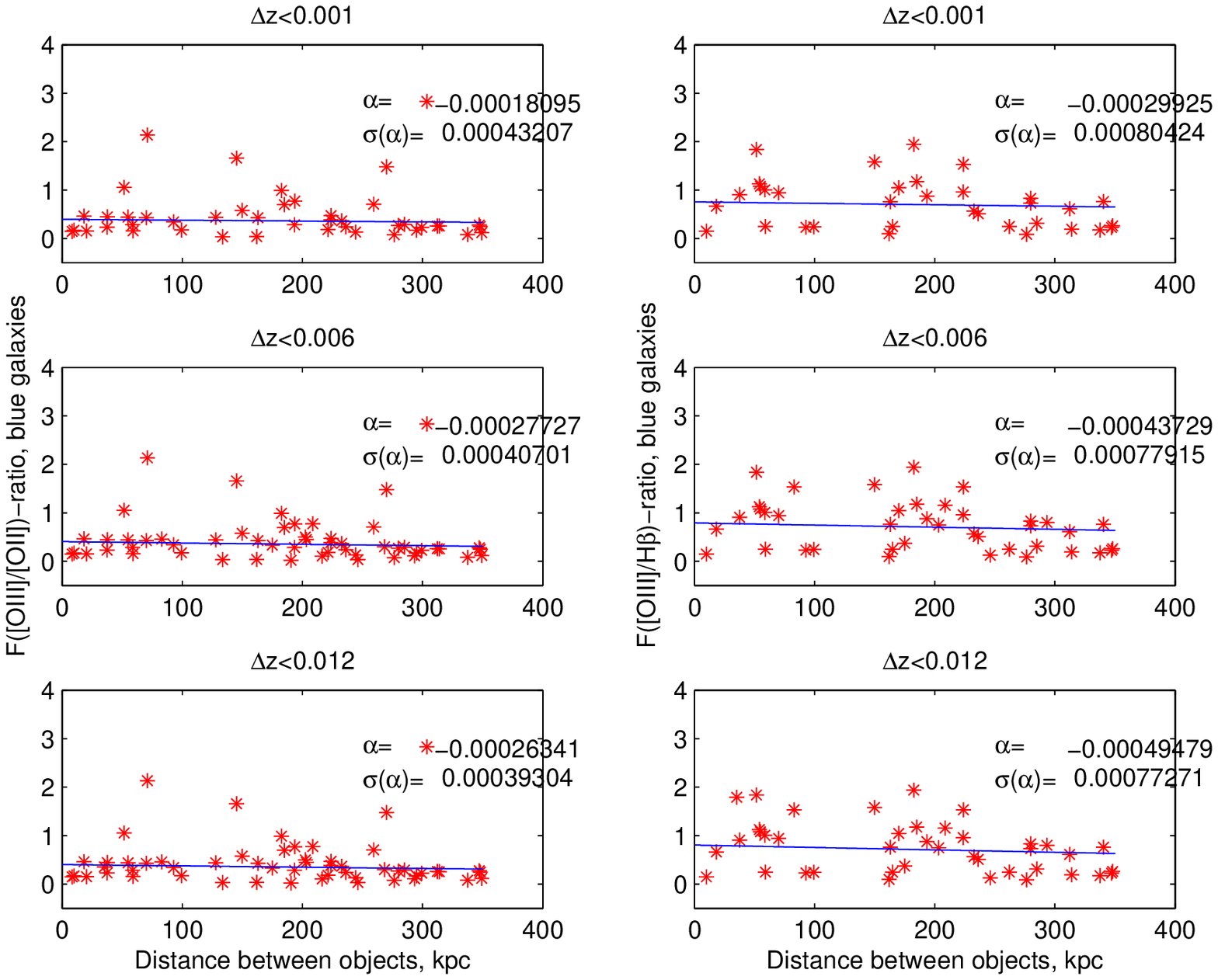}
     \caption{Ionization in blue companion galaxies. Right panel shows red companion galaxies. Three different redshift difference cuts are pictured for the pairs. Galaxies with individual errors in oxygen abundance$>$1000 are excluded. The individual errors are weighted with w$\sim$ 1/($\mu$+$\sigma)^{2}$.}
               \label{LesIonization4}
     \end{figure*}

\begin{figure*}[htb!]
 \centering
   \includegraphics[width=16cm,height=10.5cm]{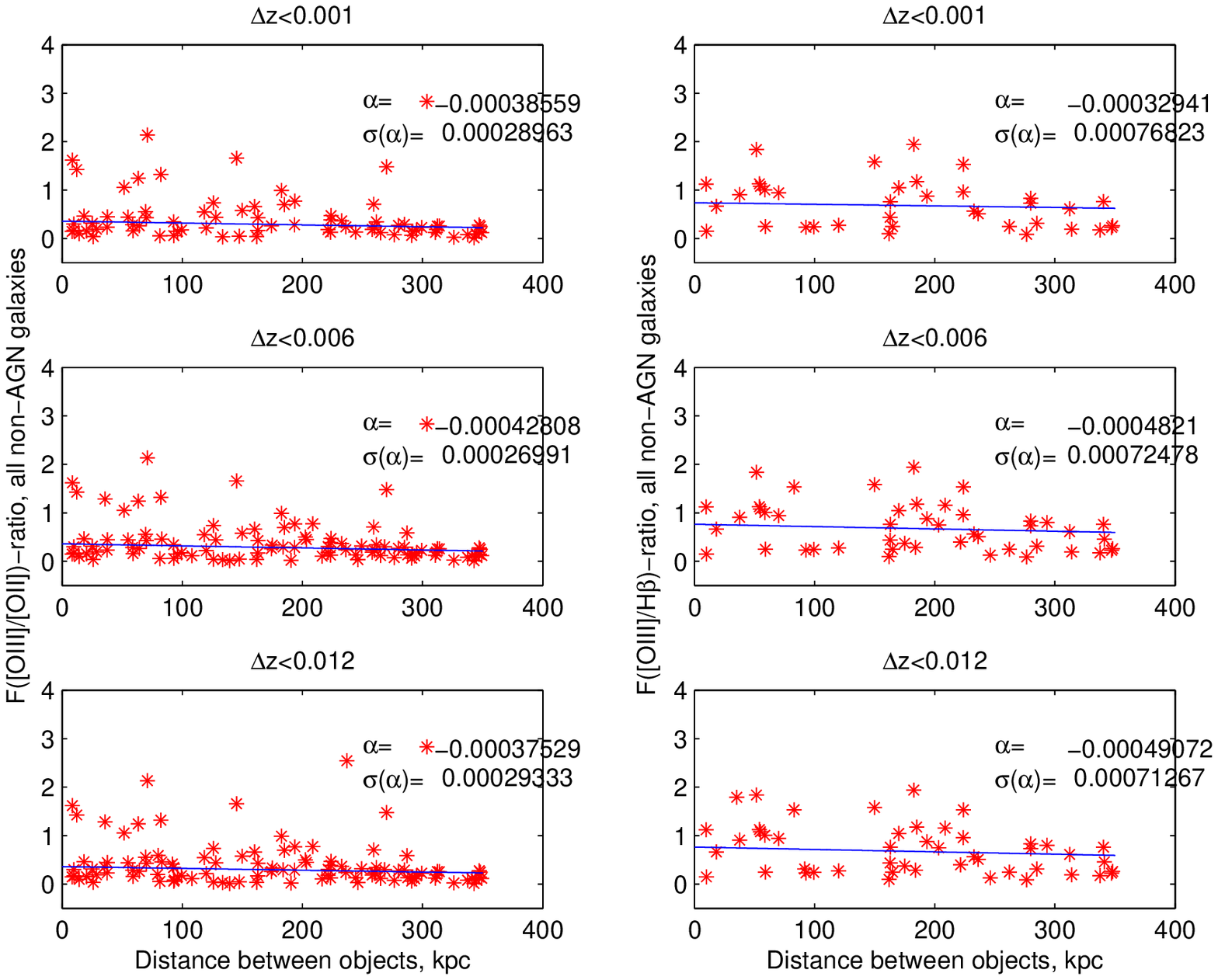}
     \caption{Ionization in all non-AGN companion galaxies. Right panel shows red companion galaxies. Three different redshift difference cuts are pictured for the pairs. Galaxies with individual errors in oxygen abundance$>$1000 are excluded. The individual errors are weighted with w$\sim$ 1/($\mu$+$\sigma)^{2}$.}
               \label{TotLesIonization4}
     \end{figure*}

\begin{figure*}[htb!]
 \centering
   \includegraphics[width=16cm,height=10.5cm]{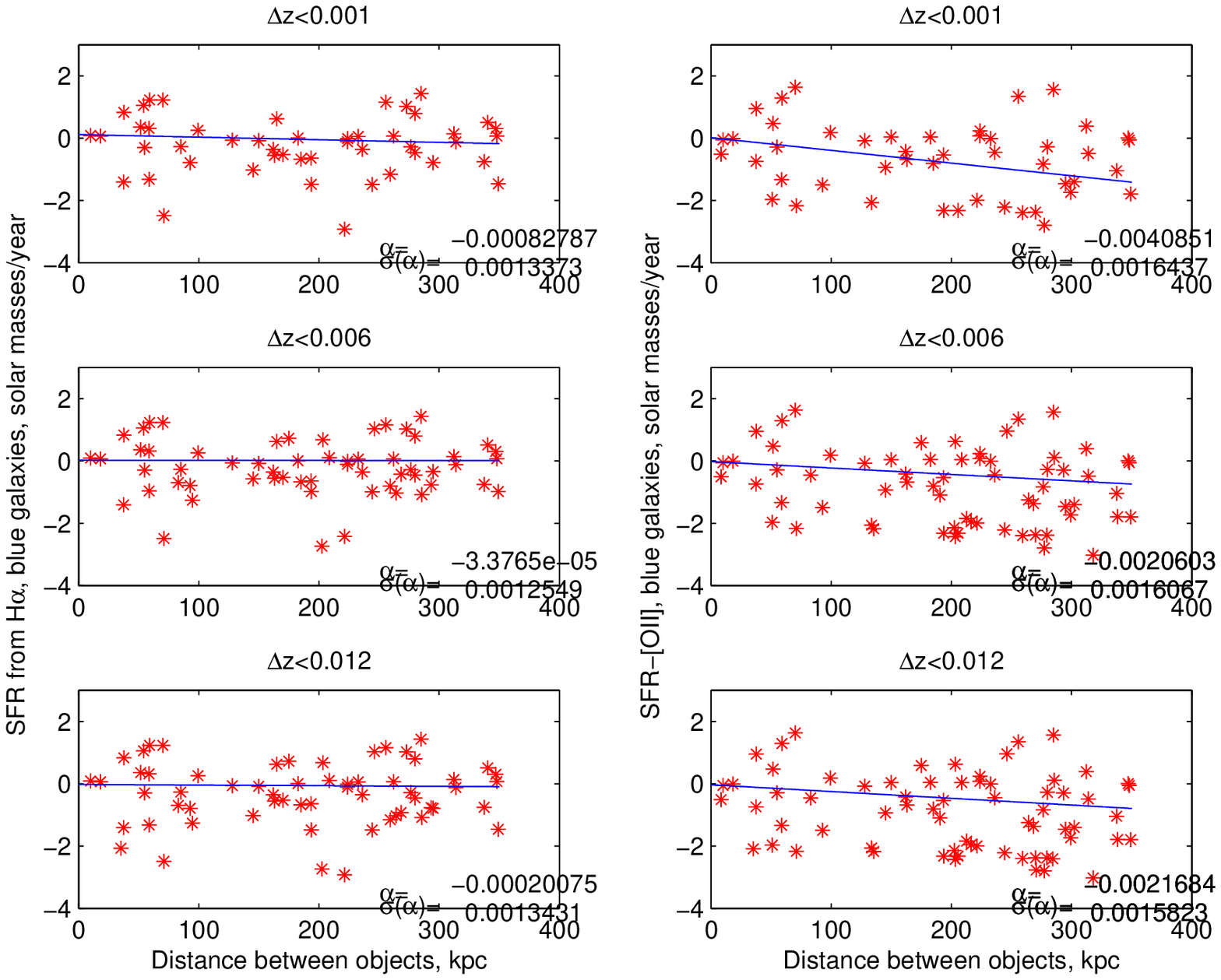}
     \caption{Star formation rate in blue companion galaxies. Left panel shows SFR from from H$\alpha$-flux. Right panel shows SFR calculated from [\ion{O}{ii}]-flux. Three different redshift difference cuts are pictured for the pairs. Galaxies with SFR$>$100 and individual errors in SFR $>$1000 are excluded. The individual errors are weighted with w$\sim$ 1/($\sigma)^{2}$. The SFR are displayed in log 10.}
               \label{LesSFR5C}
     \end{figure*}

\begin{figure*}[htb!]
 \centering
   \includegraphics[width=16cm,height=10.5cm]{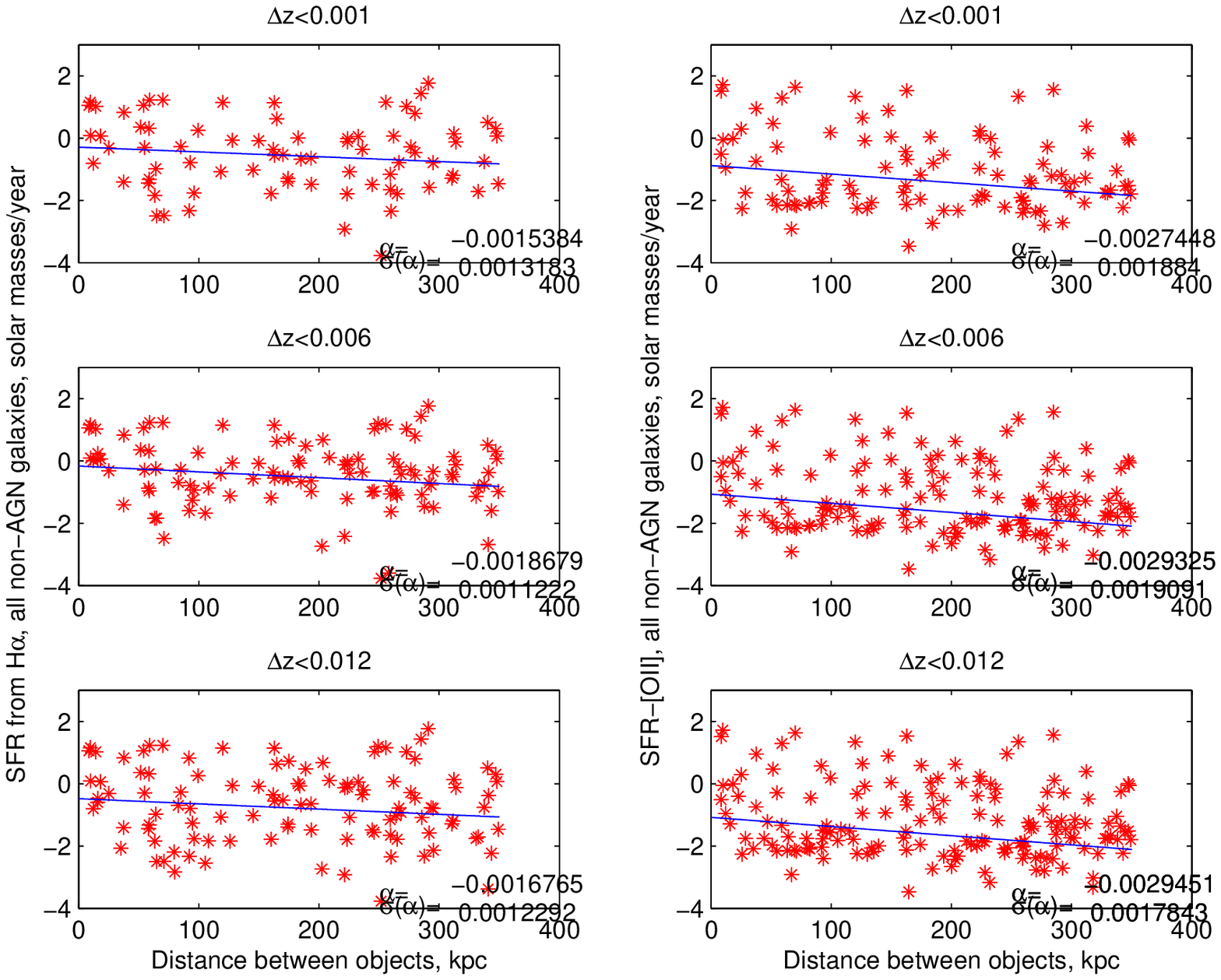}
     \caption{Star formation rate in all non-AGN companion galaxies. Left panel shows SFR from from H$\alpha$-flux. Right panel shows SFR calculated from [\ion{O}{ii}]-flux. Three different redshift difference cuts are pictured for the pairs. Galaxies with SFR$>$100 and individual errors in SFR $>$1000 are excluded. The individual errors are weighted with w$\sim$ 1/($\sigma)^{2}$. The SFR are displayed in log 10.}
               \label{TotLesSFR5C}
     \end{figure*}

\begin{figure*}[htb!]
 \centering
   \includegraphics[width=16cm,height=10.5cm]{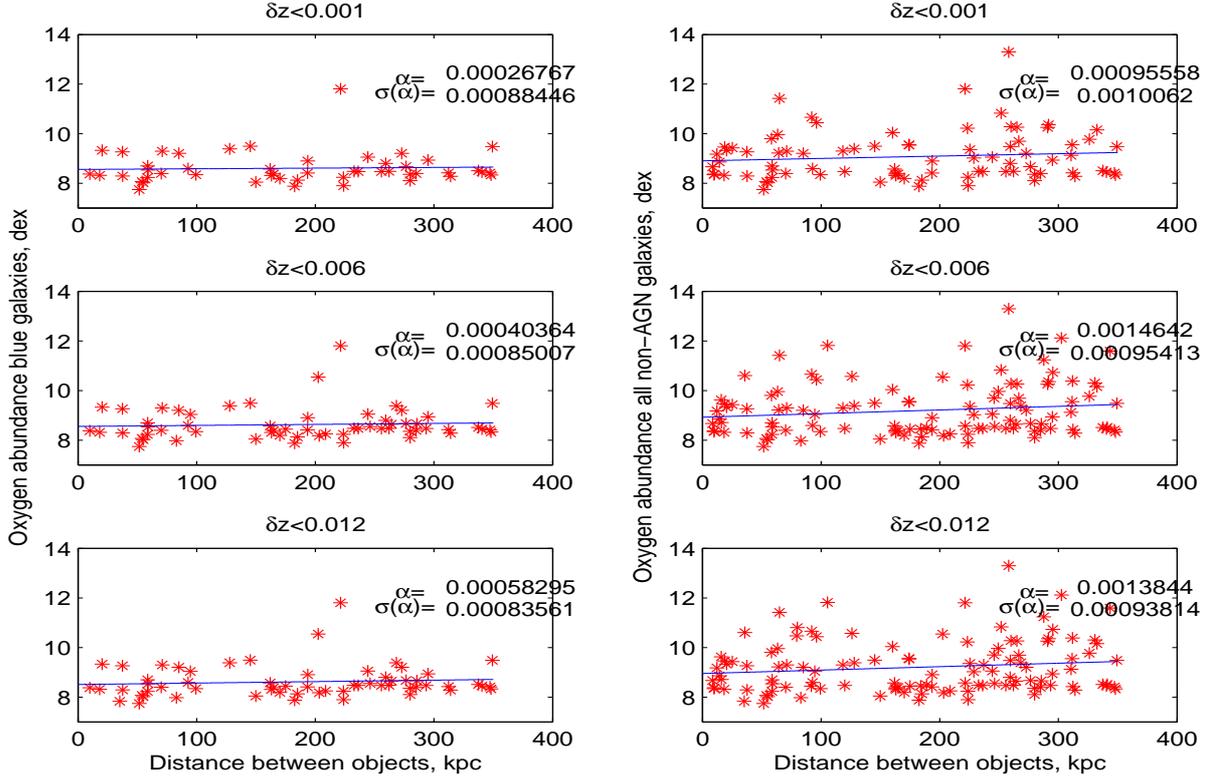}
     \caption{Oxygen abundance in companion galaxies, N2-method. Left panel shows only the blue non-AGN companion galaxies. Right panel shows all non-AGN companion galaxies. Three different redshift difference cuts are pictured for the pairs. Galaxies with individual errors in oxygen abundance$>$1000 are excluded. The individual errors are unweighted in the calculations of mean values and error bars since their relative errors could not be determined accurately.}
               \label{TotLesNFig77C}
     \end{figure*}

\begin{figure*}[htb!]
 \centering
   \includegraphics[width=16cm,height=10.5cm]{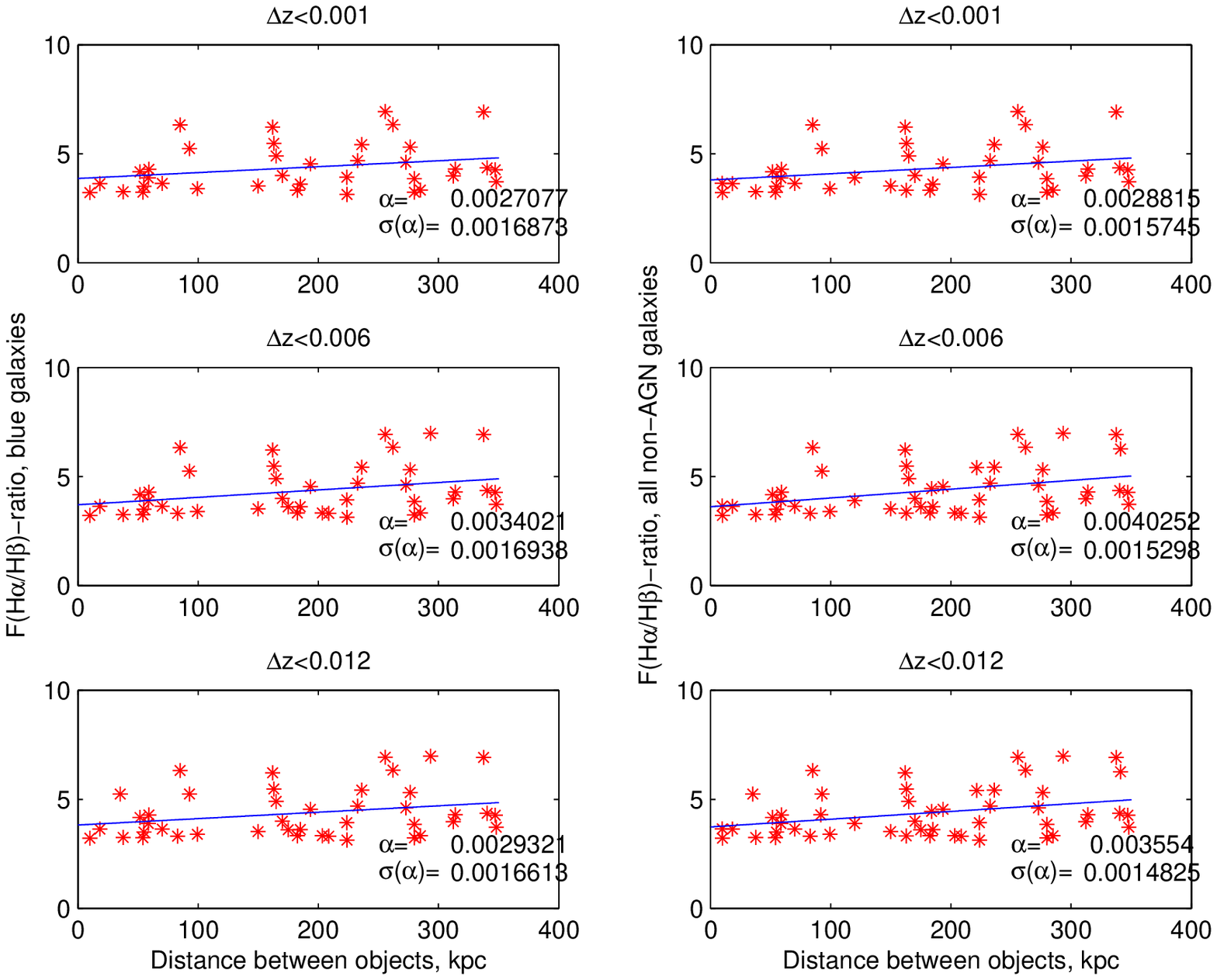}
     \caption{Dust extinction: the F(H$\alpha$/H$\beta$ ) in neighbor galaxies to quasars. Left panel show all blue galaxies with $U_{e}-R_{e}$$<$2.2. Right panel shows all non-AGN companion galaxies. Three different redshift difference cuts are pictured for the pairs. Galaxies with individual errors in oxygen abundance$>$1000 are excluded. The individual errors are weighted with w$\sim$ 1/($\mu$+$\sigma)^{2}$.}
               \label{TotLesNFig70F}
     \end{figure*}

\subsection{Surface-densities of non-AGN companions}

The surface density is the easiest way of imagining how the number of galaxies changes 
at the distance from the quasar. The density of the environment around the quasar could support possible merger 
scenarios if small over-densities around the objects are observed.

We calculated the surface densities using

\begin{equation}
\rho=N_{distance}/((d_{1})^2-(d_{2})^2)\pi
\end{equation}\label{eq3}

with N, being the number of galaxies in one bin (or "annuli") with outer radius $d_{1}$ and inner radius $d_{2}$.
The error bars were generated with bootstrapping \citep{Tibby}.

\begin{figure*}[htb!]
 \centering
   \includegraphics[width=16cm,height=10.5cm]{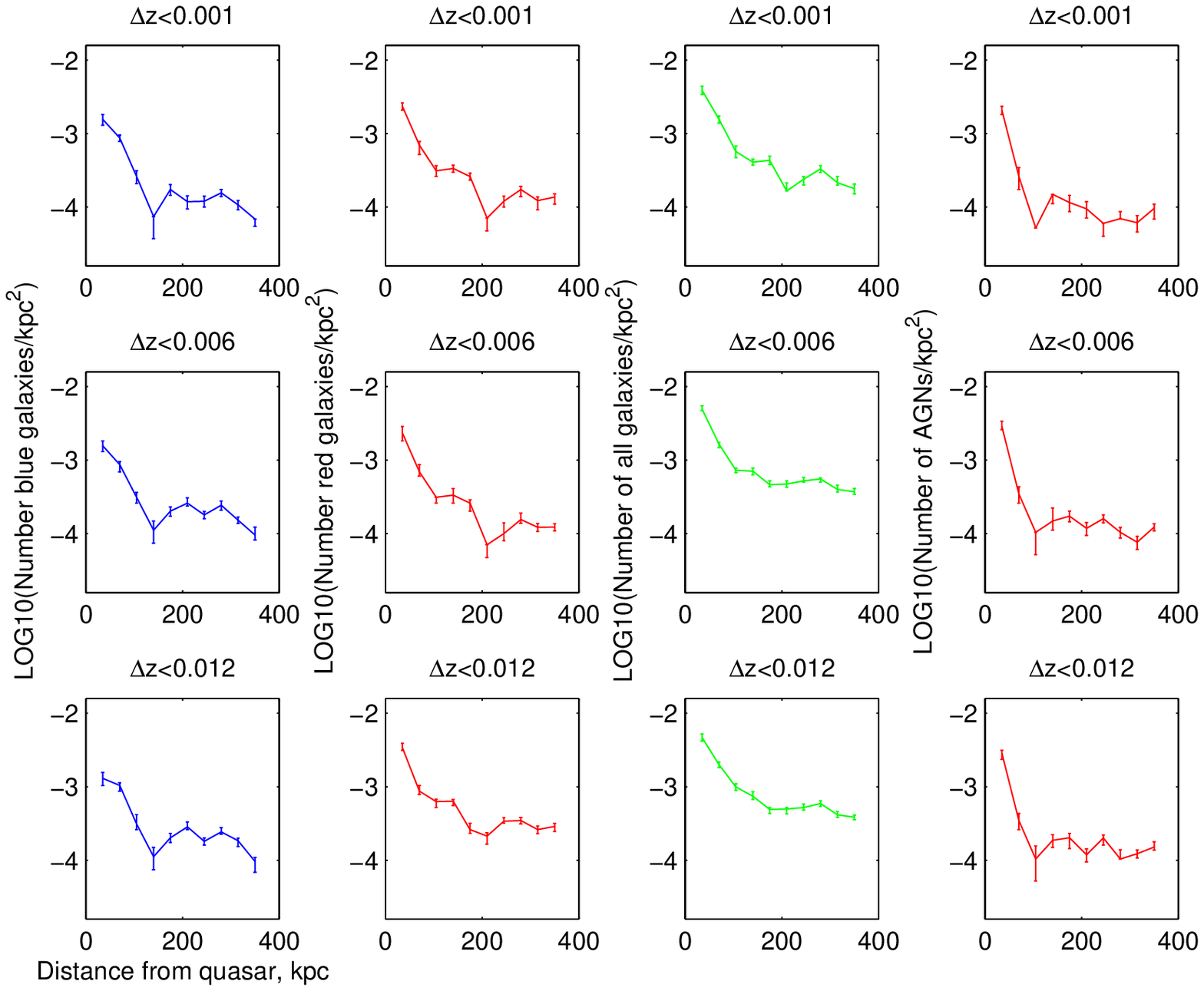}
   \caption{Annular surface densities of different types of galaxies around quasars. Surface density distribution 
     of galaxies. In the a) left panel, blue galaxies with $U_{e}-R_{e}$$<$2.2, in b) first middle, red galaxies, $U_{e}-R_{e}$$>$2.2 and
     in c) the right middle panel, all galaxies d) AGN. Three different redshift difference cuts are
     pictured for the pairs.}
               \label{endNFig85}%
     \end{figure*}

First were the companion galaxies binned up into 10 pieces each of 35 kpc (to cover the whole projected distance
between quasar and companion galaxy) for each bootstrapping round. We used annular surface areas, i.e. the
number of galaxies as a function of distance between quasar and galaxy were divided up in different 'annuli' 
(rings) around the quasars. When this was done, we performed bootstrapping on our sample 100 times to calculate 
surface densities in each bin and accompanying error bars. The error bars in the surface densities are plotted 
in fig \ref{endNFig85} with a 68.3\% confidence level, corresponding to 1 $\sigma$.

In general, the relative number of red background galaxies at distances further than d $>$ 100 kpc 
decreases with the smaller $|\Delta z|$ we use. Could this show the need to use very small 
$|\Delta z|$ cuts when we search for galaxies that have a projected separation of d $>$ 100 kpc? Indeed, 
previous studies \citep[e.g.][]{Yee1983} have shown that galaxies that lie within 100 kpc in projected 
distance have the largest chance to be a physical pair.

We see in figure \ref{endNFig85} large overdensites of galaxies around the quasars. This has been seen before, 
but only in samples with large contamination of background and foreground galaxies \citep[e.g.][]{Serber2006}. 
This is the first study with spectroscopic galaxies with well-defined redshifts made on such a small-scale environment 
near quasars (d $<$ 350 kpc) that yields the same result. However, the results presented herein 
are much more clear cut. This gives support to quasar formation either via collapse of materia
into huge dark matter halos, or a merger-driven scenario behind the birth of quasars, where massive objects 
get created via hierarchical assembling of smaller parts. 

Another curious thing about the surface density plots is the sudden drop of blue 
galaxies in the bin around 150 kpc. The same is seen in the AGN surface density.
The statistical significance of these drops can be questionated, but we do however
wish to point them out for the reader.

\subsection{AGN activity}

AGN are interesting to study since they also are hypothethized to be products of interactions. 
If this is true, a larger amount of active galactic nuclei and a larger accretion rate on the black hole 
in galaxies at small projected separations would be expected. Still, the question remains whether AGN 
could be triggered by the interactions in general, or only close to another AGN(for instance, a quasar). 
AGN usually exhibit stronger [\ion{O}{iii}]-emission due to the photoionization caused by the continuum radiation 
from the AGN \citep[e.g.][]{Lamastra}. Other emission lines that are used to separate AGNs from starburst galaxies are 
the [\ion{N}{ii}]$\lambda$6731 and the H$\beta$ line and H$\alpha$ line. 
The method to select neighbor galaxies hosting AGN is decribed in section 3.4.1.

\subsubsection{Surface density distribution of AGN neighbors}

Fig. \ref{endNFig85} shows a clear and sudden increase in AGN surface density closer to a quasar. 
These results have not previously been shown in quasar environments. For example Coldwell \& Lambas (2006)
found no difference in the amount of AGN around a quasar. Contrary to them we see that there are more 
AGN at very close distances to the quasar, although the background of AGN is very low. 
Perhaps this is a chance result, but we would need a much larger number of quasar-galaxy pairs 
using the same sensitivity in the present analysis to confirm this.

Could this perhaps mean that a quasar triggers an AGN? Or could it be that the AGN were produced in 
the same interaction that created the quasar?

\subsubsection{Some calculated properties of AGN neighbors}

\begin{figure*}[htb!]     
 \centering
   \includegraphics[width=16cm,height=10.5cm]{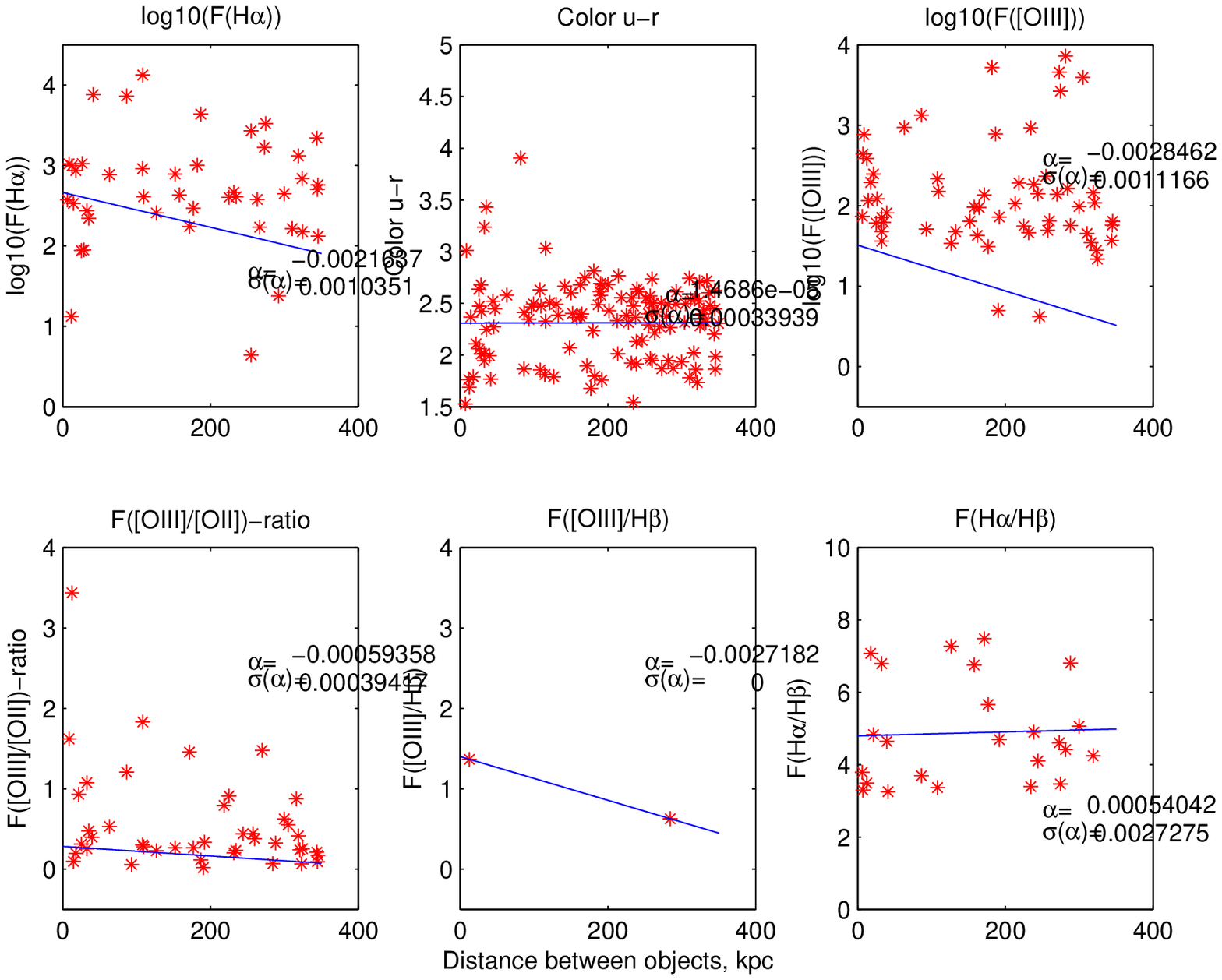}
     \caption{Properties with $\delta z<$ 0.006 for AGN. a) SFR(H$\alpha$) b) Color U$_{e}$-R$_{e}$ c) accretion rate F([\ion{O}{iii}])
     d) F([\ion{O}{iii}]/[\ion{O}{ii}]) e) F([\ion{O}{iii}]/H$\beta$) f) F(H$\alpha$/H$\beta$})
               \label{AFig47}%
     \end{figure*}

To carefully investigate whether some clear relations exist between such properties as 
accretion rate, F[\ion{O}{iii}], SFR or colors I have plotted some of them for neighbor AGN only. The results 
are depicted in fig. \ref{AFig47}.

It seems that no clear relation exists; no increased accretion rate as expected close to a quasar, no 
increased SFR, no changes in color. This could mean that while the quasar itself is not affecting the 
presence of an AGN, the same event that created the quasar could have created the AGN. The number of AGN is
probably too small to be significant, but hopefully in future we will be able to redo this study with a much larger number of quasar-AGN associations.

\subsubsection{Accretion rate in non-AGN neighbors}

As an additional investigation, I decided to plot the F([\ion{O}{iii}]) that is also a measure of accretion rate.
The drawback is that it requires corrections for the contribution from star formation which I have not done. 
The large error bars show clearly that any trend cannot be trusted. Still, from these plots one could wonder whether 
one would not observe a slightly increased F([\ion{O}{iii}]) near the quasar, since both blue and red galaxies
(with the exception of the large error bar due to deficit of blue galaxies in the blue plot) tend
to get a slightly increased F([\ion{O}{iii}]) near the quasar. 

\begin{figure*}[htb!]
\centering
\includegraphics[width=16cm,height=10.5cm]{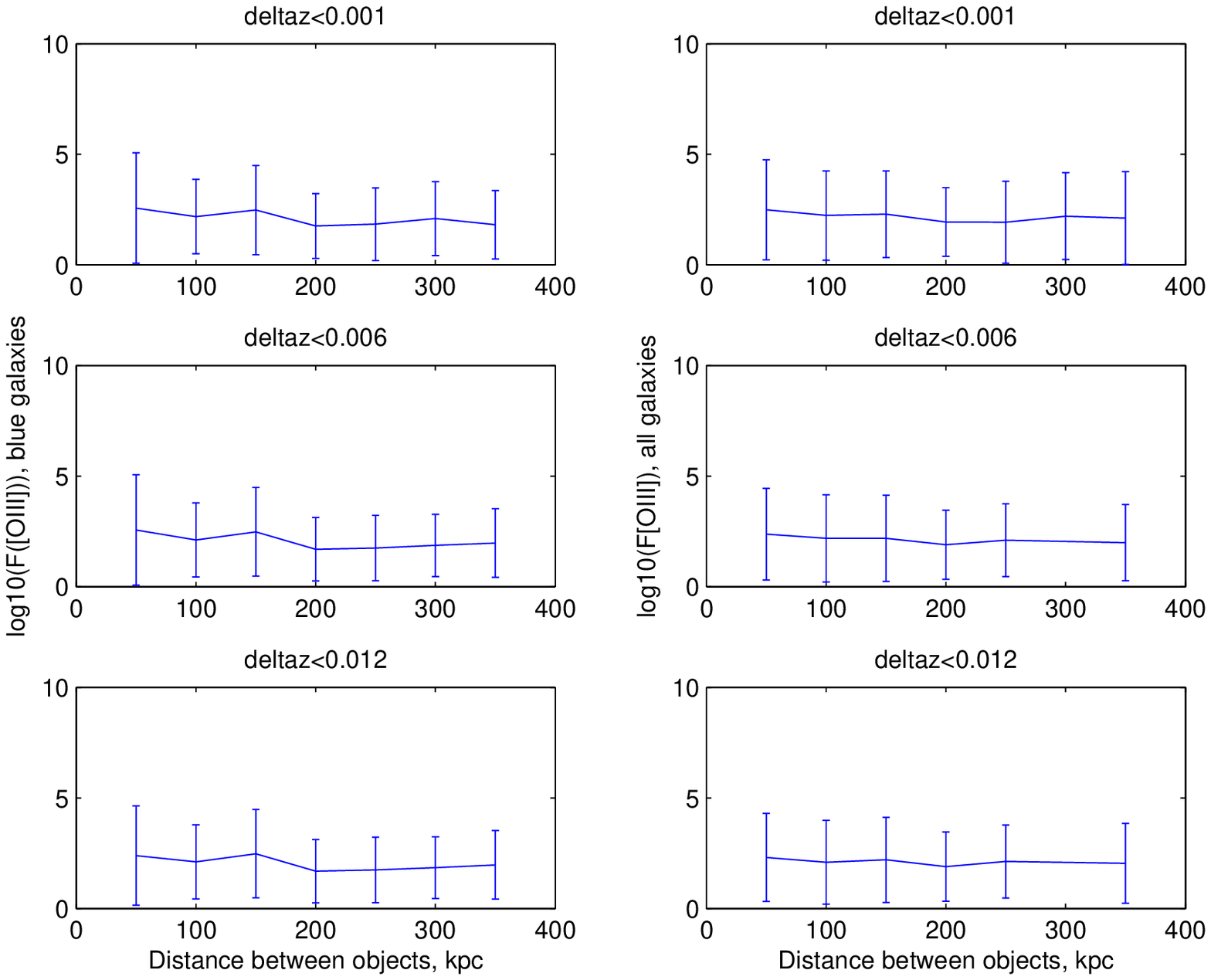}
\caption{Accretion rate F([\ion{O}{iii}]) in companion galaxies.}
\label{NFig79}%
\end{figure*}

\subsection{The $|\Delta z|$  cut}

The $|\Delta z|$  cut was introduced in the introductory chapter and will now have an explanation. 
The cut is used to select quasar-galaxy associations by claiming that all those associations
that have a $|\Delta z|$  $<$ X (where X is the chosen number) are a physical couple located at the
actual redshift of the galaxy. Many studies use photometric redshifts of galaxies that have redshift errors
$\delta$z $>$0.025, hence including both foreground and background galaxies. 

How could such a cut as ours influence the results? Could we observe an increased SFR or changes in color
for our companion galaxies if we have a very large $|\Delta z|$  ? Could the choice of method thus bias our results?

\begin{figure*}[htb!]
\centering
\includegraphics[width=16cm,height=10.5cm]{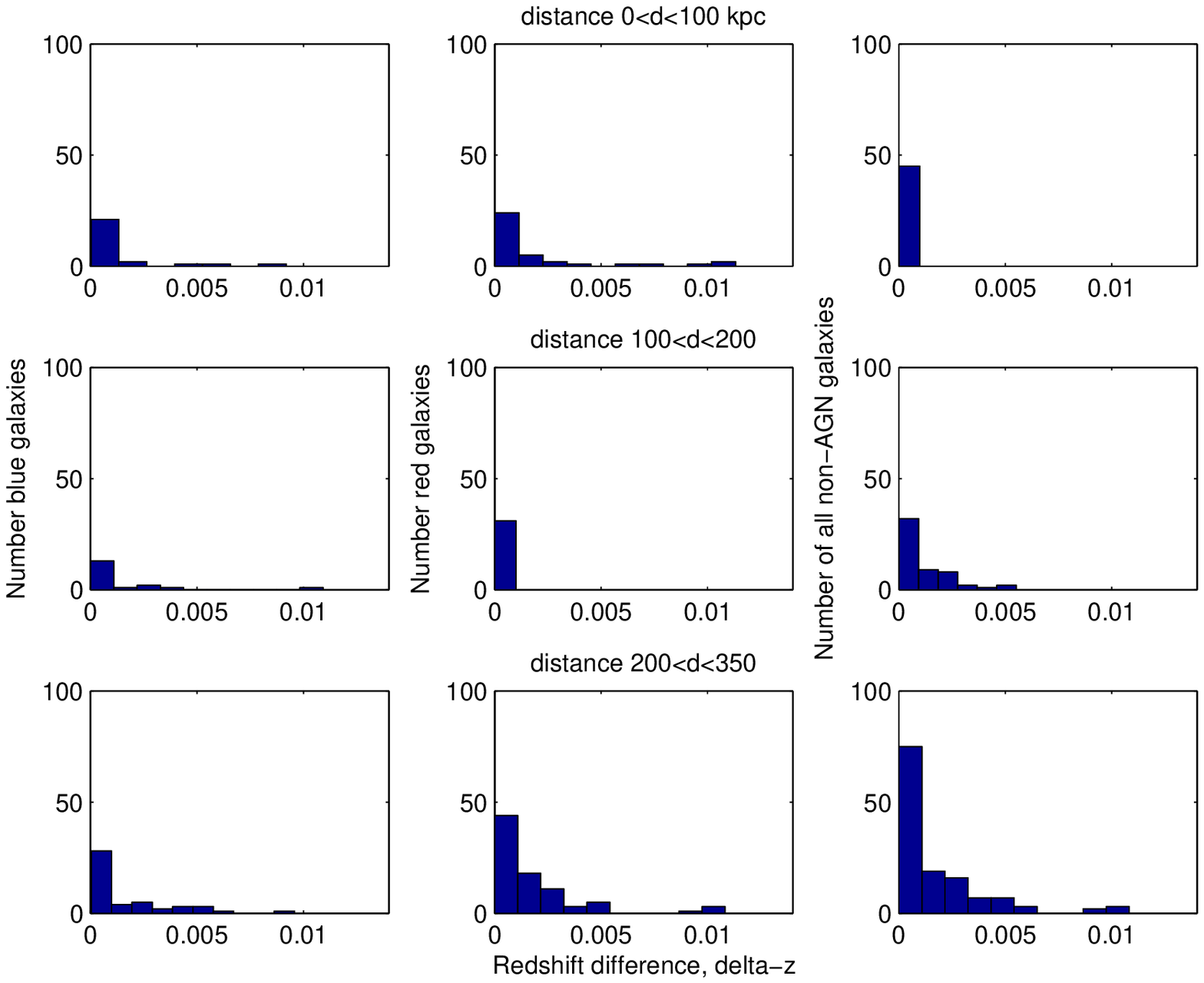}
\caption{Number of companion galaxies in different $|\Delta z|$  bins. The $|\Delta z|$  = 0.001 corresponds to a physical distance
of 4.2 Mpc, $|\Delta z|$  = 0.006 to 26 Mpc and $|\Delta z|$  = 0.012 to 52 Mpc. Left panel: blue galaxies. Middle panel: red galaxies.
Right panel: all galaxies.}
\label{NFig83}%
\end{figure*}

\begin{figure*}[htb!]
\centering
\includegraphics[width=16cm,height=10.5cm]{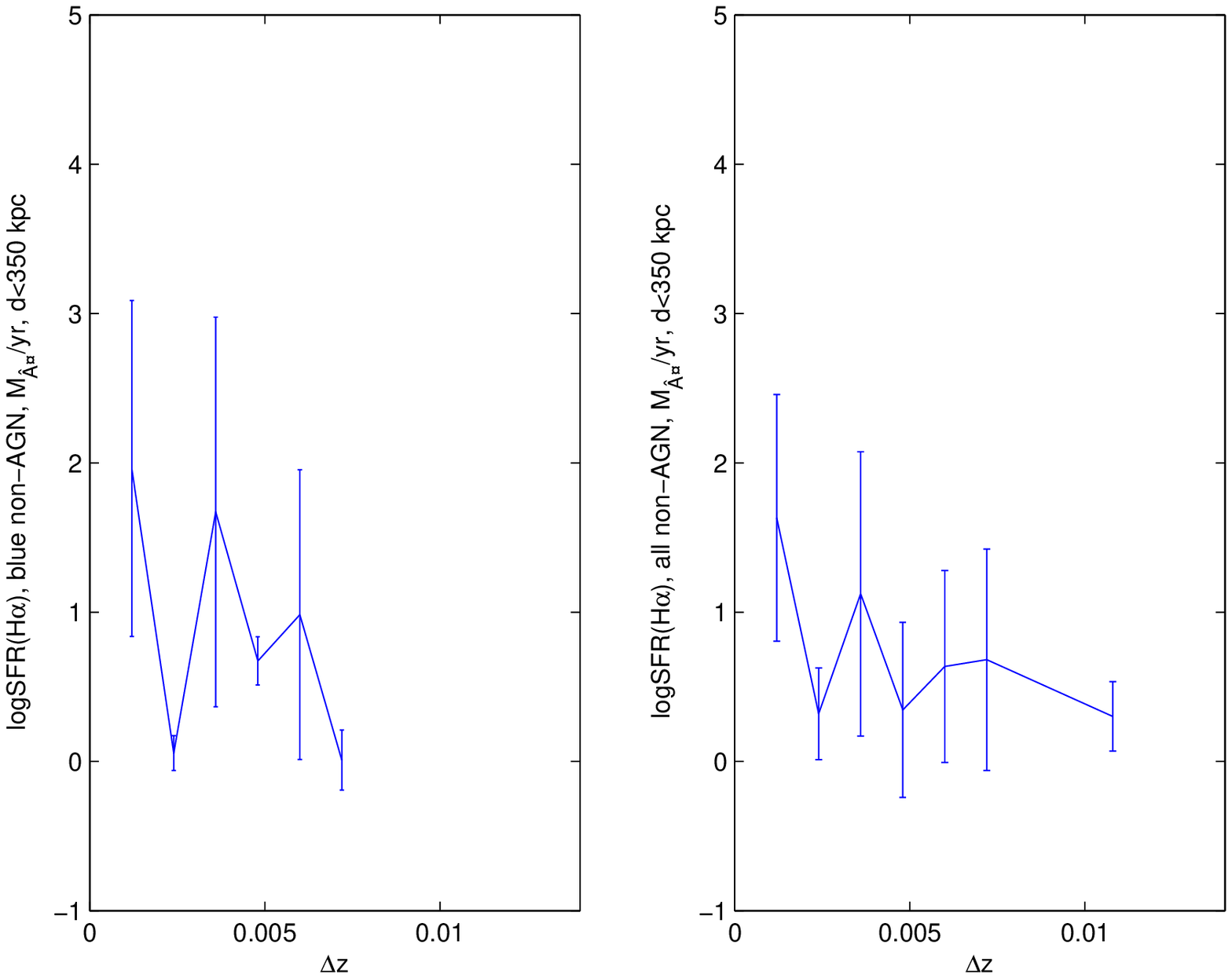}
\caption{SFR from H$\alpha$ depending on $|\Delta z|$ . The $|\Delta z|$  = 0.001 corresponds to a physical distance
of 4.2 Mpc, $|\Delta z|$  = 0.006 to 26 Mpc and $|\Delta z|$  = 0.012 to 52 Mpc. Left panel: blue galaxies. Right panel: red galaxies.}
\label{NFig65}%
\end{figure*}
     
\begin{figure*}[htb!]
\centering
\includegraphics[width=16cm,height=10.5cm]{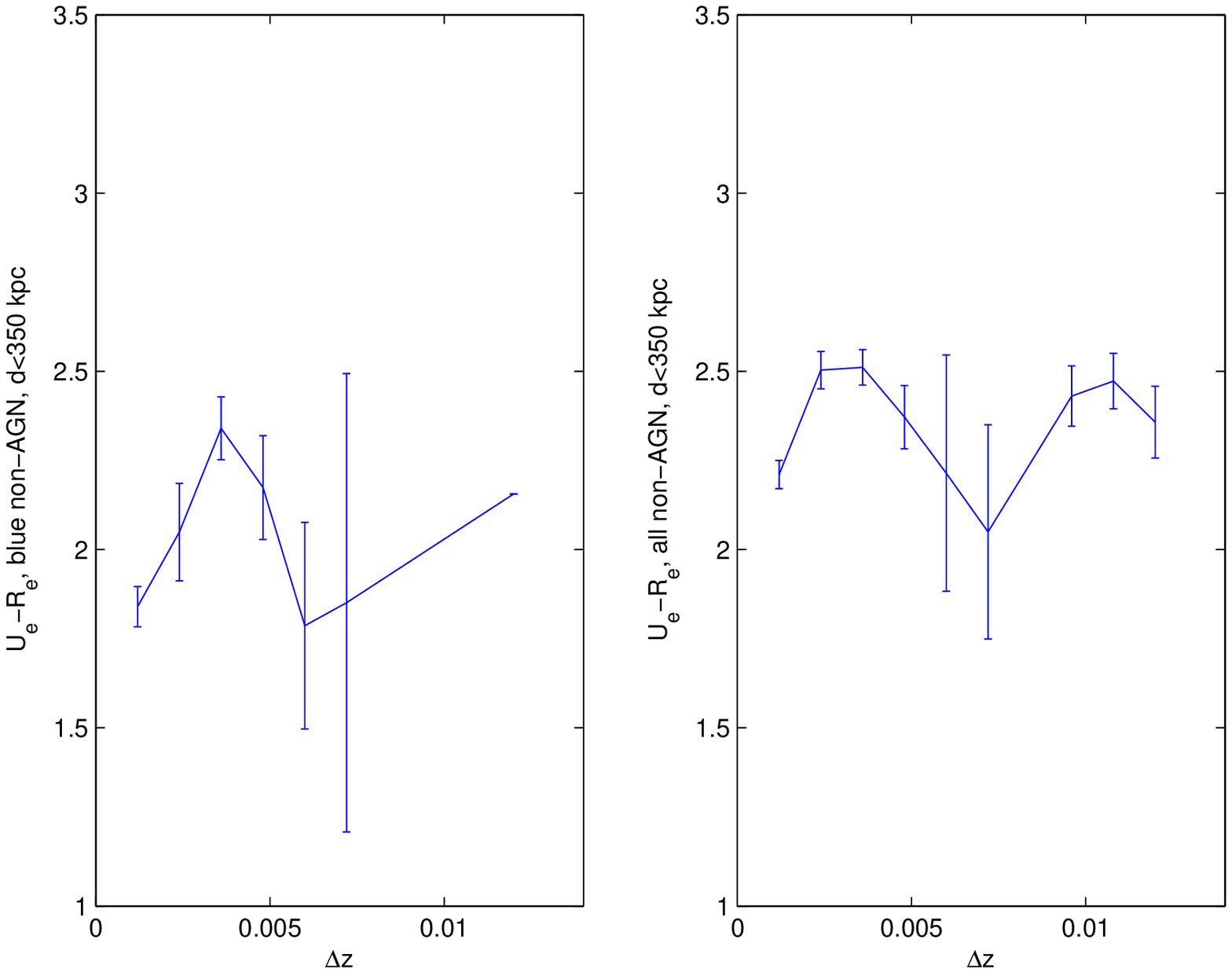}
\caption{Color of companion depending on $|\Delta z|$ . The $|\Delta z|$  = 0.001 corresponds to a physical distance
of 4.2 Mpc, $|\Delta z|$  = 0.006 to 26 Mpc and $|\Delta z|$  = 0.012 to 52 Mpc. Left panel: blue galaxies. Right panel: red galaxies.}
\label{NFig64}%
\end{figure*}

\begin{figure*}[htb!]
\centering
\includegraphics[width=16cm,height=10.5cm]{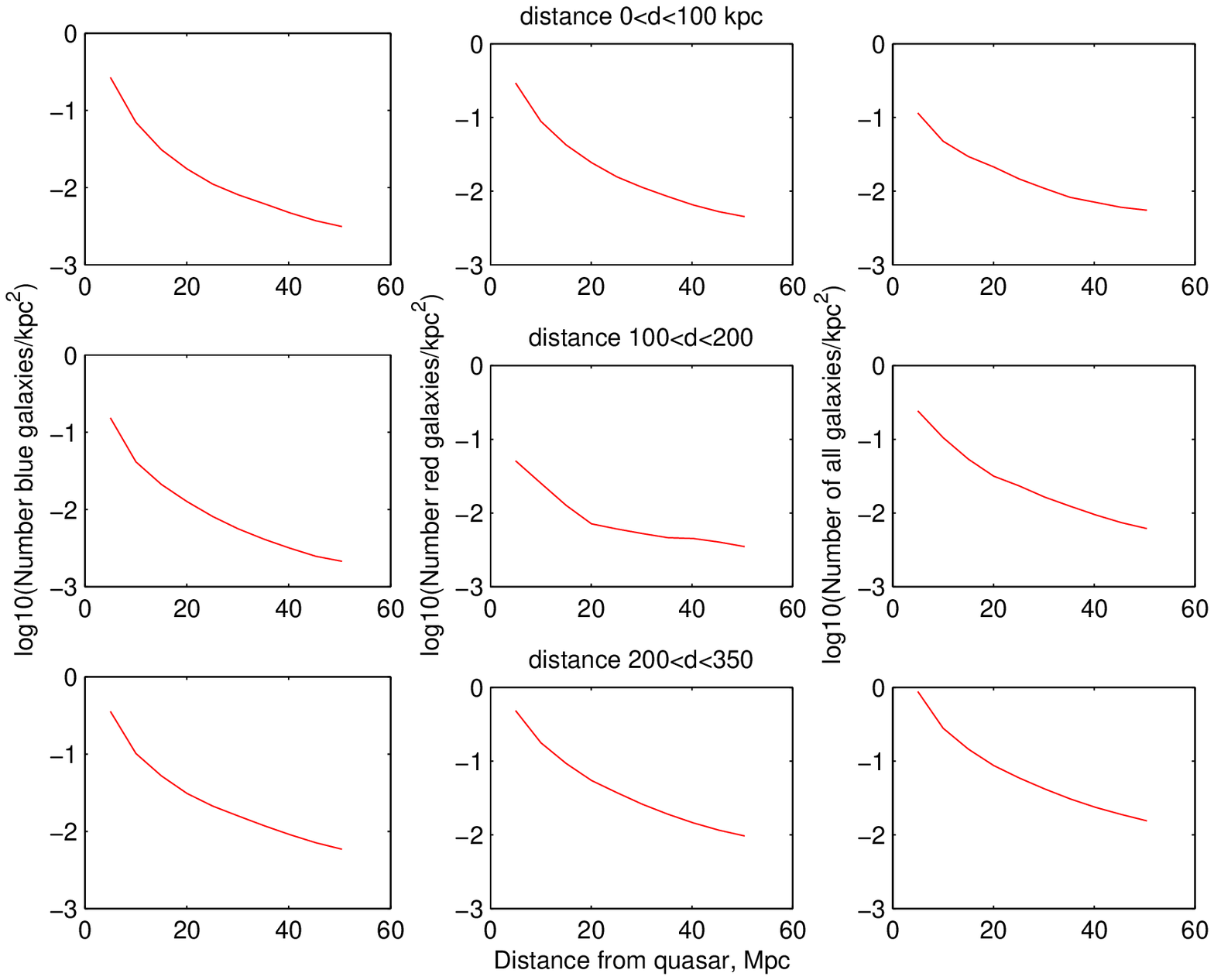}
\caption{Surface density $|\Delta z|$ . The $|\Delta z|$  = 0.001 corresponds to a physical distance
of 4.2 Mpc, $|\Delta z|$  = 0.006 to 26 Mpc and $|\Delta z|$  = 0.012 to 52 Mpc.
Left panel shows blue galaxies. Middle panel shows red galaxies. Right panel shows all galaxies.}
\label{NFig61}%
\end{figure*}

If watching the intrinsic properties of each plot one finds a very small dependence of SFR and color with 
the $|\Delta z|$  cut when we are examining quasar-galaxy associations that are assumed to lie within a projected
distance of 350 kpc. Both for the blue non-AGN neighbour galaxies, as well as for all non-AGN neighbours,
one sees that for small $|\Delta z|$ , neighbour galaxies appear both slightly bluer and have slightly higher SFR
than those with $|\Delta z|$ =0.012. On the other hand the color differences are too small to be important on small scale studies with projected distances reaching maximum d $<$ 350 kpc.

But how about the many studies that investigate neighborhoods up to several megaparsecs from the quasars?
Here the small differences could become significantly large. Coldwell and Lambas (2006) observed a higher 
fraction of star-forming galaxies over large scales around quasars using a very refined $|\Delta z|$  $<$ 0.007. 
Could their interesting observations maybe be interpreted not as particularly increased star formation 
around quasars but rather that at larger scales they simply had much higher contamination of projected galaxies?

\section{Discussion and conclusions}

We have used the Sloan Digital Sky Survey (SDSS) and utilized spectral lines, 
dereddened apparent magnitudes, redshifts and angular distances of 305 quasar-galaxy 
associations at redshifts 0.03 $<$ z $<$ 0.2 and projected distances less than 350 kpc, to study 
quasar neighborhoods.

We corrected the colors and spectral lines for dust extinction in the non-AGN companion galaxies. 
We calculated colors, surface densities, ionization ratios 
F([\ion{O}{iii}]/[\ion{O}{ii}]) and F([\ion{O}{iii}]/H$\beta$), oxygen abundances and star-formation rates (SFR) from H$\alpha$ and [\ion{O}{ii}] 
fluxes for the non-AGN galaxies. We also estimated the surface density distribution of AGNs in our galaxy sample using BPT-diagram combined 
with Kauffmann's criteria \citep{Kauffmann2003} and the H$\alpha$ width. To see how different properties of the companion galaxy 
vary as function of distance from the quasar, we binned the surface and number densities in bins of 35 kpc
and the rest of the above parameters in bins of 50 kpc, the latter in order to be easier comparable 
to similar studies with galaxy-galaxy pairs \citep[e.g.][]{Nikolic}. We excluded the AGN in our non-AGN-galaxy neighbor sample, 
and calculated the average parameter value and error bar for all galaxies within each bin. We plotted the 
average values of color, SFR, ionization parameters, dust extinction and oxygen abundance in each bin as function of distance between 
galaxy and quasar. Finally, we plotted LSAs of the same parameters as function of distance from the quasars, and estimated the maximum 
correlation from three different weights. Some conclusions are:

\begin{enumerate}

\item No correlation between color and distance in the binned plots, while the LSAs give a maximum 1$\sigma$
decrease of $U_{e}-R_{e}$ towards the bluer near the quasars.

\item We cannot see any difference in dust extinction depending on distance from quasar in the bins. Partially, 
this might be a problem of having few galaxies with well-measured H$\beta$, which stresses the importance of further studies.
However, the LSAs show a maximum 2$\sigma$ decrease of dust near the quasars.

\item There is a non-agreement between the H$\alpha$- and [\ion{O}{ii}]-derived star-formation rates in both 
the binned plots and the LSAs. In the LSAs, no change of H$\alpha$-SFR is seen at decreased distance from the quasars,
while the [\ion{O}{ii}]-derived SFRs show an increase 1$\sigma$ $<$ trend size $<$ 2$\sigma$ near the quasars.

\item No difference in ionization is seen whatsoever, neither by investigating F([\ion{O}{iii}]/[\ion{O}{ii}]) nor
F([\ion{O}{iii}]/H$\beta$) in the bins, while LSAs can show a 0 $<$ trend size $<$ 4$\sigma$ increase near quasars.

\item There is a too large scatter in the oxygen abundance plot in order to see any correlation
of the oxygen abundance with distance. A max 1$\sigma$ decrease can be seen in the LSAs.

\item There is a gap at the distance of 150 kpc in the number of AGN and as well 
as in the number of blue galaxies, something that might depend on too low number 
to make good statistics from, or more interesting physics. Further studies with 
increased number of galaxies are of very high importance.

\item The most important conclusion we have is a large increase in the surface density 
of all galaxies closer than 100 kpc to the quasars in both samples, 
which could support a merger scenario behind quasar activity. In case mergers are responsible 
for quasar activity, perhaps could some of the AGNs in the galaxy sample can be secondary products 
from the same mergers?
\end{enumerate}

\subsection{Greatly enhanced surface densities}

We investigated the number densities of non-AGN companion galaxies. In the 
closest bins in column 1 we see a significant increase in the number of galaxies, especially
blue galaxies. From the same figure we can also see that when using larger $|\Delta z|$ cuts (0.006 and 0.012) 
the relative number of galaxies in the furthest bins increase significantly. This is a result of including more 
background and foreground galaxies, which has a considerably higher probability for greater 
projected distances \citep{Nikolic} at the larger $|\Delta z|$ cuts. We had more red galaxies than 
blue galaxies in our non-AGN-neighbor sample, especially further out in projected distance.

Histogram binning is sensitive to the number of bins chosen. Using a wrong number
of bins can make ordinary noise appear as patterns in histograms and giving space for misinterpretations
in the analysis process. To avoid this we tried using a histogram binning program that calculates the optimal bin size 
\citep{Knuth}, but since we have too few galaxies it turned out to be not applicable for this problem.

In a previous chapter we plotted the surface densities of the blue neighbor galaxies, 
the red galaxies and all galaxies combined and found a huge increase in surface density of all galaxy types at bins 
closer than 150 kpc to the quasar. Especially high is the increase in blue 
gas-rich galaxies and AGN if one compares to the background of galaxies further 
away from the quasar (d $>$ 150 kpc). In case mergers are responsible for quasar activity, 
perhaps the AGN and blue galaxies could be secondary products from the same mergers? 

Other authors \citep[e.g.][]{Strand2008, Serber2006} also found an increase at close 
projected separations in the number of galaxies around quasars at low redshift. 
However, the large scales used in those studies (up to 2 Mpc) give a large 
contamination of non-physical pairs at projected separations larger than 150 kpc 
\citep{Nikolic} that increases steadily at greater separations, and with larger 
$|\Delta z|$ between the objects in the pair \citep{Strand2008}.

At the same time an increased density around the quasars does not necessarily imply 
a merger scenario, even though the temptation to assign this explanation to the result 
might be overwhelming. Monolithic collapse models can predict the formation of smaller 
objects around more massive ones when both smaller and larger objects collapse in the 
gravitational well of the same halo \citep[e.g.][]{Press}. The LCDM model itself predicts a 
large number of satellite galaxies around massive galaxies, something that is still difficult to know
whether it is the case or not. 

How could we test whether our surface densities are products of hierarchical structure formation or 
of monolithic collapse? One way would be to construct a sample of field quasars at low-redshifts 
(0.03 $<$ z $<$ 0.2) but that have no companion the closest 350 kpc. In case 
of merger scenario being the secret of quasar activity, these quasars should have an 
average luminosity that is fainter than in the sample with quasars that have companion galaxies. 
An second way is by using modelling to predict the number of lower-mass 
objects around quasars of the average luminosity in our quasar sample \citep[e.g.][]{Carlberg1990}. 
Similar attempts of investigating the clustering of quasars have been done \citep[e.g.][]{HopkinsCluster} 
where an excess of clustering could be shown, but where the results could not be clearly put in context 
of any given structure formation scenario. A third way, is by using spectral synthesis modeling for 
tracing the star formation histories of the companion galaxies. This could reveal signs of previous mergers.

\subsection{Comparison to galaxy-galaxy pairs}

One way of identifying if our observed phenomenons are related to the quasars themselves or
are normally present for all kind of galaxies, is by comparing to the environment of field galaxies.

We also generated surface and number density plots of field galaxy environments with similar absolute magnitude distribution to our quasars, in order to directly compare the environment of galaxies with same luminosity as our quasars and in the same kind of environment as the quasars namely galaxies in the field. Unfortunately, our sample size of field galaxy pairs was far too small to make any detailed or firm conclusions, and only the number and surface densities of the $|\Delta$z $|<$ 0.012 that are independent of the colors of the satellite galaxies can be used for comparison. From these plots we can see that there seem to be a small, but no great increase of satellite galaxies around the main field galaxies, indicating that they were probably not formed in the same ways as the quasars. A similar increase at close distances in galaxy-galaxy number density would have supported that any galactic objects have tendencies to merge sooner or later or that the pair of galaxies got simultanously created in the same halo. Our sample size
is although far too small to be reliable.

\begin{figure*}[htb!]
\centering
\includegraphics[width=16cm,height=10.5cm]{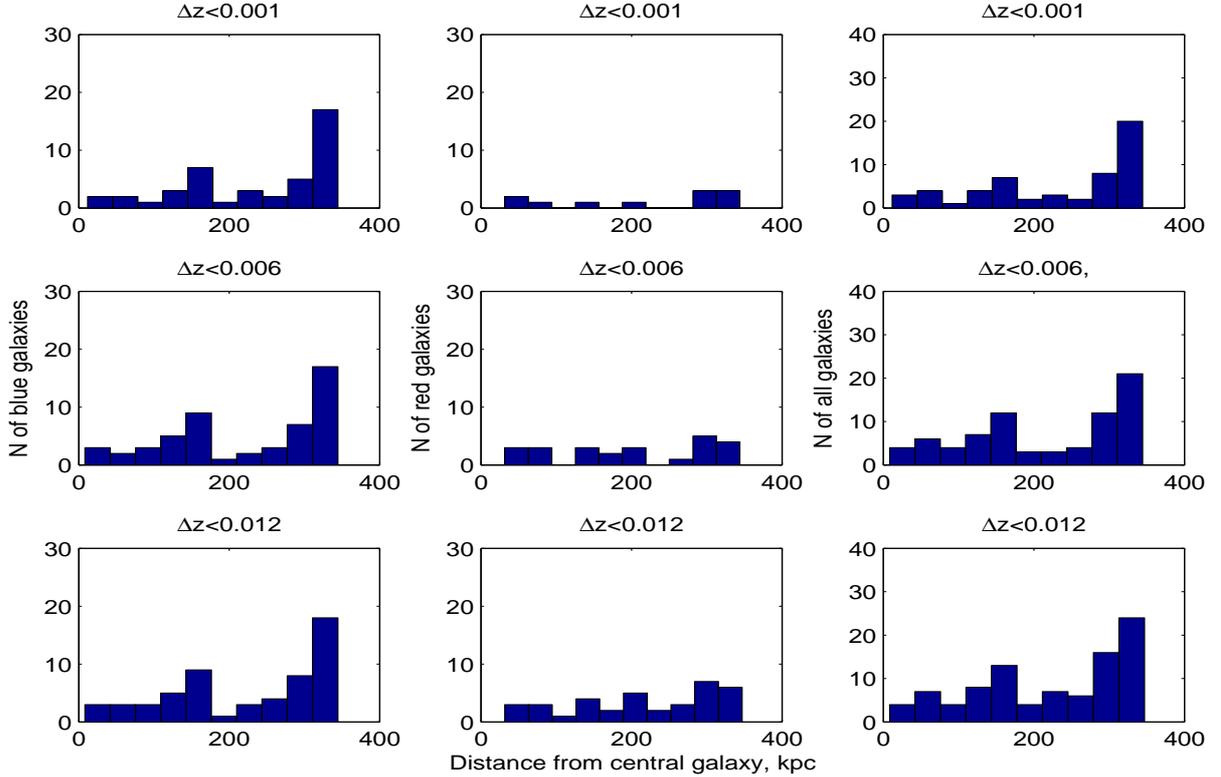}
\caption{Number distribution of galaxies around the central field galaxies. 
Left panel shows blue galaxies. Middle panel shows red galaxies. Right panel shows all non-AGN galaxies. Three different redshift difference cuts are pictured for the pairs.}
\label{ggNDtot}
\end{figure*}

\begin{figure*}
 \centering
   \includegraphics[width=16cm,height=10.5cm]{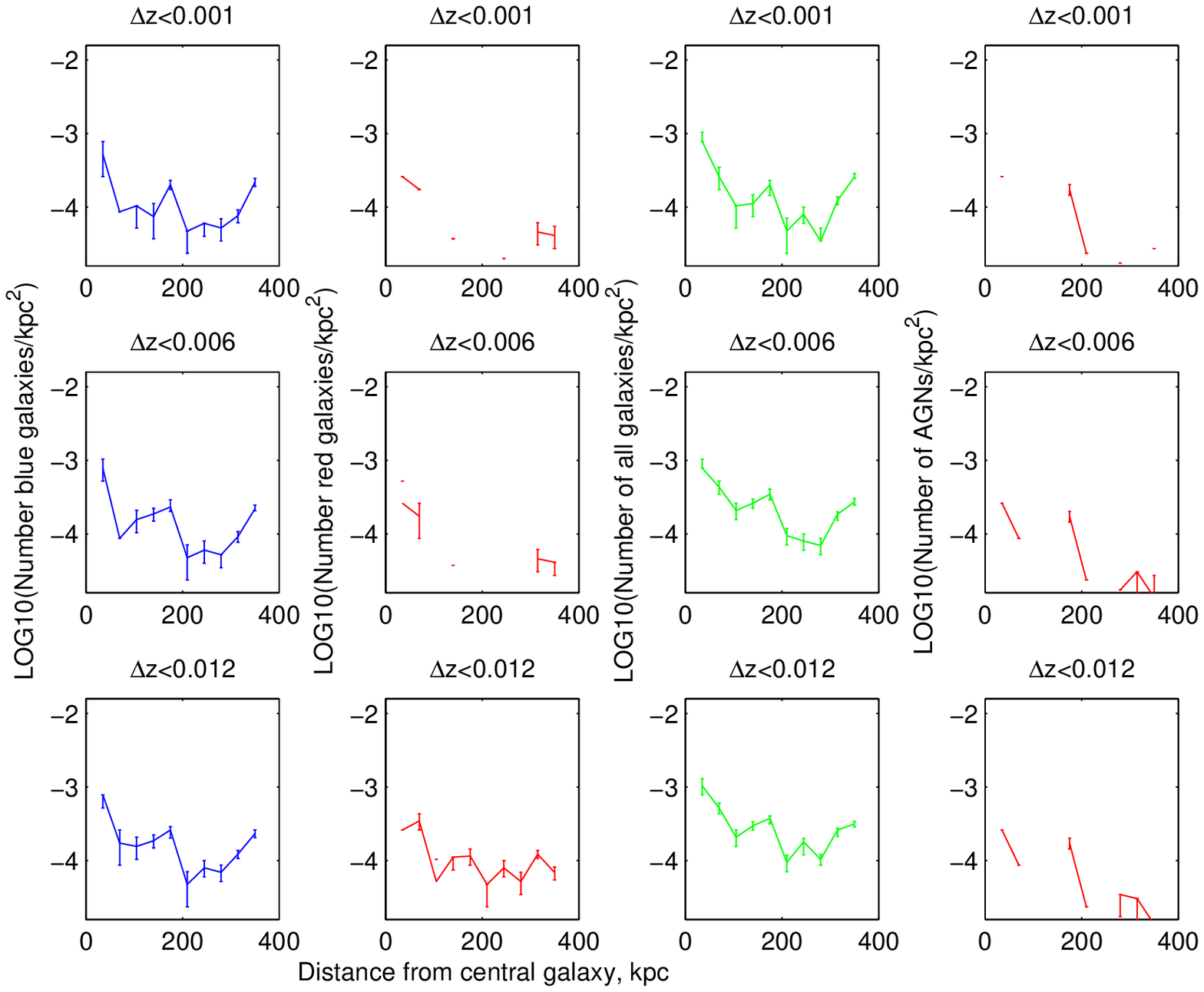}
   \caption{Annular surface densities of different types of galaxies around central field galaxies. Surface density distribution of galaxies. In the a) left panel, blue galaxies with $U_{e}-R_{e}$$<$2.2, in b) first middle, red galaxies, $U_{e}-R_{e}$$>$2.2 and in c) the right middle panel, all non-AGN galaxies d) AGN. Three different redshift difference cuts are pictured for the pairs.}
               \label{ggNFig85}%
     \end{figure*}

A study with galaxy-galaxy pairs in the SDSS \citep{Ellison} showed that in low-density
environments the number density distributions are different from the number distributions
of our quasar companions. This group used 5784 close galaxy pairs from the SDSS data release 4 (DR4) in different density environments with $zconf$ $>$ 0.7, projected distances within 80 kpc, redshift cut $|\Delta$z $|<$ 0.017 and a redshift $<$ 0.1. Their number density distribution behaved differently from our surface densities from the quasar-galaxy pairs. 
In fact the numbers of companion galaxies in Ellison et al. were dropping off at closer distances, contrary to our case with non-AGN galaxy companions to the quasars. If only comparing their bins at 35 kpc and 70 kpc with ours, one can see that while they had a remarkable difference with a decrease in the closest bin (N$_{35}$/N$_{70}$=2/3), we have no difference in the number of galaxies in these two bins. The general increase in number density around quasars seems to be specific for quasars, which might be explained by the fact that quasars need to surround themselves with fuel to survive.

Maybe it is not so surprising that the environments of quasars and field galaxies look different. Coldwell \& Lambas did study the surface density distributions of galaxies close to clusters, close to field galaxies and close to quasars. They could in their paper show that galaxies in the neighborhood of field galaxies have a lower density environment than galaxies around quasars do have. 

\subsection{Missing AGN and blue galaxies?}

Another curious thing about the surface density plots is the sudden drop of blue 
galaxies and AGN in the bin around 150 kpc. This could be a product of noisy data
or low number statistics, but at the same time we see a slight increase of red galaxies 
in the 35 kpc closer surface density bin. Could this mean that those missing blue galaxies got 
trapped by the gravitational field of the quasar and got their gas stripped off 
(hence became red) as they fell into circular orbits around the quasar? 

We remind the reader that the surface densities are shown on a logarithmic scale, and that it actually 
rather seems like the loss of blue galaxies in the 100-150 kpc bin is larger than the gain 
of red galaxies. Maybe some galaxies lost the gas when falling into circular orbits and got too faint to be 
detectable by the SDSS. Perhaps there is an unseen population of very faint galaxies 
in this bin. What if the galaxies simply not had a chance even to form any stars in their recent
star formation history, before getting trapped by the quasars? A study of Milky Way satellites \citep{Grcevich}
has revealed that within 270 kpc of the Milky Way, almost all satellite galaxies have lost their gas
and are spheroidal. This is strong support for tidal stripping processes within the distance 
range in our own study. What if something similar has happened to our blue non-AGN companion galaxies? 
Highly speculative scenarios would however need careful measurements of the velocity 
fields of the close quasar companion galaxies in order to follow how the dynamics of the gaseous content
of the companion galaxies changes around the gap.

Many thoughts remain as pure speculations until we find a way to increase our 
neighbor galaxy sample without losing precision in redshift estimates. It is difficult to know whether
the gap is real or simply an artefact from having too few quasar-galaxy pairs.

\subsection{Influence on environment?}

There is an overall lack of clear correlations in color, F([\ion{O}{iii}]/[\ion{O}{ii}]), accretion
rate F([\ion{O}{iii}]), SFR, dust and F([\ion{O}{iii}]/H$\beta$). Overall, it looks like there might be a small
increase in ionization and SFR at short projected separations as well as a very small
decrease of color and dust. What makes this to be a less likely conclusion is the two
deviating SFRs from H$\alpha$ and [\ion{O}{ii}]. We tried obtaining GALEX-data of 
FIR emission to get a third probe for star-formation in our galaxies, but FIR-emission data only existed for 27 of 
our neighbor galaxies, which was not usable in our study. A deviation as the 
one demonstrated for SFRs from H$\alpha$ and [\ion{O}{ii}] is likely to happen
either if the dust extinction and/or the oxygen abundance changes with the distance
to the quasars $or$ if we have more hidden AGN who could not be properly excluded
at short projected separations. Since we have an overall increase of the AGN surface density,
it is also likely that the number of hidden AGN is higher at close separations. This 
could lead to all the small trends observed: less dust (AGN can blow it away), no
direct increase in SFR from H$\alpha$, while a deviation from the SFR from the [\ion{O}{ii}],
and increase in ionization and a slightly bluer color. As we saw no dramatical changes
of the dust extinction or oxygen abundance with the distance to the quasars, the second
option with contamination of AGN seems more likely.

To try out the hypothesis, we tested by contaminating our sample with AGN and perform
LSAs. Doing this can generates a 2$\sigma$ decrease of $U_{e}-R_{e}$, up to 5$\sigma$ increase
of SFR from H$\alpha$ (while still disagreeing with the SFR from [\ion{O}{ii}]. This
is especially the case when analysing with the weighting w$\sim$ 1/($\sigma)^{2}$, which
means correlation analyses must be carried out with carefulness of both weighting and
a proper AGN exclusion.

One could argue that any correlations might be buried in the redshift-dependence of e.g.
SFR. We tried this by doing LSAs with different weighting models for SFR, ionization, color
and dust extinction and divide it with the surface area, but could still not find any significant
correlations.

The general lack of trends could also be a result from having a sample consisting of faint quasars. 
At the same time we would need to do a similar study with higher redshift quasars with galaxy companions
to know whether a possible influence could be luminosity-dependent. Everything can however turn 
out to be different when one does a study with only very strong high-redshift quasars, 
and maybe only is AGN feedback important for these objects. Since the present day cataloges of quasars have too few observable companion galaxies we cannot do a similar high-redshift study now, but simply have to wait for the future. At the moment the SDSS DR7 does not allow either study due to the far too small sample size
one obtains from the present sample and the lack of objects with spectroscopic redshifts above 0.2. 

To predict whether any change in the degree of ionization actually should be observed with our faint
quasars, one could estimate the radiation field from the quasars to see how the flux of 
high-energy photons changes at distances up to 350 kpc. We tried the alternative
way of constructing a field galaxy sample of the same luminosity range to see
how the satellites to these get influenced as a way to compare with, but the rather
small sample made it difficult to state any conclusive arguments.

Another thing to take into consideration might be that the corrections for internal extinction
we have used might be insufficient to show us any trends. We tried to perform
the extinction corrections as accurate as possible by correcting the colors for extinction dependent of 
angle of view of the galaxy and by adjusting emission lines for extinction according to Whitford \citep{Whitford}.
Something else that might affect the results, is that a correlation might be buried in the redshift evolution
of the parameters. We tried to take that into consideration by using the $|\Delta z|$ cuts, with
the lowest corresponding to the minimum spectroscopic redshift error.

The most important problem in the study was however still the small sample we had with many 
big individual errors in the fluxes we used, which caused many problems when investigating how
different properties could eventually vary with the projected distance from the quasars. We tried 
surpassing the problem by increasing our sample with the 2MASS catalogue, but 
only three more quasar-galaxy associations could be found there, which was not helpful for the problem.
We can only hope that the future will bring greater surveys with many more objects with careful
line measurements, and therefore wait and hope for the DR8 and DR9 to fulfill our expectations.

\section{Acknowledgements}

I would like to thank...

... my supervisor Michael J. Way (Mike) for providing inspiration, much help, many papers and 
practical solutions to problems I have met on my way.

...my supervisor Nils Bergvall for always sharing his huge wisdom and invaluable experience,
great inspiration and lots of help.

...the galaxy group: Elisabeth, Kjell, Thomas, \AA sa, Brady, Staffan, Phillip, Michael K., 
Johannes and Bertil for helping me with many technical things during my stay here, and for creating
a pleasant working atmosphere.

...Mart\'\i n L\'opez-Corredoira (Instituto de Astrof\'\i sica de Canarias, Spain) for 
introducing me to the mystery of quasars.

...the Department of Physics and Astronomy, Uppsala University (Sweden) for
making me feel welcomed, and for having wonderful Monday fika.

...lastly, but not least, my boyfriend J\'an, for always supporting me and for many times helped me 
killing goblins in both my Matlab code, in my project-related ideas and in many sonatas.

This research was done with help of the SDSS. Funding for SDSS-II has been provided by the 
Alfred P. Sloan Foundation, the Participating Institutions, the National Science Foundation, 
the U.S. Department of Energy, the Japanese Monbukagakusho, and the Max Planck Society.


\end{document}